\PassOptionsToPackage{unicode}{hyperref}
\PassOptionsToPackage{hyphens}{url}
\PassOptionsToPackage{dvipsnames,svgnames,x11names}{xcolor}
\documentclass[
  11pt,
  letterpaper,
]{article}
\usepackage{amsmath,amssymb}
\usepackage{iftex}
\ifPDFTeX
  \usepackage[T1]{fontenc}
  \usepackage[utf8]{inputenc}
  \usepackage{textcomp} 
\else 
  \usepackage{unicode-math} 
  \defaultfontfeatures{Scale=MatchLowercase}
  \defaultfontfeatures[\rmfamily]{Ligatures=TeX,Scale=1}
\fi
\usepackage[notextcomp]{stix2}
\ifPDFTeX\else
\fi
\IfFileExists{upquote.sty}{\usepackage{upquote}}{}
\IfFileExists{microtype.sty}{
  \usepackage[]{microtype}
  \UseMicrotypeSet[protrusion]{basicmath} 
}{}
\makeatletter
\@ifundefined{KOMAClassName}{
  \IfFileExists{parskip.sty}{%
    \usepackage{parskip}
  }{
    \setlength{\parindent}{0pt}
    \setlength{\parskip}{6pt plus 2pt minus 1pt}}
}{
  \KOMAoptions{parskip=half}}
\makeatother
\usepackage{xcolor}
\usepackage[margin=0.88in]{geometry}
\usepackage{longtable,booktabs,array}
\usepackage{calc} 
\usepackage{etoolbox}
\makeatletter
\patchcmd\longtable{\par}{\if@noskipsec\mbox{}\fi\par}{}{}
\makeatother
\IfFileExists{footnotehyper.sty}{\usepackage{footnotehyper}}{\usepackage{footnote}}
\makesavenoteenv{longtable}
\usepackage{graphicx}
\makeatletter
\def\maxwidth{\ifdim\Gin@nat@width>\linewidth\linewidth\else\Gin@nat@width\fi}
\def\maxheight{\ifdim\Gin@nat@height>\textheight\textheight\else\Gin@nat@height\fi}
\makeatother
\setkeys{Gin}{width=\maxwidth,height=\maxheight,keepaspectratio}
\makeatletter
\def\fps@figure{htbp}
\makeatother
\ifLuaTeX
\usepackage[bidi=basic]{babel}
\else
\usepackage[bidi=default]{babel}
\fi
\babelprovide[main,import]{american}

\def\languageshorthands#1{}
\usepackage{float}
\usepackage{needspace}
\usepackage{microtype}
\usepackage{cite}
\DeclareUnicodeCharacter{2212}{\ensuremath{-}}
\DeclareUnicodeCharacter{03A9}{\ensuremath{\Omega}}
\DeclareUnicodeCharacter{00B5}{\textmu}
\ifLuaTeX
  \usepackage{selnolig}  
\fi
\usepackage{bookmark}
\hypersetup{bookmarksdepth=3,bookmarksnumbered=true,citecolor=blue}
\IfFileExists{xurl.sty}{\usepackage{xurl}}{} 
\hypersetup{
  pdftitle={The Syncytial Mesh Model: Identifying Independent Astroglial Transport Geometry in Neuronal--Ionic Dynamics},
  pdfauthor={Andreu Ballús Santacana},
  pdflang={en-US},
  colorlinks=true,
  linkcolor={blue},
  filecolor={Maroon},
  citecolor={Blue},
  urlcolor={blue},
  pdfcreator={LaTeX via pandoc}}

\title{The Syncytial Mesh Model: Identifying Independent Astroglial Transport Geometry in Neuronal--Ionic Dynamics}
\author{Andreu Ballús Santacana}
\date{Department of Philosophy, Universitat Autònoma de Barcelona}

\begin{document}
\maketitle

\section*{Abstract}\label{abstract}
\addcontentsline{toc}{section}{Abstract}

Astroglial networks regulate neuronal excitability, but whether their transport geometry can be distinguished from extracellular transport remains an open inverse problem. The first foundational study of the Syncytial Mesh Model (SMM-I) formulates this question in terms of independent spatial operators. In calibrated passive dynamics, complete extracellular potassium transfers uniquely determine the weighted glial operator, despite severe practical ill-conditioning. Chemical transport and quasi-static electrical buffering exhibit intermediate topology-sensitive mobility regimes. Conservative electrodiffusive simulations identify axial electrical migration as a determinant of the sensitivity window; an independently motivated conductivity estimate predicts a conditional short-pulse maximum near 30 µm, corroborated by the full model. In a separate neutral KCl challenge, twelve-weight shared-support rivals fitted at four sites retain potassium and neuronal-voltage RMS discrepancies of 0.044--0.059 mM and 0.032--0.035 mV at three withheld sites under native mobility, and 0.112--0.121 mM and 0.079--0.098 mV at reduced mobility. Heterogeneous shared-support controls are recovered nearly exactly. Residuals persist in all-site fits; all four independent-target fits meet a strict numerical stationarity criterion and have positive-definite estimated objective Hessians, independently audited against forward simulations. Redistributing fixed total glial edge weight reveals topology sensitivity even with weak additional edges. Aggregate CA1 screening underestimates experimental uncoupling effects. Together these results establish conditional mathematical identifiability and a computational discrimination protocol for a falsifiable hypothesis of independent astroglial transport, not its biological confirmation.

\textbf{Correspondence:} andreu.ballus@uab.cat

\textbf{Keywords:} astroglial networks; Syncytial Mesh Model; potassium buffering; graph Laplacians; electrodiffusion; neuronal excitability; structural and practical identifiability; experimental design.

\section{Introduction}\label{introduction}

Neuronal connectivity specifies a principal organization of interactions between neurons, but it does not exhaust the spatial organization of the conditions under which those interactions take place. Extracellular ionic concentrations, membrane transport and astrocytic networks affect excitability, and the spatial routes through which these quantities change need not coincide with synaptic connections. The Syncytial Mesh Model (SMM) advances the hypothesis that a partially independent astroglial geometry contributes to the organization of neuronal responses. In this sense a glial network acts as a spatial controller of the conditions of neuronal activity, not necessarily as a separate generator of electromagnetic activity.

SMM develops the spatial architecture introduced in the Brain-Mesh preprint (version 1) \cite{ballus2024} and refined in the Syncytial Mesh preprint (version 3) \cite{ballus2026preprint}, building on the subsequent conceptual account of glial syncytial control \cite{ballus2026resonome}. This foundational study (SMM-I) restricts attention to a potassium-mediated realization (SMM-K), leaving the original calcium-wave and low-frequency resonome hypotheses to separate investigations.

Several research lineages motivate this architecture. The classic observations of Orkand and colleagues \cite{orkand1966}, the analysis of current-driven potassium transport by Gardner-Medwin \cite{gardnermedwin1983}, and its analytical and numerical extension by Chen and Nicholson \cite{chen2000} distinguish intracellular electrical spatial buffering from extracellular diffusion and local uptake. Experiments and reviews establish the role and limitations of astroglial coupling in ionic regulation, synaptic activity and plasticity \cite{kofuji2004,wallraff2006,pannasch2011,giaume2010,djukic2007}. Houades et al.~\cite{houades2008} demonstrated spatially structured astrocytic coupling in barrel cortex; Lallouette et al.~\cite{lallouette2014} showed computationally that changing gap-junction network topology can alter intercellular calcium-wave propagation even when cell positions and biophysical parameters are held fixed. Ma et al.~\cite{ma2016} showed that electrical coupling promotes syncytial isopotentiality and potassium clearance. Thus neither the importance of topology in astroglial signaling nor the distinction between electrical and chemical transport is a new claim of SMM.

A complementary functional lineage establishes that astrocytes can organize neuronal activity at the network level. Cahill et al.~\cite{cahill2024} described network-level astrocytic encoding of local neurotransmitter signals; Visser et al.~\cite{visser2024} showed that disrupting astroglial gap-junction connectivity alters spatial visual representations and associated behavior in the mouse superior colliculus. Cooper et al.~\cite{cooper2026} (first circulated as a preprint \cite{cooper2025preprint}) demonstrate selective, plastic, long-range astrocytic networks whose architecture can diverge substantially from neuronal projections. Their tracer measurements demonstrate anatomical and molecular connectivity, not a potassium conductance matrix. Oliveira et al.~\cite{oliveira2026} synthesize evidence for astrocytic functional units across scales, and Zhang et al.~\cite{zhang2026} explicitly discuss the emerging astrocytic connectome. A cortical astrocytic territory may contain numerous synapses \cite{halassa2007}; regional anatomy, subcellular morphology and micrometric ionic transport graphs are related but distinct levels of description.

The functional effect of coupling is dependent on preparation, intensity and timescale. Breithausen et al.~\cite{breithausen2020} found a limited acute gap-junction contribution to ordinary CA1 stratum radiatum potassium transients but a clearer contribution when highly localized extracellular loads produced elevations on the order of 10 mM or greater. Bojovic et al.~\cite{bojovic2025} found that interference with Kir channels and astrocytic coupling alters neuronal excitability during sustained cortical activity; their results also indicate compensation by other mechanisms. Neither finding requires coupling to dominate all physiological potassium clearance. Together they motivate experiments in which the delivered ionic load, spatial profile, and duration are independently measured rather than treating the contribution of a glial network as a fixed fraction of any neuronal response.

The modeling ingredients are also established. Coupled neuronal--ionic models include potassium-mediated feedback and regime transitions \cite{cressman2009,ullah2009}. Halnes and colleagues and Sætra and colleagues have developed increasingly complete electrochemical models of neuron--extracellular--glia interactions \cite{halnes2013,saetra2021,saetra2023,saetra2024}. The two-module microcircuit model of Cafiso et al., first circulated in June 2026 \cite{cafiso2026preprint} and published in November 2026 \cite{cafiso2026journal}, shows how local and syncytial astrocytic buffering can regulate synchronization and pathological excitability through extracellular potassium. Nakatani and De Schutter \cite{nakatani2026} investigate the dependence of astrocytic membrane responses on stimulus scale and intracellular electrical compartmentalization, showing why somatic recordings can miss local peripheral depolarization. These studies address coupling strength and synchronization, or within-cell electrical response geometry, complementing the present question of identifying intercellular transport support from extracellular and neuronal observations. General methods for compartmental systems, dynamical-network identification, profile likelihood and optimal discrimination design are likewise established \cite{jacquez1993,raue2009,atkinson1975,gevers2017,hendrickx2019,bortner2022,pant2018,liu2026}.

What distinguishes SMM-K is the problem addressed by combining these ingredients: given extracellular and neuronal alternatives, can the \emph{geometry} of the glial transport operator be recovered, and can a preparation be shown to require connections outside the support of extracellular transport? This is stronger than detecting a difference after gap-junction blockade. A change in local uptake, global diffusivity, membrane excitability, or a hidden feedback process may account for such a difference without requiring a distinct spatial organization. A meaningful test must compare independently constrained model families under matched inputs, observations and uncertainty.

This article therefore makes three connected contributions. It first derives exact implications of a distinct glial operator in a calibrated passive model, including conditions for recovery and sensitivity. It then calculates both the conditioning of that inverse problem and finite separation from shared-support alternatives under an explicit experimental design. Finally, it tests the graph representation against an independent conservative electrodiffusive model, identifies the dependence of the topology-sensitive spatial window on axial electrical migration, examines dose-dependent nonlinear contrast and distal current mechanisms, and derives an independently informed electrical parameter constraint. Under explicit conductivity constraints, we first investigate sparse source coverage and then fit twelve independently adjustable nearest-neighbor glial and extracellular weights at four sites, predicting three further sites without refitting. Heterogeneous shared-support controls test whether the same identification pipeline recovers a reference that is actually in the fitted family. All-source fits test whether the mismatch persists in the obtained candidates when no stimulus is withheld, and a constant-total-weight sweep separates the effects of glial edge redistribution from overall coupling. These comparisons distinguish mathematical identifiability, the numerical performance of a specified rival-fitting procedure, and experimental identification of an actual biological transport geometry. The original neuronal-population, metabolic, stability and EEG audits are retained in the supplementary material because they constrain the physiological interpretation, but the theory is organized here around its distinctive prospective test.

\section{Model and hypothesis}\label{model-and-hypothesis}

\subsection{Three spatial organizations}\label{three-spatial-organizations}

Let \(V=\{0,\ldots,N-1\}\) index spatial compartments. We distinguish a neuronal interaction operator \(W_n\), which may be directed; an extracellular transport Laplacian \(L_e\); and an astroglial transport Laplacian \(L_g\). For a symmetric ionic graph with nonnegative weights \(w_{ij}\),

\[L=B\operatorname{diag}(w)B^\top=\sum_{\{i,j\}\in E}w_{ij}(e_i-e_j)(e_i-e_j)^\top,\]

where \(B\) is an oriented incidence matrix and \(e_i\) is the \(i\)th coordinate vector. Edge orientation is bookkeeping; flux is determined by gradients. The Laplacian is positive semidefinite and annihilates the constant vector. A nonnegative symmetric ionic edge is not the same physical quantity as a directed synaptic weight. Distinct organizations may differ by edge support, relative weights, or both.

\begin{figure}
\centering
\includegraphics[width=0.95\textwidth,height=\textheight]{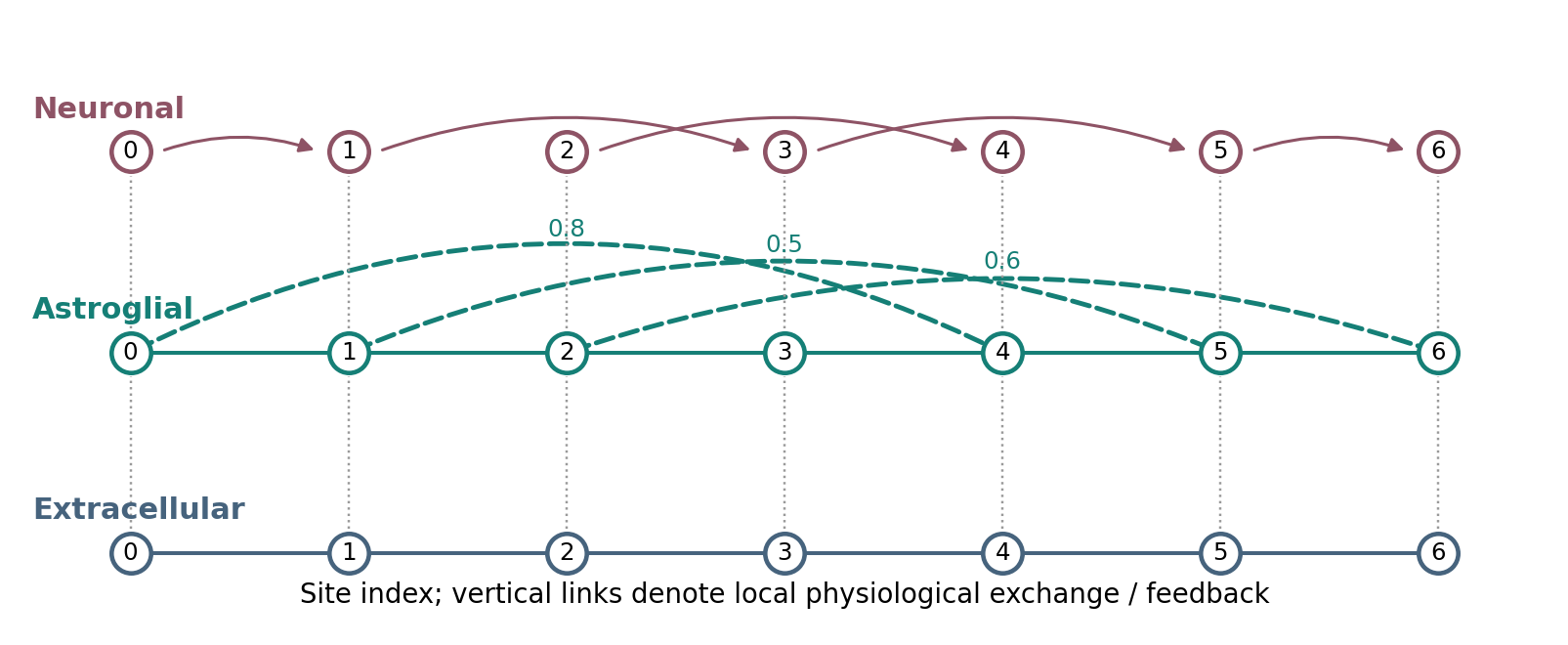}
\caption{Illustrative three-layer SMM architecture. Solid horizontal links in the ionic layers denote the shared path; dashed curves denote additional astroglial edges. The directed neuronal links are schematic, not the synaptic graph used in the full electrodiffusive simulations, where synapses were disabled. This is an abstract transport network, not an anatomical reconstruction.}
\end{figure}

A graph on spatial compartments does not have to correspond to a literal graph of individually identified astrocytes. It can describe a coarse-grained transport operator at a specified scale. At that scale, a meaningful comparison requires the same sites and a common spatial embedding or independently defined operational support. The analysis below uses a symmetric nonnegative graph as a restricted first model; changing support or coarse-graining changes the hypothesis to be tested.

\subsection{Potassium-mediated passive realization}\label{potassium-mediated-passive-realization}

Let \(k,a\in\mathbb R^N\) denote extracellular and astroglial excess potassium concentrations in mM, relative to a reference state. Let \(\phi_e,\phi_a>0\) be the respective volume fractions. With homogeneous coefficients, the local and spatial deviation balances are

\[\begin{aligned}
\phi_e\dot k &= J_K^n-(\kappa_e+\kappa_N)k+\kappa_a a-D_eL_e k+u_k,\\
\phi_a\dot a &= \kappa_e k-\kappa_a a-D_gL_g a+u_a .
\end{aligned}\]

Here \(\kappa_e\) and \(\kappa_a\) describe effective uptake and return, \(\kappa_N\) a local clearance slope, \(J_K^n\) excess neuronal release, and \(u_k,u_a\) externally specified inputs in mM of tissue-volume-equivalent potassium per second. The \(\kappa\) coefficients have units s\(^{-1}\) in this convention. If \(D_e,D_g\) are in m\(^2\)/s, \(L_e,L_g\) carry m\(^{-2}\). Specifically, for nominal spacing \(d\), write \(L_r=d^{-2}\mathcal L_r\), where \(\mathcal L_r\) is a dimensionless weighted graph Laplacian. We denote the weighted glial operator in s\(^{-1}\) by \(G=D_gL_g\) and the extracellular operator by \(E=D_eL_e\).

The passive system sets \(J_K^n=0\) to isolate transport, with \(\kappa_e,\kappa_a>0\) and \(\kappa_N\geq0\). It is a model of deviations, not an absolute ionic inventory law for a full tissue. When \(\kappa_N=0\) and the external inputs vanish, the summed potassium inventory \(\mathbf1^\top(\phi_e k+\phi_a a)\) is conserved; otherwise its change is governed by the specified sources and clearance. The two-compartment model is useful precisely because it provides a tractable conditional identification problem. It does not represent membrane capacitance, electrical migration, variation in volumes, all ionic species or the feedback of neuronal spikes. Its connection to complete physiology is tested below, rather than asserted by matching the names of parameters.

For homogeneous passive coefficients the weighted quadratic energy

\[\mathcal E=\tfrac12\big(\kappa_e\phi_e\|k\|^2+\kappa_a\phi_a\|a\|^2\big)\]

satisfies, in the unforced passive system,

\[\begin{aligned}
\dot{\mathcal E}={}&-\|\kappa_e k-\kappa_a a\|^2-\kappa_e\kappa_N\|k\|^2\\
&-\kappa_e D_e k^\top L_e k-\kappa_a D_g a^\top L_g a\le0 .
\end{aligned}\]

For \(\kappa_N>0\) the inequality is strict away from the origin and the finite-dimensional linear system is asymptotically stable. For zero clearance, neutral inventory modes may remain. This dissipation result does not require commuting graphs and does not imply stability of a closed neuronal feedback loop.

\subsection{Distinctive hypothesis and nested alternatives}\label{distinctive-hypothesis-and-nested-alternatives}

The broad SMM hypothesis is that the control geometry of a non-neuronal network can be partially independent of neuronal and extracellular geometries. A stronger, preparation-specific prediction for SMM-K is that the weighted glial operator contains functionally non-negligible edges outside the support \(E_e\) of an independently established extracellular transport operator. Define the restricted family

\[\mathcal H_0=\left\{\sum_{\{i,j\}\in E_e}w_{ij}(e_i-e_j)(e_i-e_j)^\top:\ w_{ij}\geq0\right\}.\]

The alternative contains a prespecified set of possible additional glial edges, with independently justified constraints on their weights. The prediction is \(G\notin\mathcal H_0\), to a practically and biologically specified tolerance. This excludes not merely a scaled extracellular path but \emph{every} weighted Laplacian with the specified shared support. Unequal weights on the same support can still constitute a distinct geometry under the broader architecture; the extra-edge prediction is a deliberately sharper test. The biological attribution also requires evidence that the inferred coupling is astroglial rather than an arbitrary unobserved controller.

\section{Mathematical results}\label{mathematical-results}

\subsection{Exact recovery without commuting graphs}\label{exact-recovery-without-commuting-graphs}

Consider initially unperturbed glia and extracellular potassium with calibrated extracellular drive \(u_k=u\), \(u_a=0\), no neuronal feedback, and homogeneous local parameters. For real \(s>0\) define

\[\begin{aligned}
P(s)&=(\phi_e s+\kappa_e+\kappa_N)I+E,\\
Q(s)&=(\phi_a s+\kappa_a)I+G,\\
c&=\kappa_a\kappa_e .
\end{aligned}\]

Laplace transformation gives the block system

\[\begin{pmatrix}P&-\kappa_a I\\-\kappa_e I&Q\end{pmatrix}\begin{pmatrix}\widehat k\\\widehat a\end{pmatrix}=\begin{pmatrix}\widehat u\\0\end{pmatrix}.\]

\textbf{Proposition 1 (conditional graph recoverability).} With the stated assumptions, the full extracellular transfer matrix \(H\) satisfies

\[\widehat k=H(s)\widehat u,\qquad H(s)=(P-cQ^{-1})^{-1}.\]

If \(P,c,\phi_a,\kappa_a\) are known and \(H(s)\) is measured exactly with full input and output coverage, then

\[\boxed{G=c(P-H^{-1})^{-1}-(\phi_a s+\kappa_a)I.}\]

\emph{Proof.} Since \(Q\succeq(\phi_a s+\kappa_a)I\succ0\), elimination of \(\widehat a\) is valid. Moreover,

\[P-cQ^{-1}\succeq\left(\phi_e s+\kappa_N+\frac{\kappa_e\phi_a s}{\phi_a s+\kappa_a}\right)I\succ0.\]

Thus \(H\) exists and \(P-H^{-1}=cQ^{-1}\) is positive definite. Rearranging gives the recovery expression. Neither elimination nor the inequalities exchange the order of the two graph operators. \(\square\)

The identified quantity is the product \(D_gL_g\): multiplying \(D_g\) by a positive scalar and dividing all glial weights by the same scalar leaves the transfer unchanged. A conductivity measurement or graph weight convention is necessary to distinguish these factors. For partial source and observation matrices \(B_k,C_k\), it is generally not legitimate to invert \(C_kHB_k\) as if it were \(H\). Unknown neuronal feedback or incorrectly specified local exchange invalidates the same reconstruction.

A further consequence follows from the high-frequency expansion

\[\begin{aligned}
H^{-1}(s)={}&\phi_e s I+(\kappa_e+\kappa_N)I+E-\frac{c}{\phi_a s}I\\
&+\frac{c}{\phi_a^2s^2}(\kappa_a I+G)+O(s^{-3}).
\end{aligned}\]

The off-diagonal coefficients of orders \(s^0\) and \(s^{-2}\) separate extracellular and glial contributions under the stated homogeneous calibration. In a joint identification problem the required scalar factors must also be identified. This expansion is an algebraic result, not a recommended estimator of small high-order coefficients from noisy data.

For \(G=D_gL_g\) with fixed \(L_g\), differentiating the transfer gives

\[\frac{\partial H}{\partial D_g}=-cH Q^{-1}L_gQ^{-1}H\preceq0.\]

The quadratic-form sign follows from symmetry and positive semidefiniteness, regardless of whether \(L_e\) and \(L_g\) commute. This is a Loewner-order statement for real positive \(s\), not an assertion that every sitewise time-domain transient decreases with coupling.

\subsection{Conditioning of the inverse map}\label{conditioning-of-the-inverse-map}

Exact recovery does not establish that the data needed for it can be measured with sufficient precision. With the other parameters held fixed, perturb \(H\) by a symmetric matrix \(Z\). The first differential of the recovered operator is

\[\delta G=-c^{-1}QH^{-1}ZH^{-1}Q.\]

Indeed, \(\delta H^{-1}=-H^{-1}ZH^{-1}\) and differentiating \(H^{-1}=P-cQ^{-1}\) yields \(\delta H^{-1}=cQ^{-1}(\delta G)Q^{-1}\). With \(A=QH^{-1}\), the differential is \(-AZA^\top/c\). Its exact induced Frobenius norm on unrestricted symmetric perturbations is \(\|A\|_2^2/c\): the upper bound is submultiplicativity, and a unit-Frobenius rank-one perturbation \(Z=vv^\top\) along a leading right singular vector \(v\) attains equality. For \(G\ne0\) the relative local condition number is therefore

\[\mathcal K(s)=\frac{\|QH^{-1}\|_2^2\,\|H\|_F}{c\,\|G\|_F}.\]

This is a worst-case local gain for the exact unconstrained map, not a variance bound for a structured estimator. Constraints on Laplacian support, positivity, reciprocity and multiple independent frequencies can reduce effective amplification. The expression provides a direct mathematical bridge between the recovery theorem and the need for calibrated forward fitting.

\subsection{Source-route dependence and uniform blindness}\label{source-route-dependence-and-uniform-blindness}

Let \(F\) be the \(2N\)-dimensional passive generator, and \(b=(u/\phi_e,0)^\top\) the input vector for a constant extracellular step switched on at \(t=0\). For zero initial deviations,

\[x(t)=\sum_{j\geq0}\frac{t^{j+1}}{(j+1)!}F^j b,\qquad x=(k,a)^\top.\]

\textbf{Proposition 2 (pathway order).} In the stated passive model the first possible nonzero extracellular response sensitivities to the two scalar diffusion coefficients are

\[\begin{aligned}
\partial_{D_e}k(t)&=-\frac{L_eu}{2\phi_e^2}t^2+O(t^3),\\
\partial_{D_g}k(t)&=-\frac{\kappa_a\kappa_e L_gu}{24\phi_e^2\phi_a^2}t^4+O(t^5).
\end{aligned}\]

For a strictly downstream neuronal state \(n\), with \(\dot n=A_nn+B_nk\), \(n(0)=0\), and no direct glial drive,

\[\partial_{D_g}n(t)=-\frac{\kappa_a\kappa_e B_nL_gu}{120\phi_e^2\phi_a^2}t^5+O(t^6).\]

For a direct constant glial input \(v\) instead of extracellular drive, the corresponding first possible extracellular term is

\[\partial_{D_g}k(t)=-\frac{\kappa_a L_gv}{6\phi_e\phi_a^2}t^3+O(t^4),\]

and the downstream neuronal sensitivity acquires one additional order of time.

\emph{Proof.} The extracellular operator acts immediately on the populated upper block of \(Fb\). The glial operator must first receive a local exchange contribution in the lower block and then return its effect to the upper block. Its first upper-block term is in \(\partial_{D_g}(F^3b)\), with coefficient \(-\kappa_a\kappa_e L_gu/(\phi_e^2\phi_a^2)\). With direct glial forcing the lower block is already populated at input, so the first upper-block term is in \(\partial_{D_g}(F^2b)\). Integrating these sensitivities through a strictly downstream first-order neuronal system adds a time power and the corresponding factorial. This is a path-length result, and coefficients can vanish for specific sources or projections. \(\square\)

\textbf{Proposition 3 (uniform blindness).} If the initial conditions and inputs are uniform and local coefficients homogeneous, then the passive solution stays in \(\operatorname{span}(\mathbf1)\oplus\operatorname{span}(\mathbf1)\) and is independent of both graph mobilities. The proof is invariance of this subspace, because \(L_e\mathbf1=L_g\mathbf1=0\). No commutativity is required. Heterogeneous coefficients or boundaries can break this invariance.

The onset hierarchy and uniform control constrain an experiment, but do not distinguish SMM from another two-compartment diffusion-with-exchange model with the same causal path length. They should be used as tests of the model scaffold, rather than the main biological falsifier. In particular, the onset result presumes a purely chemical forcing in the passive model; electrical buffering can open a faster pathway even under charge-neutral KCl forcing, because a local change in potassium concentration changes the local reversal potential. Charge-separating membrane stimulation can additionally generate a direct electrical footprint. The passive expansion is not a general onset theorem for electrodiffusion.

\subsection{Intermediate topology-sensitive mobility}\label{intermediate-topology-sensitive-mobility}

Consider two distinct connected symmetric glial Laplacians on the same sites, identical extracellular and local coefficients, zero initial glial deviation, and a bounded extracellular source. Multiply each glial transport operator by \(\mu>0\).

\textbf{Proposition 4 (quantitative mobility window for chemical transport).} Let \(G_1,G_2\) be connected symmetric glial Laplacians with positive gaps \(\lambda_1,\lambda_2\) on \(\mathbf 1^\perp\). Fix a common extracellular operator \(E\), homogeneous local coefficients, common initial conditions \((k_0,0)\), and an extracellular input \(u\) with \(\sup_{0\le t\le T}\|u(t)\|\le U\). For the passive responses with glial operators \(\mu G_i\), write \(\|\Delta k_{12}\|_{\infty,T}=\sup_{0\le t\le T}\|k_1(t)-k_2(t)\|\). Then

\[\|\Delta k_{12}\|_{\infty,T}\le\min\left\{C_T\mu,\frac{\kappa_a\kappa_e K_T}{\kappa_e+\kappa_N}\left(\frac{1}{\kappa_a+\mu\lambda_1}+\frac{1}{\kappa_a+\mu\lambda_2}\right)\right\},\]

where \(M_T=\sqrt{\kappa_e\phi_e}\|k_0\|+T\sqrt{\kappa_e/\phi_e}U\), \(K_T=M_T/\sqrt{\kappa_e\phi_e}\), and \(C_T=TM_T\|G_1-G_2\|/(\phi_a\sqrt{\kappa_e\phi_e})\). Thus the contrast is \(O(\mu)\) at zero mobility and \(O(\mu^{-1})\) at infinite mobility. A nonzero contrast attains a positive maximum at a finite intermediate mobility.

\emph{Proof.} In coordinates \(z=(\sqrt{\kappa_e\phi_e}\,k,\sqrt{\kappa_a\phi_a}\,a)\) the passive homogeneous generator is symmetric negative semidefinite by the energy identity, so its semigroup is contractive uniformly in \(\mu\). The forcing norm is at most \(\sqrt{\kappa_e/\phi_e}U\); hence \(\|z_i(t)\|\le M_T\). The generators differ only in their glial block, by \(-\mu(G_1-G_2)/\phi_a\). Duhamel's formula bounds \(\|z_1(t)-z_2(t)\|\) by \(T\mu\|G_1-G_2\|M_T/\phi_a\), giving the first term.

With \(\Pi=\mathbf 1\mathbf 1^\top/N\), the nonuniform glial component satisfies

\[\phi_a\dot a_{i,\perp}=\kappa_e(I-\Pi)k_i-(\kappa_a I+\mu G_i)a_{i,\perp}.\]

The spectral gap and zero initial load give \(\|a_{i,\perp}\|_{\infty,T}\le\kappa_e K_T/(\kappa_a+\mu\lambda_i)\). The spatial means of \(k_i\) and \(a_i\) satisfy identical averaged equations, since both Laplacians annihilate constants. Therefore the extracellular difference obeys \(\phi_e\dot{\Delta k}=-[(\kappa_e+\kappa_N)I+E]\Delta k+\kappa_a(a_{1,\perp}-a_{2,\perp})\) with zero initial difference. The extracellular semigroup decays at rate at least \((\kappa_e+\kappa_N)/\phi_e\). Integrating its norm gives the second term. Continuity in \(\mu\), the two endpoint bounds and positivity at some intermediate \(\mu\) prove the maximum statement. \(\square\)

The positive maximum is conditional on a discriminating input, graph pair and observation; it need not be unique. Disconnected graphs can have different nullspaces and different limiting operators. Neither this theorem nor the maximum of a relative norm implies that glia become the dominant potassium transport route. It is a statement about the information carried by internal graph shape. In the source-reconciled parameterization below \(D_g/D_e\simeq0.218\) at nominal mobility, independently of spacing.

\subsection{Electrical buffering on a graph}\label{electrical-buffering-on-a-graph}

Current-driven spatial buffering \cite{orkand1966,gardnermedwin1983,chen2000} provides a complementary, quasi-static electrical limit. Consider a frozen-concentration two-conductor network with connected symmetric glial and extracellular conductance Laplacians \(A_g=mG_c\) and \(A_e=E_c\), and a uniform positive membrane slope conductance density \(g_m\). The axial operators have the same conductance-per-volume units as \(g_m\). Let \(e\) denote a prescribed vector of perturbations of the effective membrane reversal voltage, \(v=\varphi_g-\varphi_e\), and let \(j=g_m(v-e)\) denote outward membrane current density. In the absence of imposed net electrical current, current balance gives

\[A_g\varphi_g+j=0,\qquad A_e\varphi_e-j=0.\]

On the mean-zero subspace \(H=\mathbf1^\perp\) the restricted Laplacians are positive definite and invertible, and \(\mathbf1^\top j=0\). Define the parallel sum on \(H\) by

\[B_m=\left[(mG_c)^{-1}+E_c^{-1}\right]^{-1}.\]

Eliminating the potentials gives \(j=-B_m v\), and therefore

\[v=g_m(B_m+g_m I)^{-1}e,\qquad j=-g_m B_m(B_m+g_m I)^{-1}e.\]

These expressions apply to the projected mean-zero perturbation; the spatially uniform reversal-voltage component induces the same uniform membrane-voltage shift and zero axial or membrane current. No commutativity of the two graph Laplacians is assumed. The equations represent a linearized frozen-concentration electrical response, not the time evolution of the full ionic model.

\textbf{Proposition 5 (quantitative electrical graph-shape window).} On \(H=\mathbf 1^\perp\), let \(G_1,G_2,E_c\) be positive definite restrictions of connected symmetric conductance Laplacians and \(g_m>0\). Write \(B_i(m)=[(mG_i)^{-1}+E_c^{-1}]^{-1}\) and \(J_i(m)=-g_m B_i(m)(B_i(m)+g_m I)^{-1}\). In the Euclidean operator norm,

\[\|J_1(m)-J_2(m)\|\le\|G_1^{-1}-G_2^{-1}\|\min\{m\|G_1\|\|G_2\|,\ m^{-1}\|E_c\|^2\}.\]

For distinct \(G_1,G_2\), these endpoint orders are sharp for the operator contrast: it is \(\Theta(m)\) as \(m\downarrow0\) and \(\Theta(m^{-1})\) as \(m\to\infty\). The norm attains a positive maximum at an intermediate mobility. For a particular prescribed \(e\) the same upper bound holds multiplied by \(\|e\|\); its leading contrast may vanish under that input or observation.

\emph{Proof.} The parallel-sum inequalities \(0\prec B_i\preceq mG_i\) and \(B_i\preceq E_c\) follow by inversion of positive definite matrices. The inverse difference identity gives

\[B_1-B_2=m^{-1}B_1(G_2^{-1}-G_1^{-1})B_2.\]

Using both parallel-sum inequalities yields the stated minimum of two bounds for \(\|B_1-B_2\|\). The current map \(J_i=-g_m I+g_m^2(B_i+g_mI)^{-1}\) is 1-Lipschitz in \(B_i\) by the resolvent identity, because \(\|(B_i+g_mI)^{-1}\|\le g_m^{-1}\). At the low-mobility endpoint \(B_i=mG_i+O(m^2)\), so \(J_1-J_2=-m(G_1-G_2)+O(m^2)\). At high mobility \(B_i=E_c-m^{-1}E_cG_i^{-1}E_c+O(m^{-2})\). Setting \(Q=(E_c+g_m I)^{-1}\) gives

\[J_1-J_2=m^{-1}g_m^2Q E_c(G_1^{-1}-G_2^{-1})E_c Q+O(m^{-2}).\]

The leading matrix at each endpoint is nonzero when \(G_1\ne G_2\), since \(E_c,Q\) are invertible. This proves the sharp orders in operator norm, without commutation. Continuity and the endpoint limits imply the intermediate maximum. \(\square\)

In a uniform one-dimensional continuum with axial tissue conductivities \(m\sigma_g\) and \(\sigma_e\), the corresponding voltage difference satisfies \(v''=g_m[(m\sigma_g)^{-1}+\sigma_e^{-1}](v-e)\) away from imposed current sources. Its homogeneous decay length is

\[\ell(m)=\left[g_m\left(\frac1{m\sigma_g}+\frac1{\sigma_e}\right)\right]^{-1/2}.\]

The electrical length saturates at \(\sqrt{\sigma_e/g_m}\) as \(m\to\infty\). There is also a distinct intrinsic-scale window in this reduced electrical problem. If both conductance-density Laplacians scale as \(d^{-2}\), with dimensionless graphs, \(m\) and \(g_m\) fixed, then \(B_m(d)=d^{-2}B_m^0\) on \(H\), and the membrane-current operator is \(-g_m I+g_m^2(d^{-2}B_m^0+g_m I)^{-1}\). As \(d\downarrow0\), it tends to \(-g_m I\) with a graph-dependent correction \(O(d^2)\); as \(d\to\infty\), it tends to zero with leading term \(-d^{-2}B_m^0\). Hence a nonzero absolute contrast between two connected graphs also achieves an intermediate maximum as intrinsic scale varies at fixed prescribed reversal perturbation. This follows from the same continuity argument as Proposition 5 and does not require simultaneous diagonalization. Neither result determines the optimum of the normalized, time-dependent full ionic response, for which the reversal perturbation itself changes with \(d\). The electrical current map can be nonlocal in the reversal-voltage perturbation. Under a neutral localized KCl step, \(e(t)=O(t)\) near onset, so this quasi-static current map can produce an \(O(t)\) remote current and an \(O(t^2)\) integrated ionic contribution when the relevant coefficient is nonzero. This is a possible faster pathway than Proposition 2, not a universal onset exponent for EDNM, which retains capacitive dynamics, multiple ions and concentration-dependent fluxes.

\section{Numerical methods: a matched experiment}\label{numerical-methods-a-matched-experiment}

\subsection{Graphs, kinetics and input convention}\label{graphs-kinetics-and-input-convention}

The seven-site experiments use the extracellular path \(0-1-\cdots-6\), with unit nearest-neighbor edge weights. Write its dimensionless Laplacian as \(\mathcal L\), and define the additional-edge Laplacian

\[\mathcal C=0.8b_{04}b_{04}^\top+0.5b_{15}b_{15}^\top+0.6b_{26}b_{26}^\top,\]

where \(b_{ij}=e_i-e_j\). The passive family is

\[E=D_e\mathcal L/d^2,\qquad G=\mu D_g(\mathcal L+g\mathcal C)/d^2.\]

The target \(g\) controls the three chords relative to the common path; \(g=0\) is the shared path and \(g=1\) the nominal alternative. The separate mobility multiplier \(\mu\) allows overall strength to change. Chord weights are effective weights at the chosen nominal discretization length, not conductances derived from the physical lengths of the long links. The full EDNM graph adapter uses the same weighted adjacency convention. No measured astrocyte geometry or topology is assumed. Changing \(d\) rescales the intrinsic tissue graph, including the physical lengths of the chords; it is not a change of electrode spacing on one fixed tissue. The three four-step chords of weights 0.8, 0.5 and 0.6 correspond to own-length effective diffusivities 12.8, 8 and 9.6 times the nominal glial diffusivity, respectively, when represented as ordinary diffusion across their entire physical lengths. They are deliberately synthetic high-conductance conduits, not anatomical measurements. Useful dimensionless groups include \(D_g/(d^2\kappa_a)\), \(D_e/(d^2\kappa_a)\), \(\kappa_e/\kappa_a\), \(\kappa_N/\kappa_a\), the volume fractions, mobility, and \(\kappa_a T\) for protocol duration \(T\).

\Needspace{19\baselineskip}

Table 1. Fixed passive parameters for the source-reconciled experiment. The exchange and transport coefficients are inherited from the earlier explicit source-reconciliation exercise; the clearance slope is inherited separately and the volume fractions are those of the EDNM resting state. The combination is a specified synthetic parameter set, not an independently fitted passive model of the seven-site EDNM trajectories.

\begin{longtable}[]{@{}lrl@{}}
\toprule\noalign{}
Parameter & Value & Units \\
\midrule\noalign{}
\endhead
\bottomrule\noalign{}
\endlastfoot
\(\phi_e\) & 0.2 & dimensionless \\
\(\phi_a\) & 0.4 & dimensionless \\
\(\kappa_e\) & 1.29629 & s\(^{-1}\) \\
\(\kappa_a\) & 0.52863 & s\(^{-1}\) \\
\(\kappa_N\) & 0.232 & s\(^{-1}\) \\
\(D_e\) & \(1.05270\times10^{-9}\) & m\(^2\)/s \\
\(D_g\) & \(2.29581\times10^{-10}\) & m\(^2\)/s \\
\end{longtable}

The passive extracellular input is a rectangular pulse at one site with amplitude \(q=0.1714683\) mM tissue/s, on from 0.08 to 0.40 s. This is obtained from the original EDNM membrane flux magnitude \(10^{-4}\) mM cm/s multiplied by the source membrane-area density \(\gamma_m=1714.683\) cm\(^{-1}\). This is a conversion of the potassium mass delivered per tissue volume, not equality of the electrical action of the inputs: the full model transfers potassium across a neuronal membrane and has a direct capacitive and electrical footprint. The pulse has tissue-volume-integrated dose \(q(0.32)\simeq0.05487\) mM. The initial full graph experiment uses 5 µm spacing and sealed boundaries with a 0.8 s horizon; the extended sweep uses the same normalized forcing and graph weights across intrinsic scales 5--150 µm.

For passive simulations, if \(F\) is the stable generator and \(B=(I/\phi_e,0)^\top\), the exact step response matrix is \(S(t)=F^{-1}(e^{Ft}-I)B\) for \(t>0\), zero otherwise. The rectangular pulse is \(q[S(t-t_{\mathrm{on}})-S(t-t_{\mathrm{off}})]\). In the matched comparison, the original EDNM time labels and strict on/off inequalities are used in a zero-order-hold propagation at the actual 1 ms step. For scale sweeps a diagonally weighted symmetric eigendecomposition evaluates the exact passive solution. These two evaluations agree to the intended temporal-resolution accuracy; there is no stochastic fitting in the scale sweep.

We use the same dimensionless contrast for both models,

\[\mathcal R(d,\mu)=\frac{\|K_1-K_0\|_F}{\|K_0\|_F},\]

where the matrices stack the stimulus-minus-control extracellular potassium responses at all seven somatic sites and the specified time points, subscripts label the two graph variants, and the denominator is nonzero. The statistic is not an absolute signal-to-noise ratio. A short-pulse grid uses 100 evenly spaced samples from 0.081 to 0.8 s; a longer step grid uses a sustained focal extracellular input starting at zero with observations at 0.5, 1, 3 and 10 s. Both grids span spacing 2--1000 µm (75 logarithmic values) and mobility multipliers \(10^{-4}\)--\(10^5\) (82 logarithmic values). The nominal-mobility one-dimensional maximum was located numerically in that spacing interval; it is not an analytically proved unique global optimum or a maximization over biological uncertainty. The illustrative grid is deliberately much broader than a measured physiological range.

\subsection{Fisher information under ideal source--sensor coverage}\label{fisher-information-under-ideal-sourcesensor-coverage}

For the new ideal passive design, each site is stimulated separately by the same calibrated pulse; all seven extracellular potassium channels are recorded in every experiment at 0.25, 0.5, 0.75, 1, 1.5, 2, 2.5 and 3 s. This yields \(7\times7\times8=392\) scalar observations per complete protocol. Measurement errors are modeled as independent Gaussian deviations of standard deviation \(\sigma=0.001\) mM (1 µM). The covariance and perfect source specificity are design assumptions, not instrument specifications. The sensitivity is to the dimensionless chord multiplier \(g\) at \(g=1\). Five nuisance multipliers, all at unity, scale glial mobility, extracellular transport, uptake, return and clearance independently. The volumes and pulse are fixed. Central finite differences with relative multiplier increments \(10^{-4}\) and \(5\times10^{-5}\) cross-check the sensitivities; relative changes in the complete whitened Jacobian are below \(8\times10^{-9}\) in the reported calculations.

For a mean data vector \(m(g,\eta)\) and parameter-independent positive-definite covariance \(\Sigma\), let \(s_g=\partial_g m\) and \(S_\eta=\partial_\eta m\). With \(\widetilde s_g=\Sigma^{-1/2}s_g\) and \(\widetilde S_\eta=\Sigma^{-1/2}S_\eta\), the efficient local information after nuisance adjustment is

\[\begin{aligned}
I_{g\mid\eta}&=\min_b\|\widetilde s_g-\widetilde S_\eta b\|^2\\
&=\|(I-\widetilde S_\eta\widetilde S_\eta^+)\widetilde s_g\|^2.
\end{aligned}\]

Known parameters give \(I_g=\|\widetilde s_g\|^2\). The formula states exactly whether the target score can be reproduced by linear combinations of nuisance scores in the specified data. It is a local measure and scales as the square of dose in this passive linear model. For an unbiased regular scalar estimator, the local Cramér--Rao lower bound is \(\operatorname{SD}(\widehat g)\geq I_{g\mid\eta}^{-1/2}\). A nominal three-standard-deviation local separation for change \(\Delta g\) and \(M\) independent complete protocols requires \(M I_{g\mid\eta}(\Delta g)^2\geq9\); this is a lower-bound planning calculation rather than a calibrated test's detection probability. Values below one complete protocol are not fractional experimental repetitions.

\subsection{Finite discrimination against shared support}\label{finite-discrimination-against-shared-support}

Because a local derivative does not settle the finite comparison of graph families, we also generate a noiseless mean data vector from the independent graph with \(g=1\) under the same 392-observation pulse design. Three restricted models fit those means: a single scalar glial path multiplier; six positive independently adjustable glial path-edge weights; and six path weights together with four multipliers for extracellular diffusivity, uptake, return and clearance. The latter is substantially more flexible than a globally rescaled extracellular path. The scalar and edge multipliers and local nuisances vary from 0.01 to 100 in logarithmic coordinates. These bounds define a broad, illustrative subset of the full nonnegative shared-support family; exact zero or unbounded weights are not included. The target source normalization and measurement covariance remain fixed.

The squared separation is

\[\mathcal D^2=\min_{\theta\in\Theta_0}\|\Sigma^{-1/2}[m_1-m_0(\theta)]\|^2.\]

Numerically we use bounded nonlinear least squares with ten deterministic or seeded starts, recording the best obtained distance, not claiming a certified global lower bound over the full mathematical cone. The same data vector is used to fit and compare synthetic means, so this is a \emph{finite design-discrimination calculation}, not held-out predictive validation or observed statistical significance. A prospective biological test must fit on training sources and predict held-out responses.

\section{Results: spatial scale and discriminability}\label{results-spatial-scale-and-discriminability}

\subsection{Conditional spatial sensitivity window}\label{conditional-spatial-sensitivity-window}

At the source-reconciled nominal coefficients, \(\sqrt{D_g/\kappa_a}=20.84\) µm, a characteristic length associated with glial exchange and diffusion in the passive equations. The spacing maximizing a graph contrast need not be that length, because the extracellular operator, input duration and observation schedule also set scales. The matched short-pulse schedule yields a numerical maximum at approximately 14.76 µm, where \(\mathcal R=0.04645\); the longer step schedule gives approximately 36.98 µm, where \(\mathcal R=0.19233\). At 5 µm and nominal mobility, the corresponding contrasts are 0.00392 and 0.00153. The two protocols differ in both the input waveform and the observation schedule; they show that changing the combined source and observation protocol can shift the optimum and need not increase relative graph-shape contrast at fixed intrinsic scale.

\begin{figure}
\centering
\includegraphics[width=1\textwidth,height=\textheight]{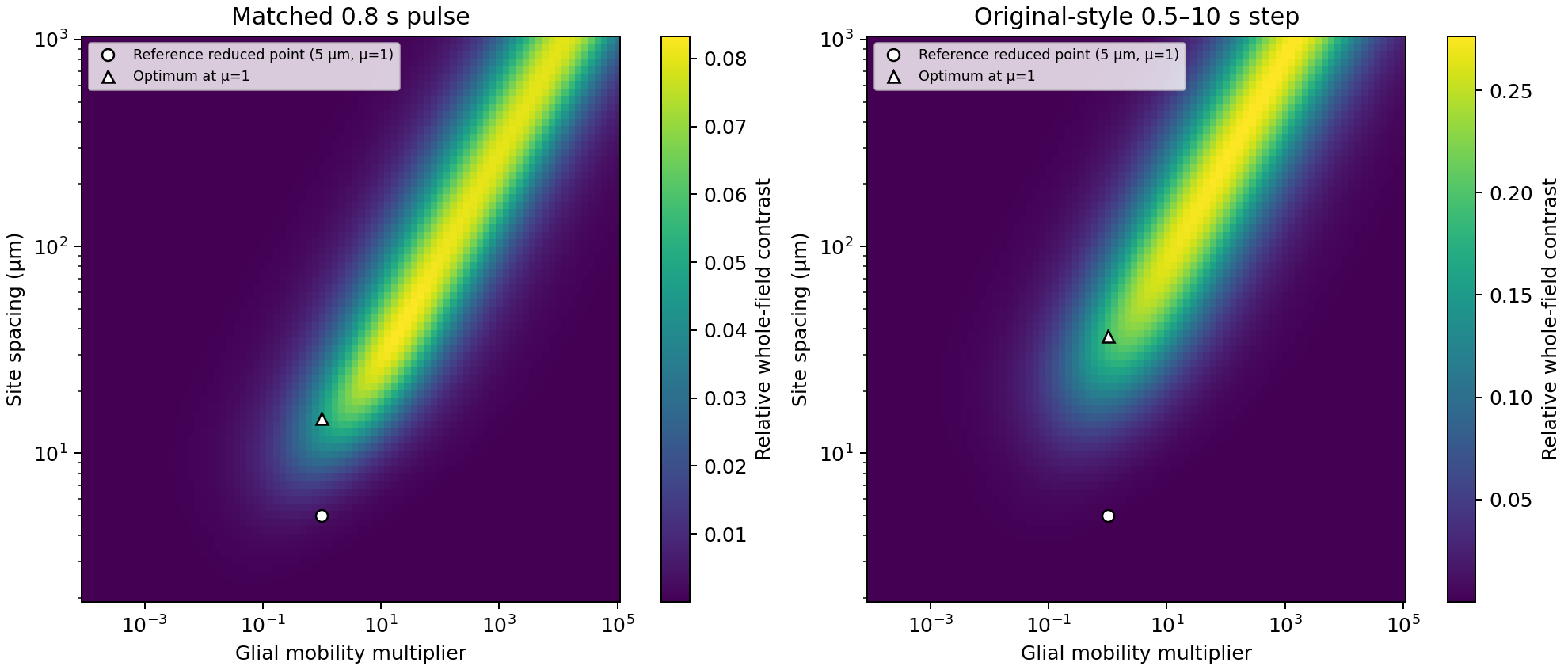}
\caption{Mobility--spacing windows for the two passive protocols. The circle marks the source-reconciled reduced point at 5 µm and unit mobility; the triangle marks the numerical spacing maximum along unit mobility. The circle is not a fitted effective location of the full multi-ion EDNM. Color scales differ by panel. The spacing and mobility axes span synthetic values, not measured tissue uncertainty.}
\end{figure}

These results make a risky but precisely delimited prediction: in this parameterized family of geometrically rescaled graphs, intrinsic scales of tens of micrometres rather than 5 µm yield greater topology sensitivity for these protocols. Changing the time protocol shifts the predicted optimum. The model does not predict a universal tissue length scale or a universal optimal astrocyte spacing; what is testable is the change in discriminability when the independently estimated coefficients and physical source--sensor schedules are specified. Varying \(\mu\) can move the window substantially. Since the sweep changes the strength as well as the weights of the alternative relative to the path, its norm alone does not isolate shape from rescaling. The shared-support comparison below performs that isolation more directly.

\subsection{Exact recovery can be severely ill-conditioned}\label{exact-recovery-can-be-severely-ill-conditioned}

For the synthetic independent graph at \(s=1\) s\(^{-1}\), the relative worst-case recovery gain \(\mathcal K\) is \(8.51\times10^5\) at 5 µm and \(2.32\times10^3\) at 14.76 µm. For \(s=0.25\) and \(4\) s\(^{-1}\), the corresponding 5 µm gains are approximately \(1.69\times10^6\) and \(4.57\times10^5\), while those at 14.76 µm are \(3.82\times10^3\) and \(2.11\times10^3\). Thus the exact inverse can amplify small response perturbations by orders of magnitude. Changing spacing improves its conditioning markedly even when the underlying relative graph weights are unchanged. This resolves the apparent tension between an exact transfer inverse and a practically weak observed effect: the hidden operator is not absent, but its imprint can be overwhelmed by the size of the extracellular transfer.

The normalized operator-space residual after optimal scalar projection of \(G\) onto the shared path is 0.27437 at these spacings. An appreciable difference between operators therefore coexists with a very small response difference. The former is not a proxy for the latter. A reconstruction algorithm that simply inverts a noisy transfer without checking uncertainty can generate spurious off-support edges even if the true graph is shared.

\subsection{Local information and finite shared-support fit}\label{local-information-and-finite-shared-support-fit}

Table 2 gives the ideal passive Fisher results. At 5 µm, all-site observation still yields weak information even when all nuisance coefficients are known. With five nuisance multipliers the local information is only \(2.40\times10^{-5}\). With just one stimulated site at 5 µm and seven sensors it falls to \(4.38\times10^{-6}\) (known-parameter information \(1.50\times10^{-3}\)). Complete source coverage is useful, but it cannot solve the intrinsic geometric suppression at this spacing. At 36.98 µm the same 392 observations yield a known-parameter information of 152.6 and nuisance-adjusted information of 17.19. The latter is about \(7.17\times10^5\) times its 5 µm value. This is a design prediction at the chosen nominal coefficients and assumed independent 1 µM noise, not a measurement of instrument feasibility.

\Needspace{15\baselineskip}

Table 2. Local information in the new passive all-site design. The last column is the formal lower bound \(9/I_{g\mid\eta}\) for nominal local three-standard-deviation separation at unit change; at least one complete protocol is required. The same short pulse and eight recording times are used at all spacings, including the spacing selected using the longer step contrast.

\begin{longtable}[]{@{}
  >{\raggedleft\arraybackslash}p{(\columnwidth - 8\tabcolsep) * \real{0.2000}}
  >{\raggedleft\arraybackslash}p{(\columnwidth - 8\tabcolsep) * \real{0.2000}}
  >{\raggedleft\arraybackslash}p{(\columnwidth - 8\tabcolsep) * \real{0.2000}}
  >{\raggedleft\arraybackslash}p{(\columnwidth - 8\tabcolsep) * \real{0.2000}}
  >{\raggedleft\arraybackslash}p{(\columnwidth - 8\tabcolsep) * \real{0.2000}}@{}}
\toprule\noalign{}
\begin{minipage}[b]{\linewidth}\raggedleft
Spacing (µm)
\end{minipage} & \begin{minipage}[b]{\linewidth}\raggedleft
Known parameters
\end{minipage} & \begin{minipage}[b]{\linewidth}\raggedleft
Unknown glial scale only
\end{minipage} & \begin{minipage}[b]{\linewidth}\raggedleft
Five nuisances
\end{minipage} & \begin{minipage}[b]{\linewidth}\raggedleft
\(9/I_{g\mid\eta}\)
\end{minipage} \\
\midrule\noalign{}
\endhead
\bottomrule\noalign{}
\endlastfoot
5.00 & 0.005692 & \(2.400\times10^{-5}\) & \(2.398\times10^{-5}\) & \(3.75\times10^5\) \\
14.76 & 19.72 & 0.3589 & 0.3502 & 25.7 \\
36.98 & 152.58 & 19.90 & 17.19 & 0.524 \\
\end{longtable}

The comparison between known and nuisance-adjusted parameters at 5 µm is informative: the scalar glial scale accounts for most of the information loss, but the raw information is already small. The bottleneck is therefore not merely insufficient nuisance calibration. At the larger spacing the target score is appreciably independent of nuisance directions even though overall glial mobility remains uncertain. This shows why a well-conditioned design can make calibration and sensor improvements much more productive than attempting to overcome an unfavorable operating regime with repetition alone.

\Needspace{14\baselineskip}

Table 3. Best obtained squared noise-normalized distance between independent-topology synthetic means and fitted shared-support rivals. The last column allows all six path weights and four local/transport nuisance multipliers to adapt. Bounds and noise are those specified in Methods.

\begin{longtable}[]{@{}
  >{\raggedleft\arraybackslash}p{(\columnwidth - 6\tabcolsep) * \real{0.2500}}
  >{\raggedleft\arraybackslash}p{(\columnwidth - 6\tabcolsep) * \real{0.2500}}
  >{\raggedleft\arraybackslash}p{(\columnwidth - 6\tabcolsep) * \real{0.2500}}
  >{\raggedleft\arraybackslash}p{(\columnwidth - 6\tabcolsep) * \real{0.2500}}@{}}
\toprule\noalign{}
\begin{minipage}[b]{\linewidth}\raggedleft
Spacing (µm)
\end{minipage} & \begin{minipage}[b]{\linewidth}\raggedleft
Scaled path
\end{minipage} & \begin{minipage}[b]{\linewidth}\raggedleft
Six free path weights
\end{minipage} & \begin{minipage}[b]{\linewidth}\raggedleft
Free path weights + nuisances
\end{minipage} \\
\midrule\noalign{}
\endhead
\bottomrule\noalign{}
\endlastfoot
5.00 & \(4.747\times10^{-5}\) & \(1.539\times10^{-5}\) & \(7.885\times10^{-6}\) \\
14.76 & 1.5176 & 0.7483 & 0.6508 \\
36.98 & 133.66 & 98.26 & 90.74 \\
\end{longtable}

At 5 µm the most flexible fitted rival leaves a maximum absolute sample discrepancy of only 0.000707 µM. At 36.98 µm that maximum increases to 2.42 µM; the squared whitened distance over the full response vector is 90.74. All ten optimization starts converge to effectively the same best cost for the reported cases. This numerical consistency is evidence against a simple local-minimum explanation, not a proof of global optimality. The larger spacing is the stronger discriminatory regime in this particular design, and the result persists when edge-specific transport weights rather than just one common strength are allowed to vary.

\begin{figure}
\centering
\includegraphics[width=1\textwidth,height=\textheight]{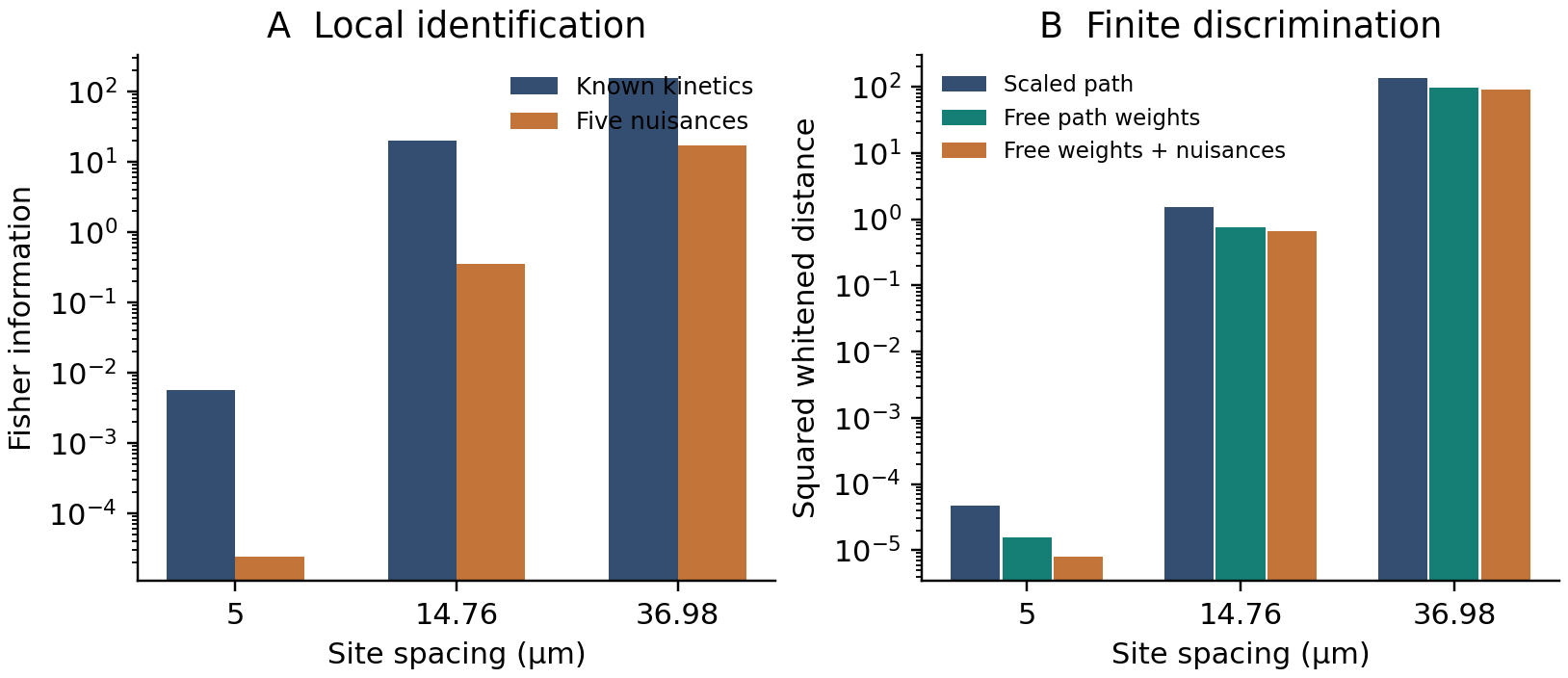}
\caption{Local Fisher information (A) and finite forward-model separation (B) under the same passive pulse design. Both panels use logarithmic ordinates. The finite comparison concerns noiseless simulated means under an illustrative noise model, not experimental p-values.}
\end{figure}

The result clarifies the difference between recovering all entries of a graph without constraints and testing a specific biologically motivated additional-edge hypothesis. The latter may require fewer identifiable parameters and can remain useful even when the full inverse has a large worst-case condition number. Conversely, rejecting one scalar rescaling of the extracellular graph is too weak a criterion: a flexible model with the same support can be much closer to the independent-topology model. The strongest numerical result is therefore the large change in separation from this \emph{more flexible} null as the spatial regime changes.

\subsection{Fixed-tissue observation design and discretization}\label{fixed-tissue-observation-design-and-discretization}

The preceding intrinsic scale sweep changes the physical graph. To isolate observation design from this transformation, we independently construct a sealed, one-dimensional finite-volume domain of length 259 µm. Each of seven 37-µm reference cells is subdivided into \(r=1,3,7,15,21\) equal cells of width \(h=37/r\) µm. The original three conduits connect the centers of the reference cells, with fixed integrated conductance \(D_g w/d\) per unit cross-sectional area. Dividing the resulting flux by control-volume length yields the chord term \(D_gw/(dh)\); local edges have the usual \(D_g/h^2\) coefficient. Each source delivers the same spatially integrated mass per unit area, and the block-average designs preserve the physical support of both sources and measurements at every refinement level. The passive equations, time samples, nominal parameters and illustrative 1 µM independent measurement noise are those of the preceding design. This construction is a synthetic fixed-tissue discretization, not a reconstruction of actual CA1 tissue.

\Needspace{14\baselineskip}

Table 4. Nuisance-adjusted Fisher information for fixed block-average inputs and measurements as the same physical domain is refined.

\begin{longtable}[]{@{}
  >{\raggedleft\arraybackslash}p{(\columnwidth - 4\tabcolsep) * \real{0.3333}}
  >{\raggedleft\arraybackslash}p{(\columnwidth - 4\tabcolsep) * \real{0.3333}}
  >{\raggedleft\arraybackslash}p{(\columnwidth - 4\tabcolsep) * \real{0.3333}}@{}}
\toprule\noalign{}
\begin{minipage}[b]{\linewidth}\raggedleft
Compartments
\end{minipage} & \begin{minipage}[b]{\linewidth}\raggedleft
Nominal passive mobility
\end{minipage} & \begin{minipage}[b]{\linewidth}\raggedleft
Reduced-mobility sensitivity case
\end{minipage} \\
\midrule\noalign{}
\endhead
\bottomrule\noalign{}
\endlastfoot
7 & 17.2117 & 2.31976 \\
21 & 9.25605 & 1.35952 \\
49 & 8.55547 & 1.22013 \\
105 & 8.43150 & 1.19015 \\
147 & 8.41456 & 1.18584 \\
\end{longtable}

The reduced-mobility case uses 0.17988 times nominal glial mobility as an illustrative cross-model sensitivity value motivated by the electrical estimate below; it is not a fitted passive reduction of the full electrodiffusive model. The coarse mesh overestimates nominal information by approximately a factor of two, but the block-average information converges toward a nonzero value. At 147 cells the relative whole-field K contrast is 0.08073 at nominal mobility and 0.02215 at reduced mobility, with maximum sample differences of 2.47 and 0.788 µM, respectively.

Holding the 49-cell tissue fixed, we next compare equal-count distributed and clustered focal source/sensor arrays. Seven sources and seven sensors distributed over the reference-cell centers yield information 12.10 (nominal) and 2.085 (reduced mobility). Seven sources and sensors clustered in the first seven cells yield 2.451 and 0.2363. Retaining the distributed sources and observing all 49 cells yields 79.57 and 12.20. These values account for the same five nuisance directions. Across a 32-corner sensitivity box combining mobility 0.1190--0.3108, extracellular diffusivity multipliers 0.7--1.3 and independent uptake, return and clearance multipliers 0.5--2, plus the central parameter point, distributed seven-source/seven-sensor information ranges from 0.124 to 18.43 (median 1.33). This is a finite illustrative sensitivity analysis, not a bound over the full continuous parameter box or a measured probability distribution. The point-source comparison holds mesh and source support fixed between designs; it is not itself a mesh-convergence claim.

Thus observation geometry remains consequential when the tissue itself is held fixed. Distributed coverage of the candidate conduits is more informative than dense sampling restricted to one end of this example. The original intrinsic-scale optimum must not be interpreted as an optimal electrode pitch.

\section{Full electrodiffusive implementation and physiological constraints}\label{full-electrodiffusive-implementation-and-physiological-constraints}

\subsection{Implementation and conservation}\label{implementation-and-conservation}

The passive analysis deliberately omits physiological pathways that can change both amplitude and sign of observed responses. To evaluate those omissions we use the independently published multicompartmental electrodiffusive network model (EDNM) of Sætra and Mori \cite{saetra2024}, pinned to commit \texttt{826c211c70d9027a4acb02202ca91a60de79a15d}. Each unit contains soma and dendrite layers with neuronal, glial and extracellular domains; it evolves Na, K, Cl and Ca concentrations, membrane and extracellular potentials, gating variables and water-dependent volume fractions. It enforces a capacitive charge relationship rather than computing extracellular potential from potassium concentration alone.

For an ionic species \(i\) in domain \(r\), the interunit flux on an oriented edge \(u\to v\) uses the centered Nernst--Planck discretization

\[j_{r,uv}^i=-\frac{D_i\upsilon_{ir}w_{uv}}{\lambda_r^2d}\left[c_v^i-c_u^i+\frac{z_i(c_u^i+c_v^i)}{2\psi}(\varphi_v-\varphi_u)\right],\]

where \(\lambda_r\) is tortuosity, \(\upsilon_{ir}\) the mobile fraction, \(\psi=RT/F_{\!A}\) and \(F_{\!A}\) Faraday's constant. The source finite-volume assembly includes the previous-step domain volume fraction at the lower-index endpoint of each edge, following the oriented convention of the parent solver, and a second factor of inverse unit spacing. This preserves equal and opposite inventory transfers but need not be invariant under relabeling when local volume fractions differ. It is a discretization convention rather than a physical directionality of the glial coupling. Consequently both chemical and electrical axial transport have the same explicit inverse-square scaling in the volume-normalized inventory equations. Membrane area per tissue volume, interlayer distance, local kinetics and normalized forcing remain fixed; absolute delivered ionic mass need not remain fixed across rescaled tissues. Each edge correction is inserted with opposite signs at its two endpoints. Glial and extracellular weights can be changed separately without changing local membrane fluxes, pumps or gating equations. The adapter replaces the original closed-path contribution by adding the difference of the new and old edge residuals and their analytic sparse Jacobian. A targeted sensitivity check below replaces the inherited oriented edge-volume factor by the symmetric mean of the endpoint volume fractions in both reference and rival graph assemblies. The compiled parent local equations are retained. Earlier finite-difference Jacobian checks and the fine-step validation are archived with the source.

This structure preserves total inventory of each ionic species under internal transfers and sealed boundaries. A separate capacitive balance must hold for charge; mass conservation alone cannot establish electrical consistency. For the neuronal soma,

\[\gamma_m C_m V_m=F_{\!A}\left(\alpha_n\sum_i z_i c_{n,i}+z_0a_n^{\mathrm{imm}}\right),\]

with constant immobile-ion inventory \(a_n^{\mathrm{imm}}\) and \(V_m=\varphi_n-\varphi_e\). We use the discrete current balance implied by this relation and the actual semi-implicit time update to diagnose the neighboring response.

The matched 5-µm full-model comparison gives a relative K contrast of 0.000438 versus 0.00393 in the source-reconciled passive approximation. This quantitative difference motivates the axial-migration intervention below; the detailed matched comparison and its flux accounting appear in \hyperref[supp-f]{Supplement F}.

\subsection{Electrical migration shifts the topology-sensitive window}\label{electrical-migration-shifts-the-topology-sensitive-window}

We extend the matched short-pulse seven-site experiment to intrinsic scales 5, 15, 37, 60, 100 and 150 µm. The native-glial sweep retains the original electrodiffusive constitutive law. A diagnostic counterfactual sets only the axial glial electrical-migration coefficient to zero in both the graph replacement residual and its Jacobian, retaining chemical diffusion, the extracellular and local membrane terms and the capacitive constraints. This changes a transport law and is not equivalent to a selective physiological blocker. In both conditions internal edge contributions remain conservative. Each statistic is calculated from paired baseline-subtracted trajectories at the same scale. Both timesteps, 1 and 0.5 ms, were tested. The interpretation record commits to a predicted shift toward smaller scales before these follow-up outputs were inspected, but after execution had started and completed; it is not a preregistration before execution (\href{https://github.com/ipsissima/SMM/blob/74a6ebbc1cddce9774335d40cdf21cd644f91b18/docs/MIGRATION_HYPOTHESIS.md}{outcome-blind interpretation record}).

\Needspace{16\baselineskip}

Table 5. Whole-field relative extracellular potassium contrast in the full EDNM (percent, 0.5 ms timestep). The maximum is on the sampled grid, not a continuous or universal tissue optimum.

\begin{longtable}[]{@{}rrr@{}}
\toprule\noalign{}
Intrinsic scale (µm) & Native migration & Glial axial migration off \\
\midrule\noalign{}
\endhead
\bottomrule\noalign{}
\endlastfoot
5 & 0.043825 & 0.397940 \\
15 & 0.728079 & 0.226676 \\
37 & 2.019322 & 0.051258 \\
60 & 2.364808 & 0.020855 \\
100 & 2.012149 & 0.007735 \\
150 & 1.479953 & 0.003472 \\
\end{longtable}

The native response confirms suppression at 5 µm and a pronounced intermediate window, but its sampled maximum is 60 µm, beyond the passive short-pulse maximum near 15 µm. At 60 µm the relative contrast is about 113 times that of the migration-off counterfactual. At 5 µm the direction of the effect reverses: migration suppresses contrast. The counterfactual shifts the sampled maximum to 5 µm, with monotonically decreasing values across the rest of the sampled grid. The refined timesteps agree closely and the extended runs satisfy the positivity, convergence and ionic inventory checks. This result identifies the axial migration term as a cause of the shift within the solver, rather than identifying its magnitude with an effective scalar diffusivity. Because the transmembrane stimulus has an electrical footprint, the primary endpoint here is extracellular K, not voltage. The passive theorem concerns mobility limits at fixed geometry and does not itself prove the existence of a full nonlinear maximum as intrinsic scale changes.

To test the length-scale interpretation independently of the zero-migration contrast, we additionally multiply only the axial glial electrical term by \(m=0.25\) or \(4\). A frozen-concentration, two-conductor approximation gives a voltage-difference length

\[\ell(m)=\left[g_m\left(\frac{1}{m\sigma_g}+\frac{1}{\sigma_e}\right)\right]^{-1/2},\]

where \(\sigma_g,\sigma_e\) are axial tissue conductivities and \(g_m\) is glial membrane slope conductance per tissue volume. In this reduction the oppositely directed axial conductor currents conserve total current and membrane exchange is proportional to their voltage difference. Eliminating the currents gives the differential relation \(v''=g_m[(m\sigma_g)^{-1}+\sigma_e^{-1}]v\) away from sources. The frozen slope and conductivities are calculated from the native EDNM resting state, independently of the graph contrast maxima. This is not the exact electrical mode of the full multilayer, concentration- and volume-dependent system.

At \(m=0.25,1,4\), respectively, the calculated lengths are 47.84, 88.34 and 140.07 µm and the sampled contrast maxima occur at 37, 60 and 100 µm. These coarse maxima motivated the scaling hypothesis \(d_*\approx0.7\ell\) under this forcing and graph family. A targeted refinement at 0.5 ms, with additional intrinsic scales bracketing each maximum, yields maxima 35, 60 and 85 µm for \(m=0.25,1,4\), with respective ratios 0.732, 0.679 and 0.607. The larger migration multiplier therefore exhibits a materially lower ratio than the initial coarse-grid calculation suggested. The electrical length predicts the ordering and approximate scale, not a universal proportionality. Each maximum is internal to the newly sampled interval; no claim of a continuous or unique global maximum is made. Current-driven spatial buffering is established physiology \cite{orkand1966,gardnermedwin1983,chen2000,ma2016}. In particular, Chen and Nicholson \cite{chen2000} derived an optimum at equality of geometric and electrotonic length for the endfoot current of an equivalent leaky cable. The present result concerns a different observable: the difference between responses generated by distinct intercellular transport graphs under matched local physiology and forcing. Thus the matching of electrical and geometrical scales in buffering is not new; the result here concerns its consequences for topology discrimination, including the noncommuting two-conductor graph operator and the protocol-specific optimum in the full electrodiffusive model. Recent within-cell electrical compartmentalization results \cite{nakatani2026} provide complementary motivation, not a prior derivation of this topology-identification result.

A separate neutral-source diagnostic tests whether the electrical pathway persists without a directly imposed membrane-current footprint. At the central empirical mobility, 30 µm intrinsic scale, and a local 3 mM tissue/s neutral KCl step, we compare baseline-subtracted independent and path geometries with glial axial migration retained or disabled. All six runs at 100, 50 and 25 µs pass the conservation and Newton checks. At the directly connected distal site (node 4), log--log slopes of the absolute K contrast at 25 µs are 2.623 versus 3.307 over 2--5 ms, 2.333 versus 2.954 over 5--10 ms, and 2.079 versus 2.559 over 20--40 ms. The slopes are consistently lower with migration retained and become close between the two finest resolutions on these finite windows. They do not establish exact asymptotic powers: in particular the migration-off response lies below the prescribed numerical floor throughout the earliest 0.5--2 ms window. The finite-time result supports an accelerated electrical contribution under neutral forcing, consistent with the reduced electrical mechanism without requiring its exact onset exponent in EDNM (\hyperref[supp-i]{Supplement I}).

The independent two-unit pump-clamp analysis and its exact current balance are reported in \hyperref[supp-g]{Supplement G}. It establishes a model-specific mechanism for voltage inversion but is not itself a topology test.

\subsection{Matched-dose neutral KCl topology challenge}\label{matched-dose-neutral-kcl-topology-challenge}

The seven-site topology comparison is repeated using a state-independent extracellular source that delivers equal K and Cl inventories to one somatic ECS control volume during 0.08--0.28 s. This is externally charge-neutral but introduces osmoles; it is not a physical reconstruction of iontophoresis. The two rates, 3 and 30 mM tissue/s, deliver 0.6 and 6 mM of each ion per source-cell tissue volume. We compare the independent graph to the ordinary path, a path whose six edge weights are all \(7.9/6\) (equal total glial edge weight and Laplacian trace), and an axially uncoupled graph. All other quantities are held fixed within a matched comparison. The intrinsic scale is 37 µm, native glial mobility is used, sources are nodes 0 and 2, and the simulation horizon is 0.8 s. Independent baseline trajectories and timestep checks at 0.5 and 0.25 ms control drift and numerical error.

\Needspace{14\baselineskip}

Table 6. Topology effects relative to the equal-total-weight path at the finer timestep. The relative K statistic divides by the norm of the rival's baseline-subtracted K response; absolute maxima use all observed nodes and times.

\begin{longtable}[]{@{}
  >{\raggedleft\arraybackslash}p{(\columnwidth - 8\tabcolsep) * \real{0.2000}}
  >{\raggedleft\arraybackslash}p{(\columnwidth - 8\tabcolsep) * \real{0.2000}}
  >{\raggedleft\arraybackslash}p{(\columnwidth - 8\tabcolsep) * \real{0.2000}}
  >{\raggedleft\arraybackslash}p{(\columnwidth - 8\tabcolsep) * \real{0.2000}}
  >{\raggedleft\arraybackslash}p{(\columnwidth - 8\tabcolsep) * \real{0.2000}}@{}}
\toprule\noalign{}
\begin{minipage}[b]{\linewidth}\raggedleft
Source
\end{minipage} & \begin{minipage}[b]{\linewidth}\raggedleft
Rate (mM tissue/s)
\end{minipage} & \begin{minipage}[b]{\linewidth}\raggedleft
Relative K contrast (\%)
\end{minipage} & \begin{minipage}[b]{\linewidth}\raggedleft
Maximum K difference (mM)
\end{minipage} & \begin{minipage}[b]{\linewidth}\raggedleft
Maximum membrane voltage difference (mV)
\end{minipage} \\
\midrule\noalign{}
\endhead
\bottomrule\noalign{}
\endlastfoot
0 & 3 & 2.15090 & 0.032603 & 0.045324 \\
0 & 30 & 3.01463 & 0.574726 & 0.339096 \\
2 & 3 & 1.13519 & 0.015345 & 0.025019 \\
2 & 30 & 1.28343 & 0.163860 & 0.256193 \\
\end{longtable}

Relative K contrast increases by 40.16\% and 13.06\% for the two sources, respectively; this dose dependence cannot occur for a linear system under identical waveform scaling. These are nonlinear contrasts relative to a particular rival, not a lower bound on topology separation after arbitrary edge fitting. Absolute differences of hundreds of µM in K and hundreds of µV in neuronal voltage are plausible targets for appropriately characterized sensors, not measured signals. The high-load source elevations are 23--26 mM, versus 2.3--2.5 mM at low load. Removing axial glial coupling from the path increases peak source K by 3.80\% and 4.38\% at low load and 4.79\% and 6.12\% at high load for sources 0 and 2. No matched-dose run reached the model's spike threshold or crossed −40 mV; there is no categorical firing or depolarization-block effect in these protocols. The separate published-source coupling-strength benchmark is reported in \hyperref[supp-e]{Supplement E}, not counted as a topology test.

At distal node 5 for source 0, the high-dose shortcut-minus-weight-matched-path difference at 0.8 s is +0.207535 mM extracellular K, +1.370738 mV K reversal potential, and −0.339096 mV neuronal voltage. Four-way subtraction of baseline and stimulated trajectories from both geometries gives integrated pump, K leak, delayed-rectifier K, Cl leak, Na leak and gated Na contributions of −12.443, +5.495, +1.897, +4.014, +1.033 and −0.567 mV, respectively, with net remaining axial terms +0.232 mV. The current sum reconstructs the voltage difference with maximum paired closure error \(1.46\times10^{-8}\) mV. The maximum somatic extracellular volume-fraction difference at this distal node is \(5.01\times10^{-6}\) in absolute fraction. The small differential swelling and current accounting support the ionic-current interpretation but do not independently exclude indirect effects of water dynamics. The earlier pump clamp (\hyperref[supp-g]{Supplement G}) establishes a separate two-site causal result; this four-way accounting is not itself an intervention. Because the residual voltage is a small difference of much larger opposing currents, we interpret its pathway and magnitude at the specified kinetics rather than predict a robust sign across uncertain pump affinity and leak conductance.

\subsection{Empirically informed electrical parameter constraint}\label{empirically-informed-electrical-parameter-constraint}

Ma et al.~\cite{ma2016} report a paired-astrocyte coupling resistance of 4.2 ± 1.5 MΩ (\(n=3\) doublets, mean pair distance 30.5 ± 3.1 µm) under special recording conditions designed to reduce membrane shunting. Independent CA1 measurements by Kiyoshi et al.~\cite[Fig. 5B–D]{kiyoshi2018} give cell density 11,238 ± 331 mm\(^{-3}\), nearest-neighbor separation 39.45 ± 1.14 µm and 8.4 ± 0.16 nearest neighbors. A simple affine effective-medium calculation combines these independent summaries. For an isotropic resistor network with density \(n\), mean degree \(z\), identical edge conductance \(1/R\) and representative edge length \(d_a\), the dissipation per tissue volume under an affine potential gradient is

\[\frac{P}{V}=\frac{nz}{2R}\mathbb E[(\nabla\varphi\cdot\mathbf d_a)^2]
=\frac{nzd_a^2}{6R}|\nabla\varphi|^2,\qquad
\sigma_g^{\mathrm{aff}}=\frac{nzd_a^2}{6R}.\]

The factor \(1/2\) avoids double counting edges, and isotropy gives the factor \(1/3\). This estimate assigns the isolated-pair resistance to all representative neighbors and neglects edge heterogeneity, anisotropy, nonaffine electrical relaxation and correlations between anatomical variables. For a specified resistor network, affine boundary-compatible voltages give a variational dissipation upper bound relative to the relaxed voltages; the estimate here is not a rigorous upper bound for real CA1. CA1 coupling can itself be anisotropic \cite{anders2014}.

The conditional estimate is \(5.83\times10^{-5}\) S/cm, or 0.180 of the native EDNM axial glial value (\(3.24\times10^{-4}\) S/cm). Changing the four inputs independently to their favorable or unfavorable reported one-SEM endpoints yields \(3.86\times10^{-5}\) to \(1.01\times10^{-4}\) S/cm, or 0.119--0.311 of native conductivity. This is an endpoint sensitivity box, not a confidence interval. With the native EDNM frozen glial membrane slope \(g_m=3.335\) S/cm\(^3\) and extracellular conductivity \(0.0013213\) S/cm, the two-conductor estimate yields an electrical length of 40.9 µm (33.5--53.0 µm in the endpoint box). Extrapolating the coarse migration-only relationship gave a prospective estimate near 29 µm (23--37 µm at the conductivity sensitivity endpoints). We tested this separately by scaling both chemical diffusion and electrical migration in the Nernst--Planck graph law, keeping the short transmembrane pulse, local kinetics and graph family unchanged. Across 22 additional constitutively scaled cases at 0.5 ms, the lower, central and upper conductivity estimates give sampled maxima 25, 30 and 40 µm, respectively; the corresponding ratios to the electrical length are 0.746, 0.733 and 0.755. The sampled maximum at the central estimate therefore corroborates the prospective approximately 30-µm prediction in the full model. The 25--40 µm range is a finite endpoint sensitivity result, not a confidence interval or a bound over continuous physiological uncertainty. This is an intrinsic scale prediction for a synthetic graph and specified input, not an electrode-spacing rule or a directly measured optimum in CA1.

\begin{figure}
\centering
\includegraphics[width=1\textwidth,height=\textheight]{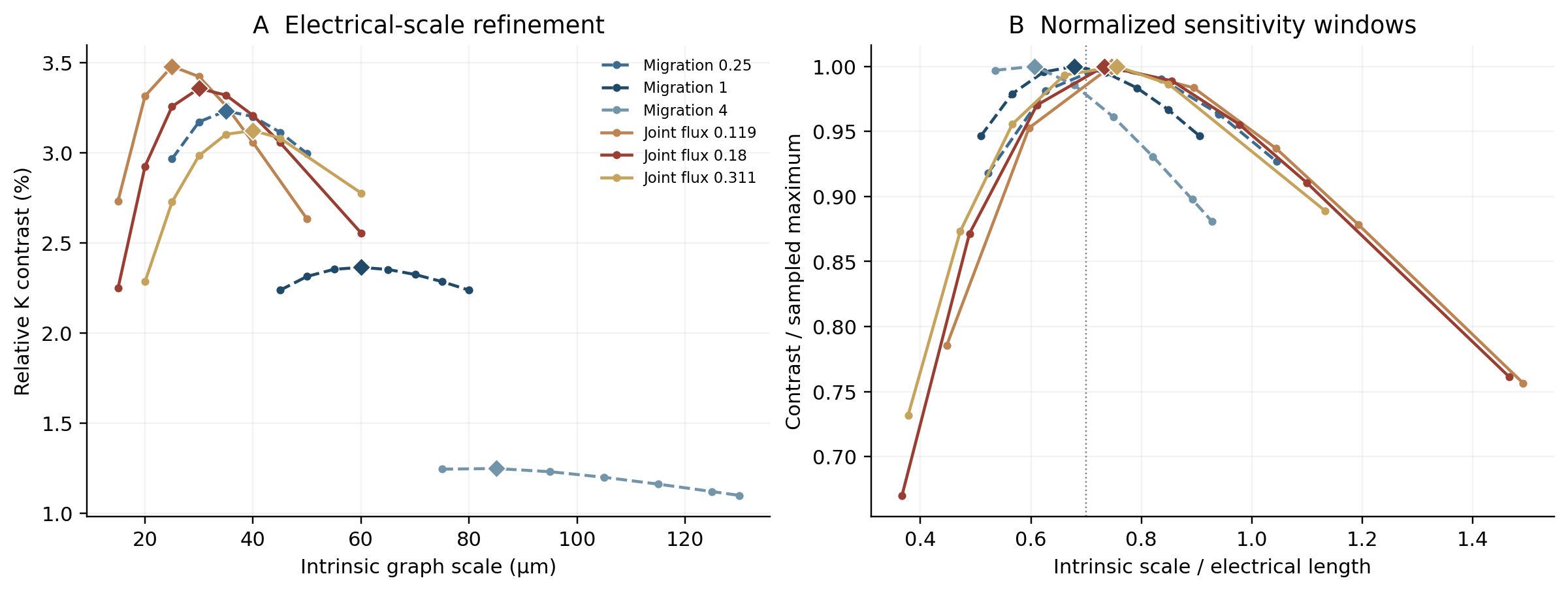}
\caption{Refined electrical topology-sensitivity windows. Left: whole-field short-pulse K contrast for migration-only interventions and jointly scaled chemical and electrical transport at the empirical conductivity estimate and sensitivity endpoints. Right: normalized contrast against intrinsic scale divided by the independently calculated electrical length; symbols mark sampled maxima. The full model does not exhibit exact universal collapse, but the three empirically informed cases cluster near 0.74. All curves use 0.5 ms steps.}
\end{figure}

Breithausen et al.~\cite{breithausen2020} also provide independent experimental targets: focal K transients with pre-blockade mean peaks 2.30 and 8.28 mM under distinct weak and strong iontophoresis protocols, and a measured spatial peak-decay length 13.08 ± 0.79 µm. The strong group showed a significant increase relative to time controls after pharmacological uncoupling; the weak group did not. The published group mean changes and time constants motivate a coupled-only source calibration followed by prediction of the uncoupled responses without fitting to them. A one-dimensional neutral-KCl screening model is not a faithful reconstruction of their three-dimensional charged iontophoresis and blocker experiment; the individual traces and recording geometry are required for quantitative trace-level validation. We do not count aggregate target extraction or a fitted source amplitude as an independent validation of SMM's extra-edge prediction.

\subsection{Topology contrast in the empirically informed conductivity regime}\label{topology-contrast-in-the-empirically-informed-conductivity-regime}

We test the independent graph with both chemical and electrical axial glial fluxes scaled together, rather than scaling only migration. Thirty-two configurations compare ordinary, independent and total-glial-edge-weight-matched paths, each with its own baseline. The design varies mobility over the effective-medium sensitivity values 0.1190, 0.1799 and 0.3108, intrinsic scale 15, 37 and 60 µm, neutral KCl dose 3 and 30 mM tissue/s, selected source sites 0 and 2, and selected extracellular axial multipliers 0.7 and 1.3. The primary timestep is 1 ms with central 37-µm cases repeated at 0.5 ms. All configurations complete the positivity, mass and Newton-convergence checks. The following results are for the total-edge-weight-matched path.

\Needspace{13\baselineskip}

Table 7. Central empirically informed mobility 0.17988 at 37 µm.

\begin{longtable}[]{@{}
  >{\raggedleft\arraybackslash}p{(\columnwidth - 10\tabcolsep) * \real{0.1667}}
  >{\raggedleft\arraybackslash}p{(\columnwidth - 10\tabcolsep) * \real{0.1667}}
  >{\raggedleft\arraybackslash}p{(\columnwidth - 10\tabcolsep) * \real{0.1667}}
  >{\raggedleft\arraybackslash}p{(\columnwidth - 10\tabcolsep) * \real{0.1667}}
  >{\raggedleft\arraybackslash}p{(\columnwidth - 10\tabcolsep) * \real{0.1667}}
  >{\raggedleft\arraybackslash}p{(\columnwidth - 10\tabcolsep) * \real{0.1667}}@{}}
\toprule\noalign{}
\begin{minipage}[b]{\linewidth}\raggedleft
Source
\end{minipage} & \begin{minipage}[b]{\linewidth}\raggedleft
Dose (mM tissue/s)
\end{minipage} & \begin{minipage}[b]{\linewidth}\raggedleft
Timestep (ms)
\end{minipage} & \begin{minipage}[b]{\linewidth}\raggedleft
Relative K contrast (\%)
\end{minipage} & \begin{minipage}[b]{\linewidth}\raggedleft
Max K difference (mM)
\end{minipage} & \begin{minipage}[b]{\linewidth}\raggedleft
Max voltage difference (mV)
\end{minipage} \\
\midrule\noalign{}
\endhead
\bottomrule\noalign{}
\endlastfoot
0 & 3 & 1 & 3.66765 & 0.062014 & 0.085459 \\
0 & 30 & 1 & 3.86802 & 0.811412 & 0.598052 \\
0 & 30 & 0.5 & 3.86830 & 0.811510 & 0.598588 \\
2 & 30 & 0.5 & 2.09264 & 0.320331 & 0.542941 \\
\end{longtable}

At the central high-dose source-0 point, the absolute differences exceed the native-conductivity results, with close agreement under timestep refinement. At the central low-dose point the relative K contrast is already 3.67\%, so high load increases it by about 5.46\%, less than the native-mobility proportional increase. This is evidence that the absolute topology contrast need not vanish near the estimated biological conductivity. It does not identify the true graph in tissue. The next section addresses predictive discrepancy after fitting freely weighted alternatives at this mobility. The upper conductivity endpoint at 37 µm and high dose gives 0.826 mM maximum K difference and 0.541 mV maximum voltage difference. The lower endpoint at 60 µm gives 0.473 mM and 0.522 mV. Across all 16 low-dose configurations (including refinement runs), maximum absolute K differences range from 0.00905 to 0.06203 mM; across the 16 high-dose configurations they range from 0.1052 to 0.8255 mM. Corresponding high-dose voltage differences range from 0.1462 to 0.6084 mV. All configurations retain nonzero separation from the restricted total-weight-matched rival; none produces spikes. The relative K contrast ratio for high versus low dose ranges from 0.785 to 1.468, increasing in nine of sixteen paired comparisons (eight of fourteen distinct parameter and source combinations, excluding the two timestep repeats). In particular, the 60-µm cases show a relative contrast decrease with dose. Nonlinearity can amplify or attenuate relative discrimination according to regime; the native-mobility amplification is not a universal monotonic law. At the central mobility and 37 µm, changing extracellular axial transport by ±30\% retains high-dose absolute K differences of 0.796--0.816 mM. These are finite sensitivity calculations rather than robustness bounds over a continuous parameter domain. No physiological transfer of the derived empirical mobility to the particular synthetic long edges is implied.

The targeted edge-volume audit repeats the central-mobility, 37-µm comparisons at 0.5 ms using both the original lower-index endpoint volume convention and the symmetric arithmetic mean of endpoint volumes. Across the four source-by-dose conditions (sources 0 and 2; rates 3 and 30 mM tissue/s), the relative change in the relative K contrast against the strength-matched path is at most 0.0106\%, in maximum absolute K discrepancy at most 0.0379\%, and in maximum absolute neuronal-voltage discrepancy at most 0.0175\%. For the high-dose source-0 case the symmetric result is 0.811817 mM and 0.598692 mV, compared with 0.811510 mM and 0.598588 mV using the original convention. These changes are relative percentages of the respective statistics, not percentage-point differences. All eight configurations completed. This audit reduces concern that the reported central matched-path contrasts are driven by endpoint indexing; it does not certify relabeling invariance of the original discretization or replace an audit of every fitted rival and physiological regime (\hyperref[supp-j]{Supplement J}).

\subsection{Spatial generalization with four training sources and three withheld sources}\label{spatial-generalization-with-four-training-sources-and-three-withheld-sources}

Exploratory two-source fits (\hyperref[supp-l]{Supplement L}) reveal that a flexible rival can fit the observed sites closely yet transfer poorly. To distinguish inadequate coverage from a support-dependent mismatch, we retain the seven-site, 37-µm geometry, local EDNM kinetics, neutral 30 mM tissue/s KCl input (0.08--0.28 s), 0.8 s horizon and 1 ms timestep, but increase intervention coverage. Synapses are disabled, the homogeneous baseline is checked for invariance, and the same edge weights scale glial diffusion and migration. Let \(Y_j^\star=(K_j^\star,V_j^\star)\) and \(Y_j(w)=(K_j(w),V_j(w))\) denote the baseline-subtracted somatic responses to source \(j\) in the independent-support reference and shared-support candidate. For the 15 sample times nearest 0.10, 0.15, \ldots, 0.80 s, across seven sites, define

\[D_j^2(w)=\sum_{t,i}\left[\frac{K_{j,i}(t;w)-K_{j,i}^\star(t)}{0.001\ \mathrm{mM}}\right]^2+\sum_{t,i}\left[\frac{V_{j,i}(t;w)-V_{j,i}^\star(t)}{0.01\ \mathrm{mV}}\right]^2.\]

The 1-µM and 10-µV scales weight the two observables; they are not experimentally estimated noise.

The training set is \(\mathcal T=\{0,2,4,6\}\); the test set is \(\mathcal H=\{1,3,5\}\). The objective and withheld prediction distance are respectively

\[D_{\mathcal T}^2(w)=\sum_{j\in\mathcal T}D_j^2(w),\qquad D_{\mathcal H}^2(w)=\sum_{j\in\mathcal H}D_j^2(w).\]

Each \(D_j^2\) uses 210 scaled residuals (15 time samples, seven sites and two observables), so the training and test totals comprise 840 and 630 residuals, respectively. All reported RMS values use the complete stored trajectories, not just these objective samples. The initial selected candidate is the best evaluated training point, including evaluation points used for finite differences. We subsequently refine the independent-target candidates using only training residuals and their derivatives (Supplements N--O); no withheld observation enters candidate selection or refinement. At each of the two mobilities, ordinary and total-edge-weight-matched starting points are used. The six extracellular path weights range independently over \([0.7,1.3]\); the six glial path weights range independently over \([0,17.2755103\mu]\) (upper limits 17.2755103 and 3.1075479 in native units for the native and reduced mobilities, respectively). The latter scales the reduced-mobility broad cap with the reference mobility, and is a sensitivity assumption rather than an empirically measured edgewise bound. The solver rejects unconverged Newton steps and checks positivity, finiteness and inventory balance. The initial optimizer uses finite-difference bounded trust-region least squares (\hyperref[supp-k]{Supplement K}); the independent-target continuation and derivative-based refinement are specified in Supplements N--O.

\Needspace{19\baselineskip}

Table 8. Multi-source fits and controls. Distances sum only the designated source sets. Independent-target rows report the final locally refined candidate (\hyperref[supp-o]{Supplement O}); four-source fitting and refinement use only training sites. All-source fits use all seven sites. The strength-matched path is not optimized. Shared-target controls use heterogeneous weights within the fitted family. The dash denotes no withheld data.

\begin{longtable}[]{@{}
  >{\raggedright\arraybackslash}p{(\columnwidth - 6\tabcolsep) * \real{0.4600}}
  >{\raggedleft\arraybackslash}p{(\columnwidth - 6\tabcolsep) * \real{0.1400}}
  >{\raggedleft\arraybackslash}p{(\columnwidth - 6\tabcolsep) * \real{0.2000}}
  >{\raggedleft\arraybackslash}p{(\columnwidth - 6\tabcolsep) * \real{0.2000}}@{}}
\toprule\noalign{}
\begin{minipage}[b]{\linewidth}\raggedright
Reference and protocol
\end{minipage} & \begin{minipage}[b]{\linewidth}\raggedleft
\(\mu\)
\end{minipage} & \begin{minipage}[b]{\linewidth}\raggedleft
Training \(D^2\)
\end{minipage} & \begin{minipage}[b]{\linewidth}\raggedleft
Withheld \(D^2\)
\end{minipage} \\
\midrule\noalign{}
\endhead
\bottomrule\noalign{}
\endlastfoot
Independent, four-source fit & 1 & \(3.177\times10^5\) & \(8.861\times10^5\) \\
Independent, fixed strength-matched path & 1 & \(8.180\times10^6\) & \(3.294\times10^6\) \\
Shared, four-source fit & 1 & \(3.714\times10^{-13}\) & \(3.650\times10^{-13}\) \\
Independent, all-source fit & 1 & \(8.269\times10^5\) & --- \\
Independent, four-source fit & 0.17988 & \(2.215\times10^6\) & \(4.777\times10^6\) \\
Independent, fixed strength-matched path & 0.17988 & \(1.612\times10^7\) & \(7.132\times10^6\) \\
Shared, four-source fit & 0.17988 & \(7.593\times10^{-11}\) & \(9.752\times10^{-11}\) \\
Independent, all-source fit & 0.17988 & \(5.069\times10^6\) & --- \\
\end{longtable}

\textbf{Four-source generalization.} In the initial fits at native mobility the two initializations produce training objectives agreeing to \(8.05\times10^{-8}\) relative; at reduced mobility the relative difference is \(4.21\times10^{-7}\). In both cases the selected fit improves aggregate withheld distance relative to the fixed strength-matched path. The improvement is approximately 73.1\% at native mobility and 33.0\% at reduced mobility; nevertheless, sizable discrepancies persist. At sources 1, 3 and 5, native-mobility K RMS discrepancies are 0.04447, 0.05870 and 0.04565 mM; voltage RMS discrepancies are 0.03170, 0.03461 and 0.03300 mV. At reduced mobility the corresponding K RMS discrepancies are 0.11304, 0.12076 and 0.11237 mM, and voltage RMS discrepancies are 0.09611, 0.07876 and 0.09849 mV. Thus the four-source fits do not simply reproduce the unfavorable transfer of the earlier sparse-source fits: increased coverage permits markedly better prediction than the unoptimized control in aggregate while leaving substantial, structured errors. The test-source sets differ between the two experiments, so their aggregate distances are not directly comparable as a coverage-effect estimate. The two mobilities also define different references and different admissible glial-weight ranges, not nested fits to identical data.

\begin{figure}
\centering
\includegraphics[width=1\textwidth,height=\textheight]{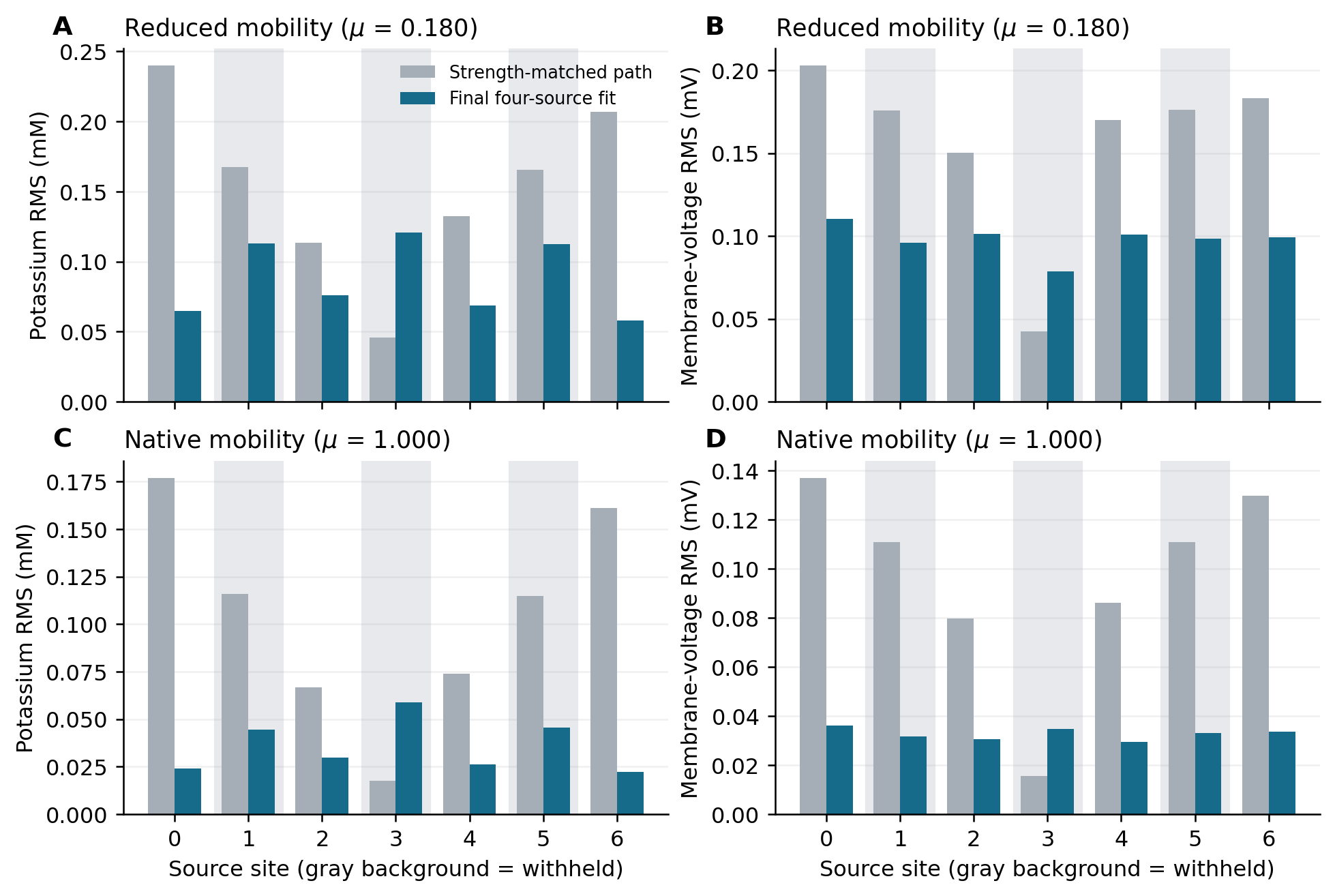}
\caption{Four-source predictive discrimination. Whole-trajectory RMS discrepancies for seven source sites at reduced (A, B) and native (C, D) mobility. The final fitted candidate (blue) is refined using only the training objective on sites 0, 2, 4 and 6; gray bands identify sites 1, 3 and 5 that are withheld from fitting. The gray bars represent the unfitted uniform path matched to the reference's total glial edge weight. Improved aggregate prediction does not eliminate source-specific discrepancies.}
\end{figure}

\textbf{In-family negative controls.} To assess whether the inverse pipeline itself generates a spurious need for different support, we replace the shortcut reference by a heterogeneous shared-support graph. Its six glial weights are proportional to \((0.7,1.4,1.1,0.9,1.8,1.2)\), normalized to sum to \(7.9\mu\), and its six extracellular weights are \((0.8,1.12,0.94,1.2,0.9,1.04)\). The generating weights are neither of the uniform starting choices, but are admissible at both mobilities. We fit the same four sources, with the same model family and objective, starting from the ordinary path. The maximum absolute weight-recovery errors are \(6.28\times10^{-10}\) at native mobility and \(2.09\times10^{-9}\) at reduced mobility, in native weight units; the corresponding withheld distances are less than \(10^{-10}\). The contrast between near-exact recovery here and the residuals for the independent target is strong internal evidence of specificity for these synthetic references. This is a noiseless same-solver control: it is not an empirical false-positive rate or a proof that every in-family reference is identifiable under noise, misspecification or alternative initializations.

\textbf{All-source fits.} We also remove the generalization requirement by fitting all seven source locations simultaneously (1,470 residuals), starting from the total-weight-matched path at each mobility. The obtained distances are \(8.269\times10^5\) and \(5.069\times10^6\), respectively, versus \(1.147\times10^7\) and \(2.325\times10^7\) for the respective unoptimized strength-matched paths over the same seven sources. Thus neither the four-source holdout nor a failure to include any particular source is necessary for the obtained residuals. The all-source fits are training fits only and provide no independent prediction test. In particular, the calculated distances are values at feasible parameter vectors and hence upper bounds, not lower bounds, on the unknown minimum training distance of the respective bounded family. A positive obtained value is not a proof of nonrepresentability.

\textbf{Numerical stationarity and local curvature.} The four independent-target fits were continued with the same physical model, bounds, data, and objective. An initial gradient-only continuation used 278--468 distinct candidate evaluations per fit and improved the training distances by less than 0.00025\%, but did not satisfy the stationarity criterion (\hyperref[supp-n]{Supplement N}). We then differentiated the full discrete implicit residual and gating updates, retaining the exact pinned forward solver; the reconstruction agrees with its compiled residual on perturbed states to \(2.5\times10^{-14}\) in absolute residual units. The final normalized objective is \(F(x)=D_{\mathcal T}^2(w(x))/(2D_{\mathcal T,0}^2)\) in the dimensionless box coordinates of \hyperref[supp-n]{Supplement N}. All four final candidates are interior. Their tangent gradient infinity norms are \(5.08\times10^{-11}\)--\(3.42\times10^{-10}\); independent whole-trajectory Richardson checks give norms of \(3.12\times10^{-9}\)--\(1.96\times10^{-8}\), below the original \(10^{-7}\) criterion. The full objective Hessians estimated at two resolutions are positive definite; their smallest eigenvalues range from 0.01692 to 0.40745, well above the inter-resolution spectral disagreements. Independent directional curvature checks also pass (\hyperref[supp-o]{Supplement O}). These diagnostics provide strong numerical evidence of strict local minima, rather than exact mathematical certificates or global bounds. The final objectives differ from those of the gradient-only continuation by less than \(5\times10^{-5}\) in unnormalized \(D^2\) units; the small changes in either direction reflect selection for numerical stationarity within a fixed loss tolerance. The withheld distances and physical RMS residuals in Table 8 and the figure use these final candidates. Thus extensive optimization, independent gradient checks and positive local curvature leave predictive separation essentially unchanged. The four-source experiment uses 1 ms integration; earlier 0.25 ms frozen-candidate refinement concerns the two-source experiments.

\textbf{Experimental scale.} The weighting in \(D_j^2\) is substantially more optimistic than an effective error scale of 0.03--0.10 mM for a prospective tissue experiment. The native K residuals are near that illustrative range, whereas the reduced-mobility residuals exceed its upper end. For context, Octeau et al.~\cite{octeau2018} report K-selective microelectrode calibration slopes of 52--58.2 mV per concentration decade, response times near 85 ms in their apparatus, and occasional spontaneous fluctuations of approximately 0.1 mM in slices they excluded. These fluctuations are biological/preparation variability, not a measured universal instrument noise floor. At 3 mM background K, a 0.05 mM change corresponds to about 0.37--0.42 mV at those slopes and a 0.12 mM change to about 0.89--0.99 mV, before temporal filtering. The neuronal voltage differences of 32--98 µV motivate repeated, time-locked averaging; achievable precision must be established from the actual voltage recordings and their covariance. The experimental comparison should propagate measured bandwidth, spatial sampling, drift and noise rather than interpret the illustrative distances as significance statistics.

\subsection{Geometry sensitivity at fixed total glial edge weight}\label{geometry-sensitivity-at-fixed-total-glial-edge-weight}

A second concern is that the original independent graph adds total glial transport weight when chords are introduced. A flexible fitted path tests an alternative support but need not preserve total weight. We therefore carried out a separate constant-strength intervention. Let \(A_p\) denote the unit-weight path adjacency and \(A_c\) the adjacency of the three shortcut edges with nominal weights 0.8, 0.5 and 0.6. For chord fraction \(g\in[0,1]\), define

\[A_g(\mu,g)=\frac{6\mu}{6+1.9g}(A_p+g A_c),\qquad A_g(\mu,0)=\mu A_p.\]

Every graph in this family has total undirected glial edge weight \(6\mu\) and hence the same weighted Laplacian trace \(12\mu\). Nonzero \(g\) redistributes rather than adds conductance. Equal trace is not equivalence of spectral structure, effective resistance or geometry; these differences are the object of the test. The extracellular graph and local physiology are fixed. We compare the response of each graph to that of \(\mu A_p\) for sources 0 and 2, using the same neutral high-dose protocol, timestep and \(D_j^2\) as above. Eight mobility values (0.05, 0.119, 0.179881685, 0.310754790, 0.5, 1, 2 and 4) and six chord fractions (0, 0.05, 0.10, 0.25, 0.50 and 1) give 96 source-specific comparisons. Here the control is not refitted independently for each chord fraction. In particular, \(g=1\) defines a constant-total-weight reference distinct from the independent reference of Table 8.

\begin{figure}
\centering
\includegraphics[width=0.92\textwidth,height=\textheight]{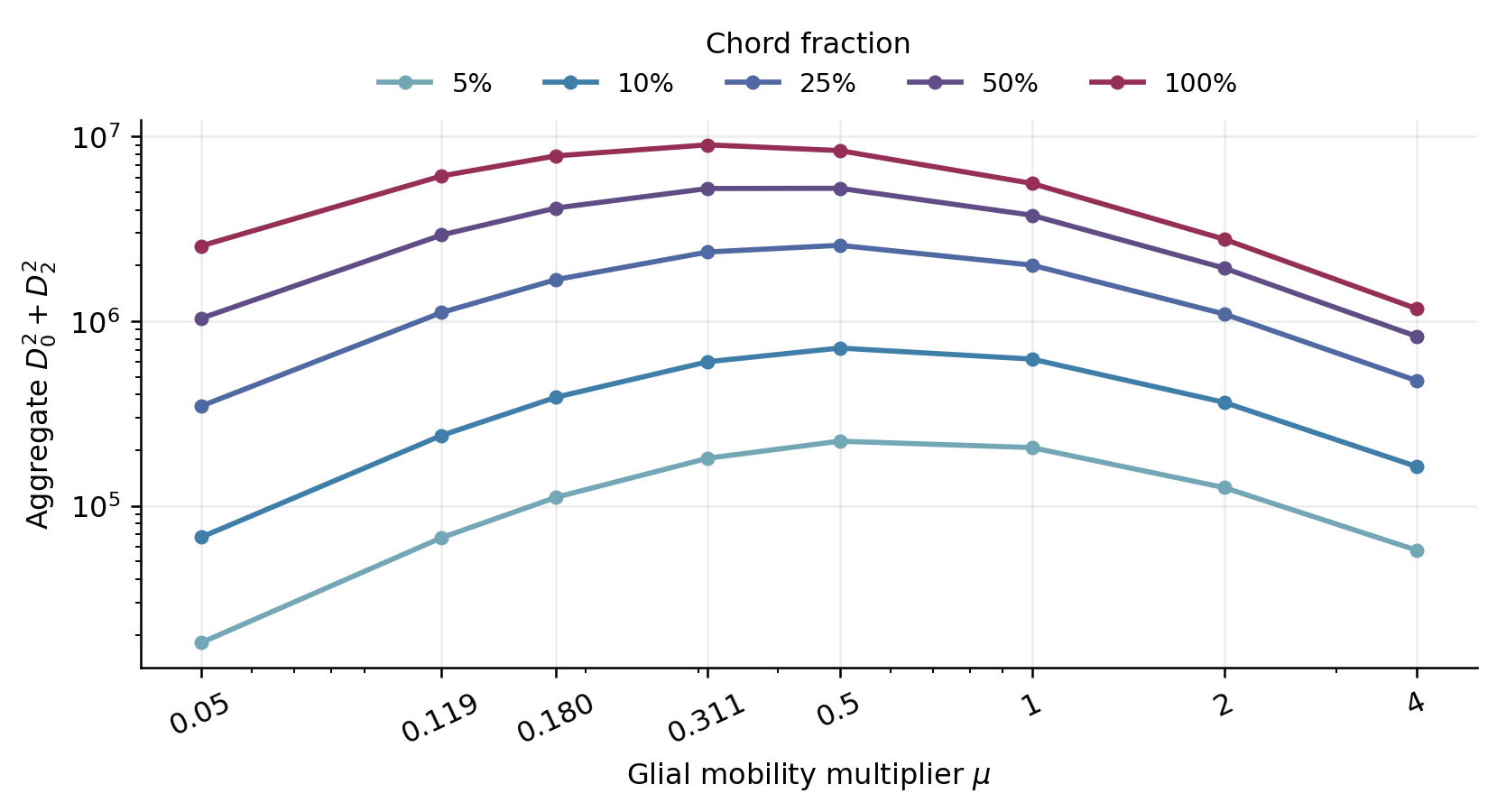}
\caption{Topology discrimination when total glial edge weight is held fixed. The aggregate squared scaled discrepancy for sources 0 and 2 is shown for five nonzero chord fractions. Zero-chord contrasts are exactly zero in all eight cases and are omitted from the logarithmic ordinate. The maxima are sampled maxima, not optimized over continuous mobility. No rival weights are re-fitted in this sweep.}
\end{figure}

All 16 zero-chord comparisons give zero discrepancy exactly, as expected. At each sampled mobility the aggregate discrepancy increases with chord fraction. The full-chord aggregate reaches a sampled maximum at \(\mu=0.310754790\) (\(9.034\times10^6\)), whereas the maxima of the four smaller nonzero chord fractions are at \(\mu=0.5\). For the 5\% chord fraction the aggregate distance is \(1.112\times10^5\) at the empirically motivated central mobility, with source-wise K RMS discrepancies of 0.02709 and 0.01291 mM and voltage RMS discrepancies of 0.02273 and 0.01625 mV. The corresponding peak sampled aggregate over all mobilities is \(2.239\times10^5\). These contrasts demonstrate that the topology-sensitive window is not a consequence solely of changing total glial edge weight or of using only strong added edges. They do not establish separation from freely optimized weak-chord alternatives: the latter experiment would require repeating the fitting procedure at each \(g\). They are also conditional on the high-dose protocol and synthetic geometry; the normalized distances remain illustrative, not experimental significance statistics. The complete grid is reported in \hyperref[supp-k]{Supplement K}.

\subsection{Aggregate CA1 screening}\label{aggregate-ca1-screening}

We fitted effective weak and strong neutral KCl source rates to the published pre-CBX means \cite{breithausen2020} in a sealed 15-site, 5-µm path, then removed glial axial coupling without changing the doses. This is an aggregate comparison of group means, not a fit to individual recordings.

\Needspace{13\baselineskip}

Table 9. Uncoupling-induced peak K change (percentage points).

\begin{longtable}[]{@{}lrr@{}}
\toprule\noalign{}
Condition & Weak pulse & Strong pulse \\
\midrule\noalign{}
\endhead
\bottomrule\noalign{}
\endlastfoot
Experiment, CBX minus control & +8.18 & +16.88 \\
EDNM native, 1 ms & +3.140 & +4.199 \\
EDNM reduced mobility, 1 ms & +2.860 & +3.756 \\
\end{longtable}

The simulations reproduce the direction but substantially underestimate the experimental effect, especially at strong loading. The native weak-pulse decay constant is 2.14 s versus 2.20 s experimentally; the native strong-pulse decay is faster. The independent near-source subgroup in Fig. 6b of \cite{breithausen2020} further emphasizes the unresolved strong-load discrepancy. The maximum coupled recorder amplitude in the separate ramps is 9.31 mM, below the reported approximately 10--15 mM regime of pronounced uncoupling effects. Geometry, source delivery, activity blockade, bath exchange and CBX selectivity differ from the experiment (\hyperref[supp-m]{Supplement M}); trace-level calibration is required.

\section{Discussion: what would identify an independent control geometry?}\label{discussion-what-would-identify-an-independent-control-geometry}

\subsection{From possibility to a model-specific biological test}\label{from-possibility-to-a-model-specific-biological-test}

The mathematics explains a potentially important dissociation: a glial graph can be structurally different and dynamically coupled yet leave almost no distinctive trace in a particular observation. At fixed homogeneous kinetics and low mobility, the graph does little redistribution. At very high mobility, connected graphs collapse toward the same uniform glial state. In between, geometry matters. Under the specified source-reconciled model, changing spatial scale can make both the local score and finite shared-support discrepancy orders of magnitude larger. This is a substantive prediction about the relationship between geometry, timescale and observability, not merely a statement that adding an edge changes an equation.

It is equally important to keep the levels separate. The passive model identifies a weighted glial operator conditional on its assumptions. The full EDNM demonstrates conservative transport, a migration-dependent shift of the sampled sensitivity window and dose-dependent nonlinear topology contrast. The sparse-source and four-source fits show why successful prediction under one intervention cannot be identified with successful recovery of transport geometry. In the bounded two-source experiments the reduced-mobility discrepancies persist across two caps and finer timesteps. In the four-source experiment the fitted paths improve prediction relative to the fixed strength-matched control, but substantial discrepancies persist at three withheld sources and in the all-source fits, including after numerical stationarity and local curvature validation. The two heterogeneous in-family controls recover both the generating weights and withheld predictions nearly exactly. Finally, the constant-total-weight chord sweep isolates a separate geometrical contribution and reveals a mobility window even for weak additional edges. The joint evidence is substantially stronger than a single-source contrast, but is still conditional on the model family, fixed local physiology, numerical solver and fitted candidates. This is an empirical-modeling caution as well as a prediction: neither rejection of one fitted rival nor successful fitting of one local intervention identifies the underlying topology. Independent aggregate physiological measurements constrain a plausible electrical parameter regime, rather than validating the particular synthetic graph. The broader architectural claim is motivated by independent anatomical and functional observations \cite{wallraff2006,pannasch2011,cooper2026,breithausen2020,bojovic2025,ma2016,kiyoshi2018}; whether the distinct organization is required to predict neuronal responses in a preparation needs a discriminating experiment. No new animal or human recordings are presented here.

A core experimental prediction should accordingly concern support. In a preparation with an independently characterized extracellular graph, the fitted glial operator should require reproducible, functionally relevant off-support coupling. The primary analysis should compare a shared-support model with free nonnegative edge weights against a prespecified independent-support model, while applying identical local kinetic, source, sensor and covariance constraints. Parameters should be trained on one subset of stimulus sites and conditions, then frozen for prediction of withheld spatial and temporal responses. A measured residual beyond uncertainty from the restricted family and a cell-specific intervention affecting the inferred route with downstream neuronal consequences would support the independent-geometry realization. A powered equivalence test placing off-support coupling and its downstream effect below prespecified biologically meaningful bounds would reject that realization in the preparation. A null result in a demonstrably insensitive experiment is not the same as such a test.

This approach is stronger than simply testing whether recovered weights are statistically nonzero: one must establish source specificity, rule out changes in effective extracellular support, and avoid confounding local membrane currents or generic latent feedback with spatial glial transport. Similarity-equivalent hidden realizations can share all recorded neuronal and extracellular transfer functions. Identifiability of the mathematical operator within a restricted model is not unrestricted identification of a cell type. Direct astroglial observations or validated interventions are essential for that final attribution.

\subsection{Experimental design and biological feasibility}\label{experimental-design-and-biological-feasibility}

The proposed sequence is first to characterize the preparation: local extracellular spread and volume fraction; astroglial coupling and relevant local uptake/return kinetics; source spatial profile and ionic/electrical dose; and neuronal synaptic/excitability controls. Potassium reporter kinetics, pH sensitivity, range and correlated noise should be measured in the exact regime, rather than assumed from published large-event detection \cite{yang2025,groschup2024}. An independently measured coupling marker can constrain the candidate graph family, but a tracer network cannot be converted directly into an absolute ionic conductivity matrix. Particularly valuable preparations would offer an astroglial route that differs from a measured extracellular transport route at the recording scale.

The rescaled-graph calculations identify candidate intrinsic transport regimes, not an electrode-spacing rule. On a fixed tissue, characterize the physical routes first and select source--sensor configurations that cover both ends of candidate conduits, with separations spanning the experimentally measured extracellular spread during the actual forcing protocol. Independently estimate effective parameters, propagate uncertainty and compare source duration and observation schedules without conflating them with changes in tissue geometry. Focal source locations and distributed inputs provide multiple spatial modes. Uniform stimulation in a sufficiently homogeneous preparation is a useful negative control; direct glial input, when selective and physically quantified, changes the pathway order and may provide a more informative design. Electrical artifacts must be monitored at onset and offset, particularly for membrane-current-driven sources. Directly applied extracellular K loads may require explicit counterion transfer and volume accounting that are different from a transmembrane K exchange.

A prospective robust design objective is

\[\mathcal D^*=\arg\max_{D\in\mathcal A}\inf_{\theta\in\Theta}I_{g\mid\eta}(D;\theta),\]

where \(\mathcal A\) is a set of experimentally feasible calibrated protocols and \(\Theta\) an independently justified parameter uncertainty set. A finite predictive-distance criterion against the flexible shared-support family is a complementary objective \cite{atkinson1975,liu2026}. The present computations do not solve this optimization over measured biological uncertainty; they show why spatial and temporal design can dominate nominal repetition counts.

For a nonlinear validation, a first step is to reproduce the contrasting low- and large-load responses of the CA1 experiments \cite{breithausen2020} with their measured focal geometry, source dose, recording positions and glial-coupling manipulation. The experimental threshold is protocol-specific; the artificial strong-glia EDNM reproduction cannot supply it. The synthetic joint-source comparisons here provide a starting design for that next step: compare independently calibrated freely weighted shared support with additional edges, test held-out responses without refitting, and quantify uncertainty in the fitted predictions. Strong-coupling or high-K regimes may increase absolute effects but may also introduce nonlinear saturation, volume changes and excitability transitions, and thus cannot simply inherit the passive information numbers.

Cell-type-specific and reversible coupling interventions with independent coupling assays and rescue are particularly useful. Connexin interventions can modify metabolite exchange, glutamate handling, volume and neuronal excitability; Kir4.1 manipulations alter more than spatial transport \cite{giaume2010,djukic2007}. Measuring those routes prevents a glial-coupling manipulation from being interpreted as a pure change in \(D_g\) when it is not. The prediction requires coupling-dependent \emph{spatial organization} beyond a generic change in total clearance.

\subsection{Implications for neuronal coordination and electrophysiology}\label{implications-for-neuronal-coordination-and-electrophysiology}

The role of glial networks in determining neuronal response geometry is compatible with the fact that neuronal membrane currents principally generate EEG and related extracellular fields. A physiologically identified glial transport graph could change the pattern of ionic states, neuronal membrane currents and thus the fields those currents generate. The required causal chain is glial transport \(\to\) extracellular and local ionic state \(\to\) neuronal current generation \(\to\) a calibrated volume-conductor or lead-field map. An EEG spectrum alone does not identify the hidden spatial controller. The original local spectral comparison omitted the glial transport operator, so it neither tests this defining prediction nor supplies evidence for it (\hyperref[supp-d]{Supplement D}). The low-frequency resonome remains a motivating hypothesis for a separately validated extension \cite{ballus2026resonome}.

\subsection{Limitations}\label{limitations}

The passive parameters are a source-reconciled reconstruction with inherited clearance, not a preparation fit or exact EDNM reduction. Synthetic graphs, fixed local physiology, volume conventions and chosen parameter bounds determine the numerical contrasts. The intrinsic-scale maxima and Fisher information are conditional design diagnostics, rather than measured optima or power estimates. The constant-strength sweep compares fixed paths, while the flexible rival fits separately test shared support. All four final independent-target fits meet the numerical stationarity and positive-curvature criteria in the stated scaled coordinates; their positive objectives remain upper bounds on the respective global minima. These results do not exclude distant competing minima or changes to the model family. The in-family controls are noiseless, same-solver tests. The four-source coverage and chord sweep use 1 ms integration, while the finer frozen-candidate tests and symmetric-volume audit concern earlier protocols. The aggregate CA1 model underestimates uncoupling effects; faithful source geometry, individual traces, independently calibrated physiology and measured noise are needed for preparation-specific validation. The conductivity constraint is an affine estimate using separate anatomical measurements and three resistance doublets. No new biological data were acquired.

\section{Conclusion}\label{conclusion}

SMM specifies independent astroglial transport support as a testable hypothesis about the spatial organization of neuronal conditions. The passive theory yields exact conditional recovery alongside severe practical ill-conditioning, and quantitative suppression of graph-shape contrast at both mobility extremes. Full electrodiffusive simulations identify axial migration as a determinant of the sensitivity window, with a conditional short-pulse maximum near 30 µm in the empirically informed regime. Four-source shared-support fits generalize better than the unfitted controls but retain structured K and neuronal-voltage errors across three withheld interventions. Noiseless in-family controls recover generating weights nearly exactly; substantial residuals survive all-source fitting, strict numerical stationarity checks and positive-definite local curvature estimates. The constant-strength experiment separates graph shape from total conductance. These results establish a quantitative discrimination protocol. The decisive biological test is whether independently characterized glial support is needed to predict spatial ionic and neuronal responses in a specified preparation.

\section{Ethics and materials availability}\label{ethics-and-materials-availability}

This study presents mathematical and computational investigations and uses published aggregate physiological observations as constraints. No new animal or human experiments were conducted, and no new biological materials were generated. The historical analysis of public de-identified EEG data is described in \hyperref[supp-d]{Supplement D}; it is not a new participant-level analysis in the present study.

\section{Generative AI assistance}\label{generative-ai-assistance}

OpenAI ChatGPT and Anthropic Claude were used to assist with language and structural editing, literature searches, code development and computational checks. The author takes responsibility for reviewing AI-assisted material and for the content, accuracy and conclusions of the manuscript.

\section{Code and data availability}\label{code-and-data-availability}

All archived SMM-I computational data and reproducibility materials supporting this study are publicly available in Zenodo, version 1.0 \cite{ballus2026dataset}, \url{https://doi.org/10.5281/zenodo.24729584}. The deposit contains 42 files (1,490,142,088 bytes), including the original simulation archives, exact driver snapshots, a selected source export and integrity manifests. The corresponding research software is available at \href{https://github.com/ipsissima/SMM/releases/tag/v2026.12.1-rc.1}{GitHub release v2026.12.1-rc.1}, pinned to commit \href{https://github.com/ipsissima/SMM/tree/83db1ffc44e5be98ebfa66ac4b9bc19efc382f6b}{\texttt{83db1ffc}}. Original numerical data are licensed under CC BY 4.0, original SMM software under MIT, and third-party materials retain their original terms. The deposit contains archival manuscript snapshots, not the current article text.

The independent EDNM implementation is available from \url{https://github.com/martejulie/electrodiffusive-network-model}, commit \texttt{826c211c70d9027a4acb02202ca91a60de79a15d}. The two-site validation and graph correction are in the SMM research repository (\url{https://github.com/ipsissima/SMM}), branches \href{https://github.com/ipsissima/SMM/tree/research/ednm-full-solver-validation-20261126}{full-solver validation} and \href{https://github.com/ipsissima/SMM/tree/research/graph-ednm-topologies-20261126}{graph topologies}. Original trajectory artifacts are associated with runs \href{https://github.com/ipsissima/SMM/actions/runs/45839123584}{45839123584}, \href{https://github.com/ipsissima/SMM/actions/runs/45840811721}{45840811721}, and \href{https://github.com/ipsissima/SMM/actions/runs/45847600616}{45847600616}. The pump clamp and high-load reproduction are in \href{https://github.com/ipsissima/SMM/tree/research/ednm-mechanism-high-load-20261126}{mechanism/high-load branch}, commit \texttt{3798c749cdbed18e4222dc805e94f28a71cde5cb}, workflow run \href{https://github.com/ipsissima/SMM/actions/runs/45859252351}{45859252351}. Relevant artifact IDs are 15137282506 and 15137766579 (pump clamp, 1 and 2 ms), 15137517741 and 15137242820 (weak and strong coupling, original 50 µs), and 15138690510 (weak coupling, 100 µs). Workflow artifacts have finite retention; the accompanying diagnostic code and machine-readable summaries identify the exact inputs and checks. The completed core trajectory arrays and checksum manifests are preserved in the associated research-data archives.

The final rival comparisons are archived in the \href{https://github.com/ipsissima/SMM/tree/research/smm-i-final-20261127/reproducibility}{reproducibility directory of the final manuscript branch}. The exact original drivers and retained response arrays are identified by the empirical fit \href{https://github.com/ipsissima/SMM/actions/runs/45881427345}{45881427345}, empirical refinement \href{https://github.com/ipsissima/SMM/actions/runs/45903481934}{45903481934}, native fit \href{https://github.com/ipsissima/SMM/actions/runs/45874768150}{45874768150}, native refinement \href{https://github.com/ipsissima/SMM/actions/runs/45904183278}{45904183278}, resumed native fit/refinement \href{https://github.com/ipsissima/SMM/actions/runs/45911020508}{45911020508}, and symmetric edge-volume audit \href{https://github.com/ipsissima/SMM/actions/runs/45887612403}{45887612403}. The two cancelled native jobs and the failed dependent refinements are retained in the provenance record; the latter failed because the cancelled jobs had not saved candidates, not because the forward integrator failed. The continuation preserves the best evaluated candidates, explicitly records nonconvergence, and completes the two frozen-candidate refinements.

The initial multi-source fits, in-family controls and chord-strength calculations are archived in \href{https://github.com/ipsissima/SMM/actions/runs/45972588944}{workflow 45972588944} at SMM commit \href{https://github.com/ipsissima/SMM/tree/9a125385cbcb02f8b0d340b3b3e936f3d131b1b9}{\texttt{9a125385}}. All 16 jobs completed. The \href{https://github.com/ipsissima/SMM/blob/9a125385cbcb02f8b0d340b3b3e936f3d131b1b9/.github/workflows/smm-i-coverage.yml}{exact workflow and embedded driver} specify the solver patch, dependency versions, graph construction, controls, optimization settings, and sampling protocol. Each original artifact includes full response arrays and machine-readable scores; fitting artifacts additionally include the evaluation trace, checkpoint, and reported terminal diagnostics. The accompanying source data include the original 16 artifacts with SHA-256 checksums, consolidated JSON and CSV records, and a script that independently recomputes the scores and regenerates the initial coverage and strength-sweep figures. \hyperref[supp-k]{Supplement K} records the exact artifact IDs and candidate-selection rules. These artifacts are computational outputs, not experimental recordings. The prior manuscript and figures are from commit \href{https://github.com/ipsissima/SMM/tree/74a6ebbc1cddce9774335d40cdf21cd644f91b18}{\texttt{74a6ebbc}}.

The gradient-only continuation is preserved in \href{https://github.com/ipsissima/SMM/actions/runs/46037902298}{workflow 46037902298}, commit \href{https://github.com/ipsissima/SMM/tree/df68b44c3f4c73f92058dd33203e587f37a7c251}{\texttt{df68b44c}}. The four final implicit-sensitivity and curvature runs are archived in \href{https://github.com/ipsissima/SMM/actions/runs/46342826627}{workflow 46342826627}, commit \href{https://github.com/ipsissima/SMM/tree/0790a96bf5a876d500de0d47af669c1d06159a0a}{\texttt{0790a96b}}. The final archives contain all seven paired response trajectories per fit, weights, exact driver and solver-patch provenance, derivative and curvature diagnostics, evaluation traces, and 15 independent directional-audit records. \hyperref[supp-o]{Supplement O} gives their identifiers. The accompanying verification script recomputes the saved response metrics, Hessian spectra, and acceptance gates from the archived records; the final predictive-discrimination figure is regenerated from the final fits and original fixed controls. The raw trajectories of the earlier experiments and exact driver snapshots are indexed in the repository validation and archival records. The archived numerical trajectories, original driver snapshots, and verification materials are available in the permanent SMM-I dataset \cite{ballus2026dataset}. Historical neuronal-population and EEG analyses in the supplementary material are not used as empirical support for spatial transport.

\clearpage
\setcounter{figure}{0}
\renewcommand{\thefigure}{S\arabic{figure}}
\renewcommand{\theHfigure}{supp.\arabic{figure}}

\section*{Supplementary material: historical model constraints and extended derivations}\label{supplementary-material}
\addcontentsline{toc}{section}{Supplementary material: historical model constraints and extended derivations}

The archived SMM-I numerical data, source snapshots and verification materials are publicly available at \url{https://doi.org/10.5281/zenodo.24729584} (Zenodo, version 1.0).

\subsection*{Supplement A. Neuronal realization and exact hidden-state reduction}\label{supp-a}
\addcontentsline{toc}{subsection}{Supplement A. Neuronal realization and exact hidden-state reduction}

The original seven-site SMM-K population model contained eight local coordinates per site,

\[X_j=(R_{E,j},V_{E,j},R_{I,j},V_{I,j},S_{EI,j},S_{IE,j},k_j,a_j)^\top,\]

for 56 states. The rate--voltage pairs are based on the Montbrió--Pazó--Roxin mean-field construction \cite{montbrio2015}. In a convenient notation the local excitatory and inhibitory populations obey

\[\begin{aligned}
\tau_b\dot R_b&=\frac{\Delta_b}{\pi\tau_b}+2R_bV_b+F_{R,b}(R,V,S),\\
\tau_b\dot V_b&=V_b^2+\eta_b(k)+F_{V,b}(R,V,S)-(\pi\tau_bR_b)^2,
\end{aligned}\]

where \(b\in\{E,I\}\), the additional functions summarize the documented synaptic and population interactions, and the source-reconciled potassium response is \(\eta_b(k)=\eta_{b0}+\chi_b\log(1+k/K_{e0})\). This response map is not independently calibrated by the mean-field equations alone. It must be related to the conductance-based parent neuron and to the operative source convention. The directed neuronal graph enters via the population interactions; the two ionic graphs enter via the balance equations.

Partition a fixed equilibrium linearization into 21 retained states \(z\) and the omitted states \(w\):

\[\begin{pmatrix}\dot z\\\dot w\end{pmatrix}=\begin{pmatrix}A_{zz}&A_{zw}\\A_{wz}&A_{ww}\end{pmatrix}\begin{pmatrix}z\\w\end{pmatrix}+\begin{pmatrix}B_z\\B_w\end{pmatrix}u.\]

Exact elimination of \(w\) gives

\[\begin{aligned}
\dot z(t)={}&A_{zz}z(t)+B_zu(t)+A_{zw}e^{A_{ww}t}w(0)\\
&+\int_0^t A_{zw}e^{A_{ww}(t-r)}[A_{wz}z(r)+B_wu(r)]\,dr .
\end{aligned}\]

Under the quasistatic approximation, if \(A_{ww}\) is invertible,

\[A_{\mathrm{DC}}=A_{zz}-A_{zw}A_{ww}^{-1}A_{wz},\qquad B_{\mathrm{DC}}=B_z-A_{zw}A_{ww}^{-1}B_w.\]

The two Schur expressions preserve the zero-frequency retained-state relation under the stated assumptions. Keeping \(B_z\) alone requires \(B_w=0\) or vanishing correction for the selected input. In general the dynamic memory kernel prevents the reduced system from preserving the entire transfer map. The earlier source reconstruction reported slow-response relative differences of approximately 0.0472\% at 1 s, 0.0122\% at 3 s and 0.000451\% at 20 s, with much larger 40 Hz errors (about 33\% for glial input and 40\% for excitatory-voltage input). Three archived values helped reconstruct the model and a fourth served as an internal holdout, not an independent experiment. The exact historical implementation's forcing matrix cannot be established from the available archive; the above formula is the corrected theoretical statement and the historical numerical results are not used to establish a frequency-valid reduction.

\subsection*{Supplement B. Physiological and energetic audits}\label{supp-b}
\addcontentsline{toc}{subsection}{Supplement B. Physiological and energetic audits}

The initial source-reconciliation exercise revealed a substantial discrepancy between the archived excitatory QIF mapping coefficient (about 0.05211 in its normalization) and the recalibrated parent-neuron estimate (about 5.04214). The corrected excitatory spike-associated excess potassium transfer was approximately \(7.0\times10^{-8}\) mol m\(^{-2}\) spike\(^{-1}\) rather than \(4.736\times10^{-8}\). These are source-model calculations, not independently measured population parameters. Inhibitory response and release remained provisional. The exchange and effective diffusivities used in the present passive calculations derive from that exercise; the original \(\kappa_N=0.232\) s\(^{-1}\) remained a separately inherited clearance slope.

The linearized neuronal feedback need not preserve the stability of the passive sector. Let \(g_E,g_I\) denote DC rate derivatives with respect to extracellular K, \(\rho_A\) the effective active membrane-area density, \(f_E\) the excitatory population fraction and \(q_E,q_I\) spike-associated release. A conditional zero-eigenvalue crossing in the homogeneous spatial mode obeys

\[\kappa_N=\rho_A[f_Eq_Eg_E+(1-f_E)q_Ig_I],\]

if the remaining block is nonsingular. This follows by setting the spatially uniform DC response to equilibrium, eliminating the local glial exchange (\(\kappa_e k=\kappa_a a\)), and equating neuronal release and clearance slopes. It is not a general Hurwitz criterion. At the historical central coefficients, the reported critical density was \(4.62424\times10^6\) m\(^2\)/m\(^3\); at \(4.929\times10^6\) the reported dominant growth rate was +0.01305 s\(^{-1}\) and at \(3.5\times10^6\) it was −0.04634 s\(^{-1}\). In later conservative extensions the reported small positive rates became +0.003232, +0.000677 and +0.000447 s\(^{-1}\) as more source pump/cotransporter terms were included, under their respective provisional closures. Introducing an uncalibrated Na-entry feedback yielded a stable illustrative point. These are historical conditional spectra, not validated tissue instability measurements. For a closed inventory-conserving model the spectrum must be checked on the fixed-inventory tangent space: if \(M x\) denotes the conserved linear inventories, restrict the linearization to \(\ker M\). The exact historical procedure is unverified; the very small reported growth rates must not be construed as verified restricted eigenvalues. This is one reason those values are supplementary and no new stability claim for the full feedback realization rests on them.

The tonic potassium release implied by the historical active membrane area and firing baseline was 1.594 mmol per L tissue/s. If all potassium is restored by neuronal Na/K ATPase with two K per ATP, the associated cost is 0.797 mmol ATP per L tissue/s. With a Na-based accounting and Na/K release ratios 1 and 1.2, the corresponding pump estimates are 0.531 and 0.638 mmol ATP per L tissue/s. Howarth et al.'s illustrative rat-cortex budget \cite{howarth2012} is 27.2 µmol/g/min, about 0.453 mmol/L tissue/s at 1 g/mL. These differing restoration assumptions require a consistent absolute flux balance before interpretation as total cost. They indicate that the original quantitative calibration cannot be accepted without resolving active surface area, spike charge and the restoration pathways. They are not an unconditional thermodynamic bound on potassium buffering.

A membrane capacitance of 1 µF/cm\(^2\) and a 100 mV voltage swing imply a monovalent capacitive charge of \(1.04\times10^{-8}\) mol/m\(^2\). The reconstructed spike potassium transfer is about 6.75 times this charge. The comparison alone does not establish sodium entry or ATP use; it motivates reconciliation of the spike-charge convention with experimentally measured low Na overlap in some mammalian pyramidal neurons \cite{carter2009}. A prior illustrative 16\% action-potential allocation reduced the original imposed 0.400 mM central K response to about 0.0201 mM and the glial-transport-specific adjacent rate contrast from 0.01322 to 0.000620 Hz. These numbers are conditional diagnostics, not a physiological calibration. A conservative absolute 91-state Na/K extension and later pump and cotransporter variants improved stoichiometry but did not independently establish the equilibrium or complete capacitive closure. This motivated use of the independent EDNM rather than treating the simplified potassium gain as sufficient physiological evidence.

\subsection*{Supplement C. General observability and earlier graph competition}\label{supp-c}
\addcontentsline{toc}{subsection}{Supplement C. General observability and earlier graph competition}

For a general linearized realization

\[\dot x=A(g,\eta)x+Bu,\qquad y=Cx+\epsilon,\qquad T(s)=C(sI-A)^{-1}B,\]

assume \(A\in\mathbb R^{m\times m}\) differentiable at the nominal point and \(B,C\) independent of \(g\). Put \(A_g=\partial_gA\). The following exact finite characterization is a useful specialization of established linear-systems realization theory \cite{gevers2017,hendrickx2019,bortner2022}. The target derivative is identically zero as a rational transfer matrix if and only if

\[M_j(A_g)=C\sum_{\ell=0}^{j-1}A^\ell A_g A^{j-1-\ell}B=0,\qquad j=1,\ldots,2m-1.\]

To prove the equivalence, define the augmented matrix \(\mathcal A=\begin{pmatrix}A&A_g\\0&A\end{pmatrix}\), augmented input \((0,B)^\top\) and output \((C,0)\). Its resolvent transfer equals \(C(sI-A)^{-1}A_g(sI-A)^{-1}B=\partial_gT\). Expansion at infinity gives the displayed Markov coefficients and a zero coefficient at power zero. Cayley--Hamilton in dimension \(2m\) gives all higher coefficients from the first \(2m\); vanishing is therefore necessary and sufficient. A nuisance direction is locally equivalent under the full transfer map if its difference from the target direction satisfies these same tests. In stiff models direct evaluation of high matrix powers is ill-conditioned; finite-window Fréchet exponential derivatives or a stable numerical alternative are preferred.

The earlier conservative seven-site GraphED realization has 70 states (nine Na/K/Cl inventories per site plus one neuronal synaptic-like state), with three distinct graphs and separate local mechanisms. Previous held-out comparisons fitted nuisance parameters on training sources and predicted withheld source protocols. The no-spatial-glia rival gave 0.00394 µM K and 0.0361 µV neuronal-voltage RMS errors relative to the synthetic reference; the shared-path rival gave 0.000711 µM and 0.00662 µV. Under the illustrative independent noise levels of 1 µM and 10 µV these are negligible. This is compatible with the unfavorable small-scale regime in the new passive experiment; the two model levels, protocols and source conventions are not numerically identical.

The earlier 70-state information design used four focal protocols and eight observation times spanning 0.25--3 s, five nuisance mechanisms and assumed 1 µM K, 10 µV voltage and 5 µM added glial K noise. Its archived nuisance-adjusted information values were \(2.9119\times10^{-10}\) (K and voltage), \(5.5835\times10^{-6}\) (added glial potassium), and \(0.0564344\) (additional idealized direct glial forcing). Nominal unit-change three-standard-deviation repetition lower bounds were \(3.09\times10^{10}\), \(1.61\times10^6\) and 159. The exact historical stimulus amplitudes are not verified in the available source, so these numbers are retained only as historical, non-dose-comparable design diagnostics. The main article uses the new fully specified passive dose and time schedule instead, and does not substitute that dose for the missing historical values.

Similarity-equivalent unobserved realizations illustrate a further limitation: for an invertible \(T_0\), the realization \((T_0AT_0^{-1},T_0B,CT_0^{-1})\) has identical transfer. The same observations cannot distinguish arbitrary interpretations of the hidden coordinates. Cellular specificity comes from separately established constraints, readouts and interventions, not names assigned to latent states.

\subsection*{Supplement D. Earlier resonome and EEG claims}\label{supp-d}
\addcontentsline{toc}{subsection}{Supplement D. Earlier resonome and EEG claims}

The original research programme's two-state cooperative calcium/IP3 kinetics do not yield a free telegraph wave simply by eliminating a variable. For \(\dot c=-\alpha c+\beta p+D\Delta c\) and \(\dot p=\gamma c-\delta p\), direct differentiation and substitution, with constant coefficients, gives

\[c_{tt}+(\alpha+\delta)c_t+(\alpha\delta-\beta\gamma)c-\delta D\Delta c-D\Delta c_t=0.\]

The spatial Laplacian term carries \(\delta D\), and the mixed space--time term is essential to the exact elimination. Discarding it is a further approximation, not an exact wave equation. Cooperative coupling by itself does not imply oscillatory poles; a passive potassium model is not a self-sustained ionic oscillator without an additional active feedback mechanism.

The earlier EEG predictor used a local two-pole ionic filter and omitted the glial interunit spatial operator. Its spectral prediction was therefore exactly independent of the defining spatial parameter. Its originally frozen primary comparison was not evaluable. The amended comparison of 535 participants gave a mean held-out score difference of approximately −0.00918 for the constrained local predictor relative to a more flexible generic two-state competitor (participant-bootstrap 95\% percentile interval approximately {[}−0.01016, −0.00823{]} in the archived score normalization). This is a negative result for that particular predictor; neither the frozen outcome nor a direct test of spatial SMM was established. Neuron-only M0 or single-feedback M1 comparative performance was not established by that analysis. The published low-frequency resonome hypothesis \cite{ballus2026resonome} therefore remains separate from the validated mathematical and computational statements in this work. The appropriate future electrophysiological model must carry the glial geometry through ionic dynamics and neuronal transmembrane currents to an independently calibrated lead field.

\subsection*{Supplement E. Independent nonlinear source-model positive control}\label{supp-e}
\addcontentsline{toc}{subsection}{Supplement E. Independent nonlinear source-model positive control}

The small-load graph and clamp experiments do not test whether the solver is capable of reproducing a large coupling-dependent neuronal transition. We therefore reproduced the published EDNM two-unit high-load protocol \cite[Table 4 and Figs. 9–11]{saetra2024}: 1 mm interunit spacing, sealed boundaries, synapses off, a depolarizing constant stimulus \(j_{\mathrm{stim}}=+2\times10^{-4}\) mM cm/s from 0.1 to 10 s, and the original 50 µs timestep. The source's strong-glia variant multiplies interunit glial Nernst--Planck fluxes by 2000, leaving the other domains and the local membrane equations unchanged. The multiplier is an artificial computational positive control, not a measurement of gap-junction conductivity. Unlike the earlier small-load negative K-transfer pulse, the positive stimulus drives neuronal depolarization and activity-associated potassium accumulation.

At 10 s, our native (weak-glia) reproduction yields an extracellular K increase of 13.644 mM at the stimulated soma, and the strong-glia reproduction yields 2.617 mM. These match the independently reported values of about 13.6 and 2.6 mM. In the native case, the voltage remains near −29.7 mV without spikes over the final second; in the strong case, voltage repeatedly traverses a spiking range (−65.9 to +9.1 mV over the final second). The strong-coupling trajectory peaks at a K elevation of 7.82 mM at 6.63 s, close to the reported peak of approximately 7.7 mM at 6.7 s. The weak-glia 100 µs run also terminates in depolarization block, with K elevated 13.677 mM; precise spike counts and the time of transition differ under timestep refinement. The original-resolution runs are used for the positive-control comparison, not for a claim of pointwise convergence of spike timing.

\Needspace{16\baselineskip}

Table S1. Original-resolution high-load reproduction against the independently published computational benchmark. Both variants use the source-published stimulus and original 50 µs timestep.

\begin{longtable}[]{@{}
  >{\raggedright\arraybackslash}p{(\columnwidth - 4\tabcolsep) * \real{0.3333}}
  >{\raggedleft\arraybackslash}p{(\columnwidth - 4\tabcolsep) * \real{0.3333}}
  >{\raggedleft\arraybackslash}p{(\columnwidth - 4\tabcolsep) * \real{0.3333}}@{}}
\toprule\noalign{}
\begin{minipage}[b]{\linewidth}\raggedright
Outcome
\end{minipage} & \begin{minipage}[b]{\linewidth}\raggedleft
Weak glia
\end{minipage} & \begin{minipage}[b]{\linewidth}\raggedleft
Strong glia
\end{minipage} \\
\midrule\noalign{}
\endhead
\bottomrule\noalign{}
\endlastfoot
Published final K elevation (mM) \cite{saetra2024} & 13.6 & 2.6 \\
Reproduced final K elevation (mM) & 13.644 & 2.617 \\
Reproduced peak K elevation (mM) & 13.644 & 7.822 \\
Voltage range in final 1 s (mV) & −29.68 to −29.62 & −65.88 to +9.13 \\
Threshold-detected spikes in final 1 s & 0 & 3 \\
Max relative ionic inventory drift & \(6.17\times10^{-9}\) & \(1.20\times10^{-8}\) \\
\end{longtable}

\begin{figure}
\centering
\includegraphics[width=0.86\textwidth,height=\textheight]{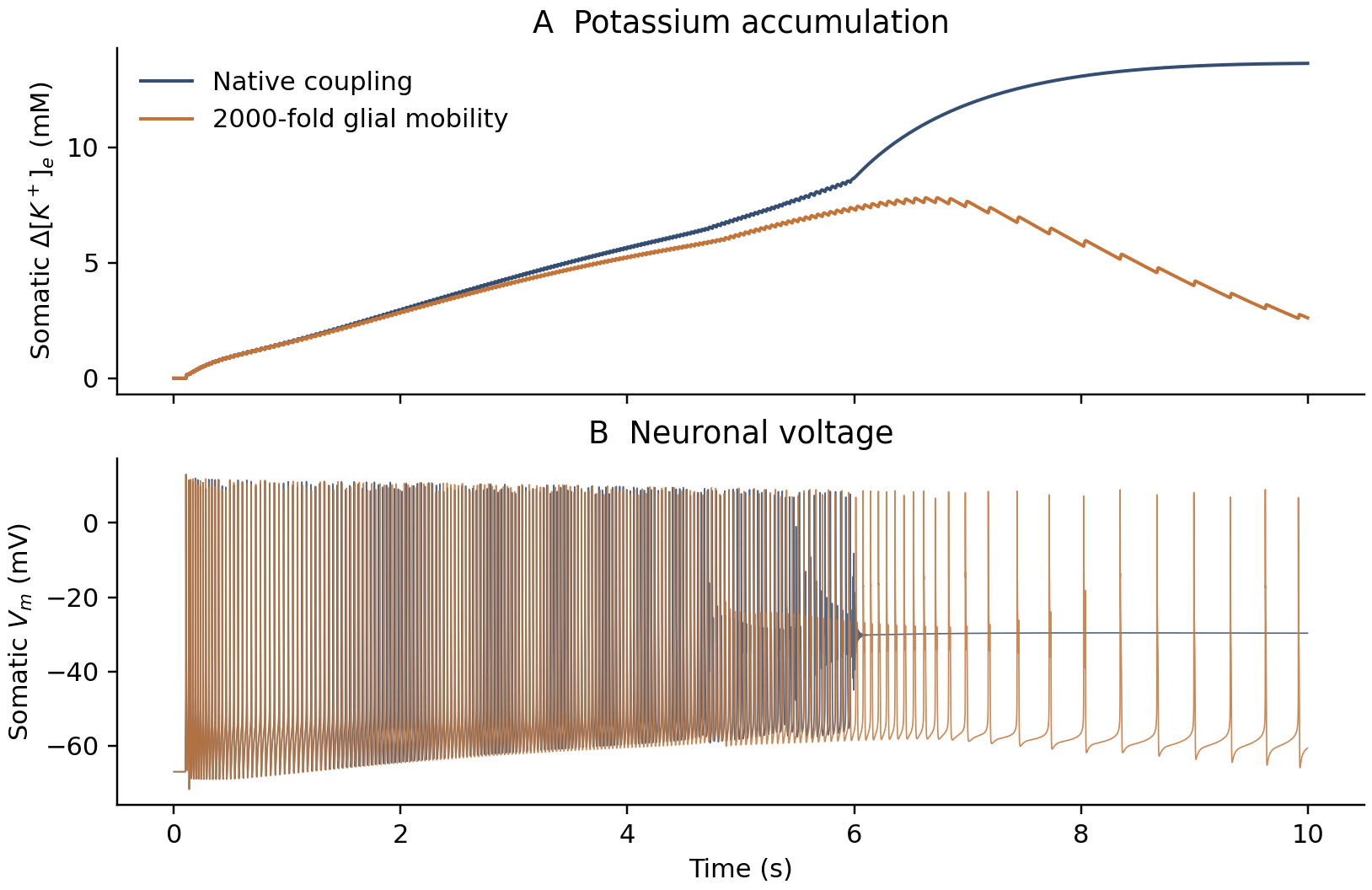}
\caption{Reproduction of the independent high-load EDNM weak/strong coupling experiment at 50 µs. Large K accumulation accompanies the native-coupling depolarization block, whereas the artificially enhanced glial mobility permits continuing spiking. Voltage traces are displayed with light decimation for legibility.}
\end{figure}

The full trajectories are finite and satisfy positive volume fractions and nonnegative concentrations (glial calcium is zero in the source). Inventory checks were repeated directly on the archived arrays. The trajectories and reconstructed values are documented in the repository validation and provenance records. These are computational source-model reproduction results. They show that the source physiology can yield strong dependence of neuronal state on glial transport under a deliberately different regime. They do not demonstrate that the synthetic shortcut edges determine that transition, that 2000-fold mobility is physiological, or that an externally applied local KCl load and connexin manipulation reproduce it. The CA1 large-load experiment \cite{breithausen2020} requires its own calibrated source, counterion/charge-balancing representation, geometry and off-target controls; it remains an experimental validation target.

\subsection*{Supplement F. Matched small-load numerical benchmark}\label{supp-f}
\addcontentsline{toc}{subsection}{Supplement F. Matched small-load numerical benchmark}

At 5 µm spacing, the seven-site full source simulation transfers potassium from the neuronal soma of site 0 into extracellular space from 0.08 to 0.40 s using \(j_{\mathrm{stim}}=-10^{-4}\) mM cm/s in the original source convention; synaptic connections are disabled. The 1 ms path run yields a stimulated-site peak extracellular elevation of 64.49 µM and a nearest-neighbor peak of 45.68 µM. Adding the three specified glial chords changes the path-distal site 6 extracellular signal by at most 0.02053 µM; the corresponding neuronal membrane contrast is about 0.021 µV. This is a very small graph-specific effect. Inventory drifts are below \(3\times10^{-13}\) relative, and the graph contrast was previously checked at 2 and 1 ms, with relative RMS discrepancies of approximately 0.51--0.52\% for K and 0.27\% for voltage. The focal membrane voltage excursion is about 32 mV despite small extracellular loads, so the full protocol is not a globally small-signal neuronal experiment.

The original passive and full model results appeared to differ by roughly two orders of magnitude when contrasts with different observation windows and normalizations were compared. Using the same 1 ms time labels, weighted graph, pulse duration, mass source and whole-field statistic removes much of that disparity. The contrast ratio between the source-reconciled passive and the complete model is now approximately 8.96. The difference remains scientifically significant. A passive model using the EDNM concentration-diffusion coefficients alone, \(D_e=\phi_e D_K/\lambda_e^2=1.53125\times10^{-10}\) m\(^2\)/s and \(D_g=\phi_a D_K/\lambda_g^2=7.65625\times10^{-11}\) m\(^2\)/s (after unit conversion), predicts an even larger contrast of 0.05219. Those coefficients omit electrical migration and are not effective fitted EDNM coefficients.

\Needspace{13\baselineskip}

Table S2. Matched relative extracellular graph contrast at 5 µm. All models use the same source-site pulse convention and full 0.8 s sampling labels; the passive model matches delivered potassium mass, not the electrical action of transmembrane injection.

\begin{longtable}[]{@{}
  >{\raggedright\arraybackslash}p{(\columnwidth - 6\tabcolsep) * \real{0.2500}}
  >{\raggedleft\arraybackslash}p{(\columnwidth - 6\tabcolsep) * \real{0.2500}}
  >{\raggedleft\arraybackslash}p{(\columnwidth - 6\tabcolsep) * \real{0.2500}}
  >{\raggedleft\arraybackslash}p{(\columnwidth - 6\tabcolsep) * \real{0.2500}}@{}}
\toprule\noalign{}
\begin{minipage}[b]{\linewidth}\raggedright
Model
\end{minipage} & \begin{minipage}[b]{\linewidth}\raggedleft
Relative contrast \(\mathcal R\)
\end{minipage} & \begin{minipage}[b]{\linewidth}\raggedleft
Peak path K (µM)
\end{minipage} & \begin{minipage}[b]{\linewidth}\raggedleft
Peak graph difference, site 6 (µM)
\end{minipage} \\
\midrule\noalign{}
\endhead
\bottomrule\noalign{}
\endlastfoot
Full EDNM & 0.0004380 & 64.49 & 0.02053 \\
Source-reconciled passive & 0.0039262 & 23.17 & 0.08865 \\
Passive, EDNM diffusion only & 0.0521922 & 45.80 & 1.31430 \\
\end{longtable}

\begin{figure}
\centering
\includegraphics[width=1\textwidth,height=\textheight]{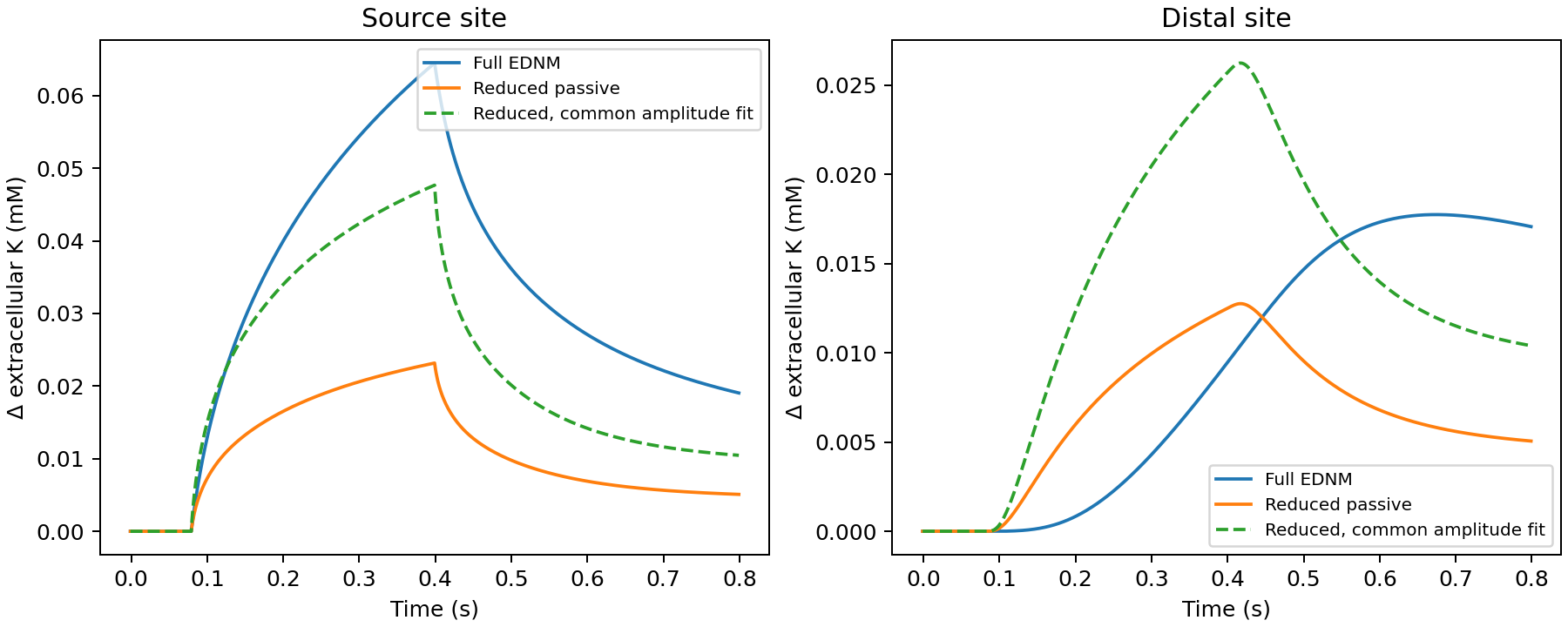}
\caption{Comparison of source-reconciled passive and full EDNM trajectories under matched small-load conditions. Different total response magnitudes and distal graph contrasts demonstrate that matching mass, geometry and protocol is insufficient to make the passive model a quantitatively equivalent physiological surrogate.}
\end{figure}

Even after fitting a single amplitude multiplier (2.057) to the passive path response, the relative whole-field RMS mismatch is 0.420. Because the graph statistic is invariant to common amplitude scaling, this mismatch is not simply a different source amplitude. The underlying glial flux also cannot be identified with a single chemical diffusion coefficient: on the three added somatic glial edges, the L2 norms of stimulus-induced electrical migration terms in the actual trajectories are about 1.97, 2.13 and 2.19 times those of their chemical-diffusion terms. The combined changes are 2.77--3.00 times the diffusion-only norms. Their correlations are positive but far from perfect. This is descriptive flux accounting, not a causal decomposition of the full passive/full-model discrepancy. In particular, migration is not a uniformly suppressive mechanism; membrane transport, electric fields, ionic feedback, and local dynamics all matter. An effective scalar mobility that would locate full EDNM on the passive sensitivity surface has not been established. The circle in the scale figure therefore marks the specified passive reference, not a pretend full-model fit.

This benchmark provides a negative quantitative constraint on strong-effect interpretations at the tested small-load protocol. The following extended sweep tests the larger-scale prediction directly in the full physiology.

\subsection*{Supplement G. Mechanism of the neighboring voltage inversion}\label{supp-g}
\addcontentsline{toc}{subsection}{Supplement G. Mechanism of the neighboring voltage inversion}

The independent two-unit EDNM challenge uses 5 µm spacing, sealed boundaries, no synapses, a 0.5 s neuronal-to-extracellular K transfer of magnitude \(10^{-4}\) mM cm/s (0.05--0.55 s), and a 2 s horizon. The original 2 ms run gives a neighboring extracellular K peak of 0.13611 mM and a peak absolute neuronal voltage response of 0.14986 mV. Turning off glial interunit transport changes those neighboring signals by at most 0.04113 µM and 0.06834 µV, respectively; relative RMS convergence of the intervention contrast from 2 to 1 ms is approximately 0.93\% and 0.69\%. Thus the overall ionic response should not be conflated with the specific contribution of interunit glial transport.

At \(t=0.4\) s in the native 2 ms run, neighboring extracellular K increases by 0.101726 mM and the K Nernst potential by 0.753641 mV, but the neighboring membrane potential decreases by 0.111094 mV. The changes in neuronal intracellular and extracellular potential are respectively \(-0.08030\) and \(+0.03080\) mV. This decomposition is a kinematic identity, not an explanation of the sign. There are two mechanistically distinguishable contributions in the original stimulus and response.

First, the imposed flux transfers positively charged ions between compartments: it is globally potassium- and charge-conserving, not locally electrically neutral. Its outward transmembrane current has magnitude \(F_{\!A}|j_{\mathrm{stim}}|=9.648\) µA/cm\(^2\) in the source units. In the 1 ms trajectories, the first stimulated timestep at the neighboring soma produces an extracellular potential change of \(+34.86\) µV while extracellular K changes by only \(+0.0220\) µM. At the onset of stimulus removal the extracellular potential falls by approximately 34.86 µV within one step, whereas K remains continuous to that temporal resolution. The rapid response is consistent with the quasistatic electrical constraints responding to imposed membrane current. It should not be interpreted as exceptionally fast chemical transport or as a topology-specific glial signal.

Second, the longer voltage response reflects a balance of electrogenic membrane processes. With outward current positive, multiplying the ionic balance by \(F_{\!A}z_i\) and summing gives at the unstimulated neighboring soma the discrete identity

\[C_m\frac{V_m^j-V_m^{j-1}}{\Delta t}=-I_{\mathrm{mem}}^{j}-I_{\mathrm{ax}}^{j}.\]

The axial term includes soma--dendrite ionic flux at the source volume and area conventions. The neuronal Na/K pump exchanges three Na for two K and generates net outward current \(F_{\!A}j_{\mathrm{pump}}\). KCC2, NKCC1 and the source two-Na/Ca exchanger are directly electroneutral by stoichiometry, although they affect the concentrations that determine other currents. Currents are reconstructed with the exact source semi-implicit convention (previous-step pump and gating factors, current-step voltage and reversal potentials). Integrating paired stimulated-minus-baseline currents over the time steps reconstructs the voltage change without fitting.

\Needspace{19\baselineskip}

Table S3. Current decomposition of the neighboring soma at 0.4 s (native 2 ms trajectory). Outward-positive current differences are in µA/cm\(^2\); integrated voltage contributions use the actual discrete capacitive balance. Totals may differ in the last displayed digit owing to rounding.

\begin{longtable}[]{@{}
  >{\raggedright\arraybackslash}p{(\columnwidth - 4\tabcolsep) * \real{0.3333}}
  >{\raggedleft\arraybackslash}p{(\columnwidth - 4\tabcolsep) * \real{0.3333}}
  >{\raggedleft\arraybackslash}p{(\columnwidth - 4\tabcolsep) * \real{0.3333}}@{}}
\toprule\noalign{}
\begin{minipage}[b]{\linewidth}\raggedright
Current component
\end{minipage} & \begin{minipage}[b]{\linewidth}\raggedleft
Instantaneous difference
\end{minipage} & \begin{minipage}[b]{\linewidth}\raggedleft
Integrated contribution (mV)
\end{minipage} \\
\midrule\noalign{}
\endhead
\bottomrule\noalign{}
\endlastfoot
Sodium leak & −0.002640 & +0.169354 \\
Gated sodium & +0.001805 & −0.116674 \\
Potassium leak & −0.021186 & +1.301747 \\
Delayed-rectifier potassium & −0.006943 & +0.436137 \\
Chloride leak & −0.009736 & +0.633226 \\
Na/K ATPase & +0.049874 & −3.019603 \\
Axial ionic current, summed & −0.010354 & +0.484720 \\
\textbf{Net} & \textbf{+0.000819} & \textbf{−0.111094} \\
\end{longtable}

The maximum difference between reconstructed and actual paired voltage across the complete trajectory is \(5.25\times10^{-9}\) mV at 2 ms and \(8.75\times10^{-9}\) mV at 1 ms. At 1 ms the integrated pump contribution is −3.01203 mV, with the potassium channels contributing +1.72835 mV and the net voltage −0.111302 mV. The pump provides the dominant negative contribution; the residual is the difference of larger opposing processes. This is exact current accounting along the trajectory, not a counterfactual on its own.

To test causality within the model, we independently modify only the extracellular K argument of the neuronal Na/K pump at the unstimulated soma, clamping it at the initial reference value 3.543992 mM while retaining sodium dependence and the pump's 3:2 stoichiometry. All other equations are unchanged; each intervention uses its own matched unstimulated baseline. The clamped neighboring response at 0.4 s becomes +0.045166 mV at 1 ms and +0.045317 mV at 2 ms, while extracellular K elevations are 0.102358 and 0.102571 mM, close to the native values of 0.101515 and 0.101726 mM. The sign reversal persists under timestep refinement. Both intervention runs finish with no flagged unconverged steps and relative ionic inventory drifts of order \(10^{-12}\) or below.

\begin{figure}
\centering
\includegraphics[width=1\textwidth,height=\textheight]{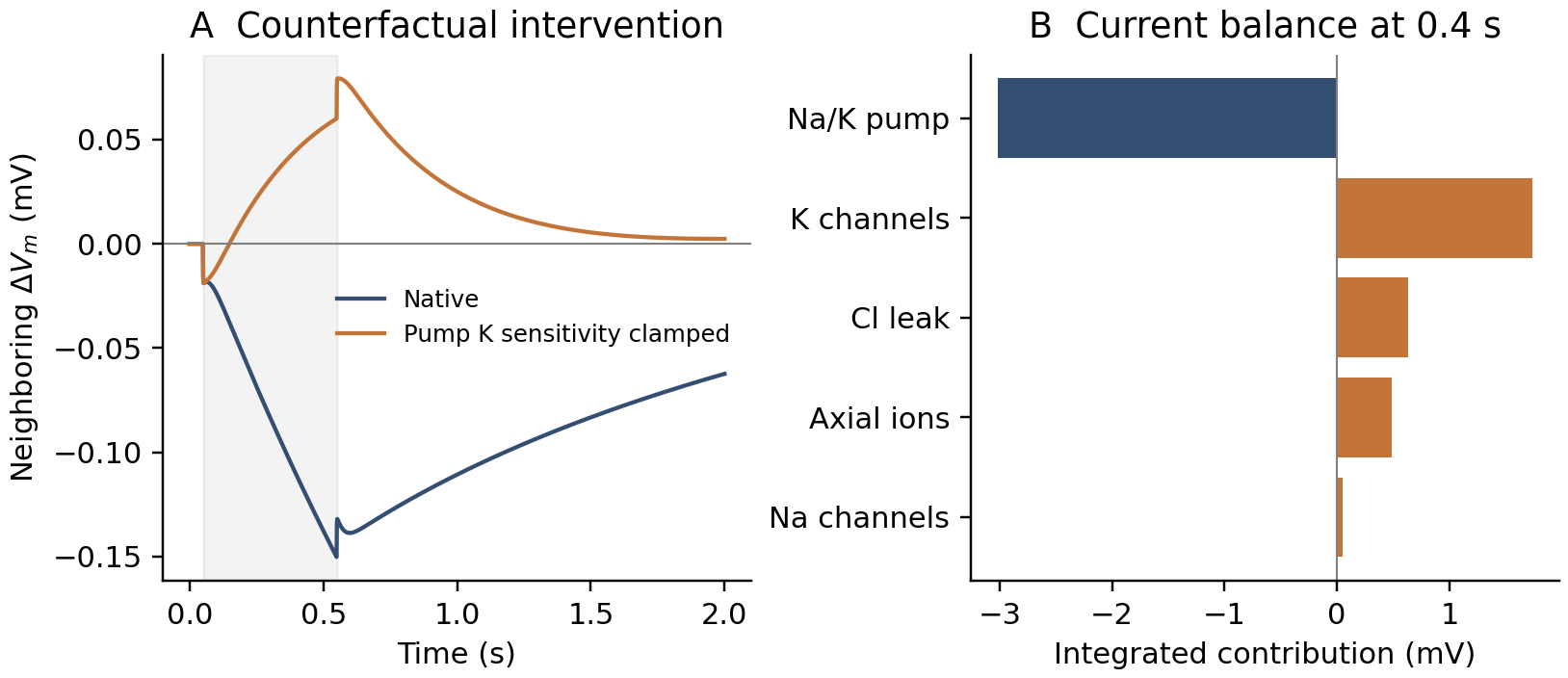}
\caption{Neighboring membrane-voltage response with native versus clamped pump K sensitivity (A), and integrated native current components at 0.4 s (B). Shading marks the stimulus interval. The clamped-pump intervention tests the causal role of K-sensitive pump activation in this model while preserving ionic stoichiometry.}
\end{figure}

The inversion is thus explained within the parent physiology by the competition between K-dependent reversal-potential shifts and electrogenic pump activation, together with other membrane and axial currents. It persists in the native glial-off control, so it is not itself evidence of distinct astroglial topology. Nor does the pump intervention prove that the identical current competition governs every biological preparation. It demonstrates why a K-only neuronal response map cannot be assumed to have the sign of the Nernst shift under all conditions.

\subsection*{Supplement H. Native-conductivity nonlinear shared-support stress tests}\label{supp-h}
\addcontentsline{toc}{subsection}{Supplement H. Native-conductivity nonlinear shared-support stress tests}

A single-source nonlinear fit using six independently adjustable nonnegative glial path weights, fixed extracellular and local physiology and bounds 0--50 is a stronger comparison than equal-total-weight normalization. At 37 µm, high load and 1 ms timestep, we fit source 0 using K and voltage at seven sites and 15 time samples, with illustrative independent scale factors 1 µM and 10 µV. Two initializations converge to the same fitted weights (approximately 1.918, 4.708, 50, 50, 50, 1.262). Three weights reach the upper bound. The training K RMS discrepancy falls from 0.177 mM for the total-weight-matched path to 0.0109 mM, while the withheld source 2 K RMS discrepancy is 0.1021 mM (0.0667 mM for the total-weight-matched path). The fitted rival therefore reproduces the training source far better but generalizes worse in K; its held-out voltage RMS is 0.0492 mV versus 0.0796 mV for the total-weight-matched path. The comparison uses the independent graph as reference. These are synthetic model-discrimination and generalization results, not independent biological validation or a certified lower bound over all shared-support models. The broader joint-source experiments below assess sensitivity to this restriction.

Allowing all six extracellular path weights to vary in addition to the six glial weights further reduces single-source training K RMS discrepancy to 0.00286 mM and voltage RMS discrepancy to 0.00512 mV. The better of two obtained fits has held-out source-2 K RMS discrepancy 0.2865 mM and voltage RMS discrepancy 0.1902 mV. Both twelve-parameter optimizations exhausted their prescribed iteration budget; the values describe obtained candidates, not converged minima. The additional flexibility illustrates a severe risk of fitting only one stimulation site.

A broader joint-source glial-only optimization fits sources 0 and 2 together and withholds source 4, with bounds 0--500. It yields training potassium RMS discrepancies 0.0422 and 0.0482 mM, and held-out potassium RMS discrepancy 0.0662 mM (maximum 0.187 mM). The held-out neuronal voltage RMS discrepancy is 0.0539 mV (maximum 0.156 mV). The fitted path weights span 2.59--217.46; the large weights are not empirically constrained. These comparisons show finite predictive separation for the fitted alternatives, not a certificate of separation from the entire shared-support cone. The joint fit allowing independent ECS weights is a further stress test, not a replacement for the more informative reduced-mobility comparison in the main article.

The reduced-mobility comparison uses the independent-support reference at \(\mu=0.17988168474408803\), \(d=37\) µm, neutral KCl rate 30 mM tissue/s and the identical timing, geometry and local EDNM equations of the main joint-source comparison. The objective is the sum of the squared distances of sources 0 and 2. Source 4 is not used for parameter or restart selection. The six rival glial weights have lower bound zero and upper bound either 0.9322643711725827 or 3.107547903908609; the six ECS weights lie in \([0.7,1.3]\). The fitting timestep is 1 ms. SciPy least\_squares uses the default bounded trust-region reflective method, relative finite-difference step 0.002, Jacobian scaling, and ftol, xtol and gtol of \(10^{-5}\), with max\_nfev 45. Evaluation counts below include finite-difference evaluations, not just optimizer major iterations. The best evaluated weight vector is retained; reported optimality and active-mask values pertain to the optimizer's own terminal point. They do not prove that the retained point satisfies a first-order optimality condition.

\Needspace{17\baselineskip}

Table S4. Original joint-source optimization records. ``T'' and ``U'' denote total-weight-matched and ordinary uniform starts. Objective values are sums over the two training sources. A budget stop is not an optimizer convergence result.

\begin{longtable}[]{@{}
  >{\raggedright\arraybackslash}p{(\columnwidth - 10\tabcolsep) * \real{0.2273}}
  >{\centering\arraybackslash}p{(\columnwidth - 10\tabcolsep) * \real{0.0682}}
  >{\raggedleft\arraybackslash}p{(\columnwidth - 10\tabcolsep) * \real{0.1932}}
  >{\raggedleft\arraybackslash}p{(\columnwidth - 10\tabcolsep) * \real{0.1591}}
  >{\raggedright\arraybackslash}p{(\columnwidth - 10\tabcolsep) * \real{0.1932}}
  >{\raggedleft\arraybackslash}p{(\columnwidth - 10\tabcolsep) * \real{0.1591}}@{}}
\toprule\noalign{}
\begin{minipage}[b]{\linewidth}\raggedright
Reference / cap
\end{minipage} & \begin{minipage}[b]{\linewidth}\centering
Start
\end{minipage} & \begin{minipage}[b]{\linewidth}\raggedleft
Training \(D^2\)
\end{minipage} & \begin{minipage}[b]{\linewidth}\raggedleft
Evaluations
\end{minipage} & \begin{minipage}[b]{\linewidth}\raggedright
Termination
\end{minipage} & \begin{minipage}[b]{\linewidth}\raggedleft
Reported optimality
\end{minipage} \\
\midrule\noalign{}
\endhead
\bottomrule\noalign{}
\endlastfoot
Reduced / 0.932 & T & 1051357.71458 & 156 & ftol, xtol & 1.164 \\
Reduced / 0.932 & U & 1051357.71437 & 169 & ftol, xtol & 0.346 \\
Reduced / 3.108 & T & 973086.73909 & 195 & ftol & 2532.98 \\
Reduced / 3.108 & U & 973086.51892 & 208 & ftol & 58.01 \\
Native / glial only, 500 & T & 493671.04870 & 63 & ftol & 3817.94 \\
Native / glial + ECS, 500 & T & 126573.55939 & 268 + 445 & runtime budget & --- \\
Native / glial + ECS, 500 & U & 126734.68599 & 270 + 437 & runtime budget & --- \\
\end{longtable}

At the lower cap the selected glial weights are approximately (0.932264, 0.459682, 0.932264, 0.766309, 0.932264, 0.932264); at the higher cap they are (3.107548, 0.312190, 1.071423, 0.652975, 0.849531, 3.107548). The corresponding extracellular vectors are (0.878753, 0.895536, 0.889534, 0.700000, 0.700000, 1.300000) and (0.821844, 0.924735, 0.877988, 0.700000, 0.700000, 1.300000). Thus the results depend partly on explicit conductance constraints, and changing those constraints is a sensitivity analysis rather than evidence of an independently measured edge distribution. The two starts yield almost identical training values and held-out responses at each cap, but the high-cap termination and reported optimality caution against calling the constrained minimum established.

For the native twelve-weight fit, the original four-hour jobs were cancelled after 268 and 270 objective evaluations. A continuation uses the best training-evaluated vector from each log, checks that its recomputed objective agrees with the original, retains the same physical model and bounds and extends the iteration allowance to 60 optimizer function evaluations and the runtime budget to 285 minutes before graceful termination. The total-matched and uniform continuations reproduce starting objectives 128347.06474 and 129483.67147 and reach 126573.55939 and 126734.68599, reductions of about 1.38\% and 2.12\%. Their final training costs differ by approximately 0.127\%. The final candidate vectors are distinct, with glial weights (90.763, 1.727, 5.735, 158.216, 3.604, 5.088) and (59.939, 1.749, 5.783, 104.284, 3.681, 5.076), respectively. Their extracellular weights remain near unity but not all within the reduced-mobility sensitivity bounds. Neither fit formally converges, and the final optimizer optimality is unavailable. These are two extended local search trajectories, not independent global optimization certificates.

\Needspace{19\baselineskip}

Table S5. Frozen-candidate refinement, source 4. Relative residuals use the independent-support reference norm. All candidates are chosen from the 1-ms optimization by training distance alone. Fine-step evaluations use unchanged weights and recheck baseline invariance, finite states, positivity, Newton convergence and ionic inventory. No refitting occurs.

\begin{longtable}[]{@{}
  >{\raggedright\arraybackslash}p{(\columnwidth - 12\tabcolsep) * \real{0.14285}}
  >{\raggedleft\arraybackslash}p{(\columnwidth - 12\tabcolsep) * \real{0.14285}}
  >{\raggedleft\arraybackslash}p{(\columnwidth - 12\tabcolsep) * \real{0.14285}}
  >{\raggedleft\arraybackslash}p{(\columnwidth - 12\tabcolsep) * \real{0.14285}}
  >{\raggedleft\arraybackslash}p{(\columnwidth - 12\tabcolsep) * \real{0.14285}}
  >{\raggedleft\arraybackslash}p{(\columnwidth - 12\tabcolsep) * \real{0.14285}}
  >{\raggedleft\arraybackslash}p{(\columnwidth - 12\tabcolsep) * \real{0.14285}}@{}}
\toprule\noalign{}
\begin{minipage}[b]{\linewidth}\raggedright
Reference / rival
\end{minipage} & \begin{minipage}[b]{\linewidth}\raggedleft
Step (ms)
\end{minipage} & \begin{minipage}[b]{\linewidth}\raggedleft
Held-out \(D^2\) (\(10^6\))
\end{minipage} & \begin{minipage}[b]{\linewidth}\raggedleft
Relative K (\%)
\end{minipage} & \begin{minipage}[b]{\linewidth}\raggedleft
Relative \(V_m\) (\%)
\end{minipage} & \begin{minipage}[b]{\linewidth}\raggedleft
K RMS (mM)
\end{minipage} & \begin{minipage}[b]{\linewidth}\raggedleft
\(V_m\) RMS (mV)
\end{minipage} \\
\midrule\noalign{}
\endhead
\bottomrule\noalign{}
\endlastfoot
Reduced / cap 0.932 & 1 & 10.706968 & 5.64914 & 13.55425 & 0.302621 & 0.145961 \\
Reduced / cap 0.932 & 0.25 & 10.699160 & 5.64858 & 13.55263 & 0.302756 & 0.146254 \\
Reduced / cap 3.108 & 1 & 11.652681 & 5.88950 & 15.11567 & 0.315497 & 0.162776 \\
Reduced / cap 3.108 & 0.25 & 11.644234 & 5.88899 & 15.11430 & 0.315642 & 0.163106 \\
Native / glial only & 1 & 0.518648 & 1.28405 & 4.68864 & 0.066213 & 0.053926 \\
Native / glial only & 0.25 & 0.518816 & 1.28474 & 4.68964 & 0.066284 & 0.054049 \\
Native / glial + ECS & 1 & 19.669857 & 7.97338 & 8.97585 & 0.411155 & 0.103236 \\
Native / glial + ECS & 0.25 & 19.660898 & 7.97364 & 8.96769 & 0.411386 & 0.103354 \\
\end{longtable}

For the native reference the ordinary path gives held-out \(D^2=1.341218\times10^6\) and the total-weight-matched path \(0.670266\times10^6\); the six-glial-weight fitted rival improves upon both, whereas the twelve-weight training-selected candidate does not. For the reduced reference the corresponding fixed controls give \(3.809459\times10^6\) and \(2.166659\times10^6\). Both bounded twelve-weight candidates predict K less accurately but voltage slightly more accurately than their total-weight-matched control at the withheld source. These adverse and favorable comparisons delimit precisely the claim: additional fitting flexibility need not transfer to a new intervention. We do not estimate the minimum over weights selected on source 4, which would violate the held-out protocol. The large illustrative whitened distances cannot be interpreted as independent experimental z-scores or p-values.

The original and resumed optimization records, including unsuccessful or budget-limited jobs, are identified in the reproducibility archive and workflow provenance. The original native refinement run had six successful forward jobs and two missing-candidate failures; the completed resumed workflow supplies the two previously missing forward evaluations. The source model, solver revision, graph patch and time protocol are pinned in the associated driver snapshots and workflow run identifiers (main article, Code and data availability).

\subsection*{Supplement I. Neutral-source early-time response}\label{supp-i}
\addcontentsline{toc}{subsection}{Supplement I. Neutral-source early-time response}

This diagnostic uses seven units at 30 µm intrinsic scale, central mobility 0.179881685, the specified independent graph and an ordinary shared path, and the neutral KCl source of 3 mM tissue/s at node 0 beginning at 0.08 s. The observation ends at 0.12 s. Glial migration is multiplied by zero or one while chemical flux and other source-model terms are preserved. The time since input is measured from the start of the first stimulated inventory step (the solver labels that first step at the stimulus onset). For each geometry the no-input trajectory is subtracted before differencing the two responses. This isolates the graph contrast, rather than the absolute response to the stimulus. Fitted slopes use samples at elapsed times greater than ten timesteps and absolute contrast above a floor of \(10^{-10}\) mM or 100 times the maximum prestimulus numerical contrast, whichever is larger; six samples are required. The floor is \(10^{-10}\) mM for all six cases. These are finite-window fits, not estimates of a rigorously established asymptotic exponent.

\Needspace{15\baselineskip}

Table S6. Least-squares slopes of log absolute extracellular K topology contrast versus log elapsed time, node 4. A dash indicates insufficient samples above the numerical floor.

\begin{longtable}[]{@{}
  >{\raggedleft\arraybackslash}p{(\columnwidth - 12\tabcolsep) * \real{0.14285}}
  >{\raggedright\arraybackslash}p{(\columnwidth - 12\tabcolsep) * \real{0.14285}}
  >{\raggedleft\arraybackslash}p{(\columnwidth - 12\tabcolsep) * \real{0.14285}}
  >{\raggedleft\arraybackslash}p{(\columnwidth - 12\tabcolsep) * \real{0.14285}}
  >{\raggedleft\arraybackslash}p{(\columnwidth - 12\tabcolsep) * \real{0.14285}}
  >{\raggedleft\arraybackslash}p{(\columnwidth - 12\tabcolsep) * \real{0.14285}}
  >{\raggedleft\arraybackslash}p{(\columnwidth - 12\tabcolsep) * \real{0.14285}}@{}}
\toprule\noalign{}
\begin{minipage}[b]{\linewidth}\raggedleft
Timestep (µs)
\end{minipage} & \begin{minipage}[b]{\linewidth}\raggedright
Migration
\end{minipage} & \begin{minipage}[b]{\linewidth}\raggedleft
0.5--2 ms
\end{minipage} & \begin{minipage}[b]{\linewidth}\raggedleft
2--5 ms
\end{minipage} & \begin{minipage}[b]{\linewidth}\raggedleft
5--10 ms
\end{minipage} & \begin{minipage}[b]{\linewidth}\raggedleft
10--20 ms
\end{minipage} & \begin{minipage}[b]{\linewidth}\raggedleft
20--40 ms
\end{minipage} \\
\midrule\noalign{}
\endhead
\bottomrule\noalign{}
\endlastfoot
100 & Retained & 2.732 & 2.541 & 2.306 & 2.160 & 2.075 \\
100 & Disabled & --- & 3.188 & 2.907 & 2.652 & 2.553 \\
50 & Retained & 2.925 & 2.594 & 2.324 & 2.167 & 2.078 \\
50 & Disabled & --- & 3.268 & 2.938 & 2.663 & 2.557 \\
25 & Retained & 3.045 & 2.623 & 2.333 & 2.171 & 2.079 \\
25 & Disabled & --- & 3.307 & 2.954 & 2.668 & 2.559 \\
\end{longtable}

All six jobs completed; maximum relative inventory error was below \(4.58\times10^{-11}\). The finite-window ordering is consistent across time resolutions. Neither the migration-off condition nor the full model without it is the strictly chemical two-compartment system of Proposition 2: extracellular migration, membrane-current pathways, local kinetics and capacitive constraints are retained. Therefore a fourth-order onset is not required in this diagnostic. The raw trajectories, exact input and analysis driver are identified by GitHub Actions run \href{https://github.com/ipsissima/SMM/actions/runs/45885207597}{45885207597}, commit \texttt{b605b58aea96e0541e2f6fb55e7e8d32b4dfc6b2}; the complete fit windows and quality diagnostics are in reproducibility/neutral\_onset\_results.json.

\subsection*{Supplement J. Edge-volume discretization sensitivity}\label{supp-j}
\addcontentsline{toc}{subsection}{Supplement J. Edge-volume discretization sensitivity}

The inherited graph flux convention uses the previous-step volume fraction at the lower-index endpoint. Replacing this factor by the arithmetic mean of the two endpoint fractions preserves equal and opposite edge inventory transfers and removes this specific index dependence. We compare both conventions at central glial mobility 0.179881685, 37 µm, 0.5 ms, neutral KCl sources 0 and 2 and rates 3 and 30 mM tissue/s. Each condition uses its own unforced baseline and the same local kinetics. The comparison concerns the independent graph versus the total-glial-edge-weight-matched path, not a refitting of the twelve-weight models. Eight configurations were completed in run \href{https://github.com/ipsissima/SMM/actions/runs/45887612403}{45887612403}, commit 87c0aa2e1b1c0c9926d1e6093bbfc08022bbc5f4.

\Needspace{13\baselineskip}

Table S7. Symmetric edge-volume audit. Percent changes are relative to the corresponding original-convention statistics, not percentage points.

\begin{longtable}[]{@{}
  >{\raggedleft\arraybackslash}p{(\columnwidth - 8\tabcolsep) * \real{0.2000}}
  >{\raggedleft\arraybackslash}p{(\columnwidth - 8\tabcolsep) * \real{0.2000}}
  >{\raggedleft\arraybackslash}p{(\columnwidth - 8\tabcolsep) * \real{0.2000}}
  >{\raggedleft\arraybackslash}p{(\columnwidth - 8\tabcolsep) * \real{0.2000}}
  >{\raggedleft\arraybackslash}p{(\columnwidth - 8\tabcolsep) * \real{0.2000}}@{}}
\toprule\noalign{}
\begin{minipage}[b]{\linewidth}\raggedleft
Source
\end{minipage} & \begin{minipage}[b]{\linewidth}\raggedleft
Rate (mM tissue/s)
\end{minipage} & \begin{minipage}[b]{\linewidth}\raggedleft
Change in relative K contrast (\%)
\end{minipage} & \begin{minipage}[b]{\linewidth}\raggedleft
Change in maximum K difference (\%)
\end{minipage} & \begin{minipage}[b]{\linewidth}\raggedleft
Change in maximum voltage difference (\%)
\end{minipage} \\
\midrule\noalign{}
\endhead
\bottomrule\noalign{}
\endlastfoot
0 & 3 & 0.00035 & 0.00238 & 0.00109 \\
0 & 30 & 0.00867 & 0.03785 & 0.01745 \\
2 & 3 & 0.00067 & 0.00150 & 0.00138 \\
2 & 30 & 0.01056 & 0.01627 & 0.01483 \\
\end{longtable}

The absolute maximum differences for the source-0 high-load condition are 0.81150981 and 0.81181700 mM K, and 0.59858773 and 0.59869215 mV voltage under the original and symmetric conventions, respectively. The complete per-site values and both rival controls are preserved in reproducibility/edge\_volume\_results.json. The changes are small for the specified comparison. This does not justify a general theorem about all variable-volume graph discretizations, and the same check would be needed for any materially different regime.

\subsection*{Supplement K. Initial expanded source coverage, in-family controls and constant-strength sweep}\label{supp-k}
\addcontentsline{toc}{subsection}{Supplement K. Initial expanded source coverage, in-family controls and constant-strength sweep}

The initial expanded-coverage workflow, \href{https://github.com/ipsissima/SMM/actions/runs/45972588944}{45972588944}, is pinned to \href{https://github.com/ipsissima/SMM/tree/9a125385cbcb02f8b0d340b3b3e936f3d131b1b9}{SMM commit \texttt{9a125385}} and \href{https://github.com/martejulie/electrodiffusive-network-model/tree/826c211c70d9027a4acb02202ca91a60de79a15d}{external EDNM commit \texttt{826c211c}}. All sixteen jobs completed successfully. The exact embedded driver is preserved in the workflow source. These calculations are separate from the earlier two-source stress tests and their frozen-candidate timestep refinement. Tables S8--S11 retain the original outputs; Supplements N--O describe subsequent optimization and the final candidates used in the main text.

The numerical protocol uses seven sites, 37 µm spacing, 1 ms timesteps, a 0.8 s horizon and paired extracellular KCl injection at 30 mM tissue/s over the discrete interval 0.08--0.28 s. The graphs are symmetric weighted adjacencies; the reference for the fitted independent-target experiments has six glial path edges of weight \(\mu\) and three chord edges with weights \(0.8\mu\), \(0.5\mu\) and \(0.6\mu\). Its extracellular graph is a unit-weight path. The shared-support negative controls instead use the six nonuniform glial weights \(\mu(7.9/7.1)(0.7,1.4,1.1,0.9,1.8,1.2)\) and the six nonuniform extracellular weights \((0.8,1.12,0.94,1.2,0.9,1.04)\). Local kinetics and all other model parameters are unchanged. Uniform and strength-matched initialization means six glial weights \(\mu\) and \(7.9\mu/6\), respectively, with extracellular weights initialized at unity. The optimizer uses the SciPy 1.12.0 bounded trust-region least-squares method, finite-difference relative step 0.002, Jacobian scaling, \texttt{max\_nfev=55}, \texttt{ftol=xtol=gtol=1e-5}, and a 285-minute graceful optimization deadline. The bounds are identical for the two starts within each mobility and kind. The trace records each distinct evaluated parameter vector and its training distance. The candidate selected is the best evaluated training point; the terminal diagnostics are taken from the optimizer's returned state, which need not be exactly that best point. No withheld source affects the objective, stopping rule or choice between starts.

The observation times are the closest stored values to 0.10, 0.15, \ldots, 0.80 s; the final stored sample is at 0.799 s. Each sampled \(D_j^2\) is the sum of squared residuals in K divided by 0.001 mM and in membrane voltage divided by 0.01 mV, with the division performed before squaring. Whole-trajectory RMS is calculated over 801 stored time labels and seven sites for each channel. These scales are illustrative weights, not estimates of experimental noise. Baseline uniformity and agreement between graph baselines are explicitly checked. Simulations reject nonfinite trajectories, nonpositive volume fractions and nonpositive concentrations for the first three ions; inventory error must be below \(10^{-7}\), and the solver rejects a Newton iteration that exhausts the iteration limit without satisfying its tolerance. Among the 56 fitted-source and 96 sweep scores independently checked against the archived response arrays, the largest reported relative inventory error is \(4.14861\times10^{-12}\). The source arrays and objective calculations were cross-checked from the downloaded artifacts; no new solver reruns or timestep-refinement runs are claimed.

\Needspace{21\baselineskip}

Table S8. Optimizer summaries. I = independent-support reference; S = shared-support reference; U = uniform start; M = total-weight-matched start. The bold starts identify the selected four-source independent-target candidates. Dashes indicate no held-out sources. The number in parentheses is the optimizer's reported number of function evaluations (distinct actual objective evaluations including finite-difference evaluations are in the archive). Optimality is the reported terminal measure. Unrounded objectives are in the source data.

\begin{longtable}[]{@{}
  >{\raggedleft\arraybackslash}p{(\columnwidth - 12\tabcolsep) * \real{0.0920}}
  >{\raggedright\arraybackslash}p{(\columnwidth - 12\tabcolsep) * \real{0.2184}}
  >{\centering\arraybackslash}p{(\columnwidth - 12\tabcolsep) * \real{0.0805}}
  >{\raggedleft\arraybackslash}p{(\columnwidth - 12\tabcolsep) * \real{0.1609}}
  >{\raggedleft\arraybackslash}p{(\columnwidth - 12\tabcolsep) * \real{0.1609}}
  >{\raggedright\arraybackslash}p{(\columnwidth - 12\tabcolsep) * \real{0.1379}}
  >{\raggedleft\arraybackslash}p{(\columnwidth - 12\tabcolsep) * \real{0.1494}}@{}}
\toprule\noalign{}
\begin{minipage}[b]{\linewidth}\raggedleft
Mobility
\end{minipage} & \begin{minipage}[b]{\linewidth}\raggedright
Reference / sites
\end{minipage} & \begin{minipage}[b]{\linewidth}\centering
Start
\end{minipage} & \begin{minipage}[b]{\linewidth}\raggedleft
Train \(D^2\)
\end{minipage} & \begin{minipage}[b]{\linewidth}\raggedleft
Test \(D^2\)
\end{minipage} & \begin{minipage}[b]{\linewidth}\raggedright
Stop (\(n\))
\end{minipage} & \begin{minipage}[b]{\linewidth}\raggedleft
Optimality
\end{minipage} \\
\midrule\noalign{}
\endhead
\bottomrule\noalign{}
\endlastfoot
1 & I / all & M & 8.269e5 & --- & ftol (7) & 717.03 \\
1 & I / four & M & 3.177e5 & 8.859e5 & ftol (13) & 57.808 \\
1 & I / four & \textbf{U} & 3.177e5 & 8.863e5 & ftol (12) & 101.47 \\
1 & S / four & U & 3.714e-13 & 3.650e-13 & xtol (13) & 0.002422 \\
0.17988 & I / all & M & 5.069e6 & --- & ftol (8) & 2559.5 \\
0.17988 & I / four & \textbf{M} & 2.215e6 & 4.775e6 & ftol (12) & 475.47 \\
0.17988 & I / four & U & 2.215e6 & 4.775e6 & ftol (12) & 987.49 \\
0.17988 & S / four & U & 7.593e-11 & 9.752e-11 & xtol (10) & 0.048227 \\
\end{longtable}

The four independent-target runs from two starts agree closely in training loss. The two all-source fits are each single-start calculations. In the native all-source case, the best evaluated and terminal parameter vectors differ by 0.01269 in maximum absolute component; for the remaining seven jobs these coincide. Although the optimizer returned success for all eight jobs, the six independent-target jobs did not satisfy the gradient criterion. Their achieved training objectives are upper bounds on the respective unknown minimum. These results do not prove that the nonlinear shared-support family cannot reproduce the reference. The in-family control weight errors are \(6.27698\times10^{-10}\) (native) and \(2.08670\times10^{-9}\) (reduced) in maximum absolute native weight units; nearly exact recovery also verifies held-out generalization in those two noiseless controls. This is a specification check using the same numerical forward solver, not a population-level false-positive estimate.

\Needspace{24\baselineskip}

Table S9. Per-source RMS discrepancies and scaled distances for training-selected independent-target four-source fits. The fixed strength-matched control and alternative-start scores are included in the machine-readable records. T = training, H = held-out.

\begin{longtable}[]{@{}rrcrrr@{}}
\toprule\noalign{}
Mobility & Source & Set & \(D_j^2\) & K RMS (mM) & Voltage RMS (mV) \\
\midrule\noalign{}
\endhead
\bottomrule\noalign{}
\endlastfoot
1 & 0 & T & 71622.7 & 0.024087 & 0.036170 \\
1 & 1 & H & 234285 & 0.044483 & 0.031711 \\
1 & 2 & T & 104215 & 0.029706 & 0.030586 \\
1 & 3 & H & 404553 & 0.058694 & 0.034609 \\
1 & 4 & T & 81044.7 & 0.026152 & 0.029343 \\
1 & 5 & H & 247463 & 0.045669 & 0.033013 \\
1 & 6 & T & 60789.1 & 0.022162 & 0.033664 \\
0.17988 & 0 & T & 526810 & 0.064784 & 0.110176 \\
0.17988 & 1 & H & 1.52833e+06 & 0.112969 & 0.096068 \\
0.17988 & 2 & T & 693717 & 0.075937 & 0.101293 \\
0.17988 & 3 & H & 1.73182e+06 & 0.120810 & 0.078788 \\
0.17988 & 4 & T & 571208 & 0.068673 & 0.100757 \\
0.17988 & 5 & H & 1.51515e+06 & 0.112326 & 0.098464 \\
0.17988 & 6 & T & 423017 & 0.057969 & 0.099042 \\
\end{longtable}

The constant-strength sweep compares the independently specified graph \(A_g=6\mu(A_p+gA_c)/(6+1.9g)\) with \(\mu A_p\) at each mobility. It conserves total undirected glial edge weight \(6\mu\), not the eigenvalues or resistances of the network. All other parameters, including the unit extracellular path, remain identical. Each row of the next table is the sum of the two source-specific distances at sites 0 and 2; every \(g=0\) comparison yields exactly zero. The complete 96-row result table, source-specific arrays and score-validation script accompany this supplement. The comparison is against a fixed strength-matched path, not against a newly optimized twelve-weight family.

\Needspace{18\baselineskip}

Table S10. Aggregate scaled distances \(D_0^2+D_2^2\) for the constant-total-weight sweep, reported in units of \(10^6\). The values are descriptive distances between synthetic predictions.

\begin{longtable}[]{@{}rrrrrrr@{}}
\toprule\noalign{}
Mobility & \(g=0\) & \(g=0.05\) & \(g=0.10\) & \(g=0.25\) & \(g=0.50\) & \(g=1\) \\
\midrule\noalign{}
\endhead
\bottomrule\noalign{}
\endlastfoot
0.05 & 0 & 0.018114 & 0.067558 & 0.346833 & 1.036812 & 2.559761 \\
0.119 & 0 & 0.067139 & 0.240524 & 1.114764 & 2.932987 & 6.126932 \\
0.179882 & 0 & 0.111216 & 0.387556 & 1.681788 & 4.108295 & 7.872950 \\
0.310755 & 0 & 0.180934 & 0.603224 & 2.371846 & 5.237286 & 9.033608 \\
0.5 & 0 & 0.223888 & 0.715543 & 2.577901 & 5.251747 & 8.407064 \\
1 & 0 & 0.206714 & 0.623662 & 2.014400 & 3.752668 & 5.581978 \\
2 & 0 & 0.125527 & 0.363514 & 1.095090 & 1.941246 & 2.784703 \\
4 & 0 & 0.057532 & 0.163304 & 0.477199 & 0.829308 & 1.171417 \\
\end{longtable}

\Needspace{26\baselineskip}

Table S11. Expanded-coverage workflow artifact index. Each ID identifies the original downloadable GitHub Actions archive. The associated checksum-indexed copies and derived summary tables are provided with the final-run source data.

\begin{longtable}[]{@{}rlcr@{}}
\toprule\noalign{}
Mobility & Protocol & Start & Artifact ID \\
\midrule\noalign{}
\endhead
\bottomrule\noalign{}
\endlastfoot
4 & Strength sweep & --- & 15186076323 \\
2 & Strength sweep & --- & 15185542680 \\
1 & Strength sweep & --- & 15185812801 \\
1 & Independent / all & M & 15198266621 \\
1 & Independent / four & M & 15202612399 \\
1 & Shared / four & U & 15202803132 \\
1 & Independent / four & U & 15203748853 \\
0.5 & Strength sweep & --- & 15185947895 \\
0.310755 & Strength sweep & --- & 15185902116 \\
0.179882 & Strength sweep & --- & 15186305617 \\
0.179882 & Independent / all & M & 15196440049 \\
0.179882 & Independent / four & U & 15199060008 \\
0.179882 & Shared / four & U & 15200707899 \\
0.179882 & Independent / four & M & 15203832567 \\
0.119 & Strength sweep & --- & 15185688415 \\
0.05 & Strength sweep & --- & 15186256009 \\
\end{longtable}

The full artifact JSON preserves unrounded quantities, complete parameter vectors, bounds, reference adjacency matrices, sample times, terminal weights, solver messages and scores for the fixed comparator graphs. The included CSV files provide source-wise access; the script \texttt{make\_coverage\_figures.py} independently recomputes sampled distances and whole-trajectory RMS from the NPZ arrays, checks the source-specific summaries, and regenerates the coverage and strength-sweep figures. The primary numerical record is the archived output, not the rounded tables. The distinction between workflow completion, solver checks, optimization termination and certified convergence applies throughout.

\subsection*{Supplement L. Exploratory predictive discrimination with two training sources}\label{supp-l}
\addcontentsline{toc}{subsection}{Supplement L. Exploratory predictive discrimination with two training sources}

A stricter comparison asks whether a rival with the same nearest-neighbor support in its glial and extracellular domains can reproduce the independent graph across \emph{different} source locations. The reference at 37 µm has the three additional glial edges specified above; local membrane kinetics, volumes, extracellular reference weights, boundary conditions and the neutral KCl source (30 mM tissue/s, 0.08--0.28 s, 0.8 s horizon) are otherwise fixed. Each candidate has six nonnegative glial path weights and, in the more flexible family, six independently adjustable extracellular path weights. Baseline invariance is checked before using the homogeneous unforced response. The numerical solver and inventory checks are those of the preceding full-model experiments.

For source \(j\), let \(Y_j^\star=(K_j^\star,V_j^\star)\) and \(Y_j(w)=(K_j(w),V_j(w))\) denote the baseline-subtracted somatic responses of the independent-support reference and candidate. At the 15 specified times nearest to the uniform grid from 0.1 to 0.8 s, across seven sites, define

\[D_j^2(w)=\sum_{t,i}\left[\frac{K_{j,i}(t;w)-K_{j,i}^\star(t)}{0.001\ \mathrm{mM}}\right]^2+\sum_{t,i}\left[\frac{V_{j,i}(t;w)-V_{j,i}^\star(t)}{0.01\ \mathrm{mV}}\right]^2.\]

The fitting objective is \(D_0^2+D_2^2\) (420 scalar residuals). Source 4 is withheld from fitting and candidate selection (210 scalar residuals); after selecting the best evaluated training candidate, its weights are frozen. The normalizations are illustrative measurement scales of 1 µM and 10 µV, not estimated experimental noise or independent samples from a biological distribution. RMS and relative trajectory residuals use all recorded times and sites; the latter divide by the independent-support reference norm, unlike the matched-path contrast convention of Table 6. We use bounded trust-region least squares with finite-difference sensitivities, ordinary uniform and total-glial-weight-matched initializations, and retain the evaluated candidate with smallest training objective. Finite objective values characterize obtained candidates; they are not certified infima of the shared-support family.

The primary comparison uses the independently motivated central mobility \(\mu=0.179881685\) for the reference, multiplying both chemical and electrical glial transport. Two sensitivity assumptions bound each rival glial weight by either 0.932264 or 3.107548 in native weight units (three or ten times the upper effective-medium endpoint); each extracellular weight is bounded between 0.7 and 1.3. The initial glial weights are \(\mu\) or \(7.9\mu/6\). These edge-wise caps are explicit modeling assumptions, not bounds inferred for individual biological junctions. Both initializations at each cap terminate on the stated objective-change criterion (the tighter cap also satisfies the step-change criterion); their training objectives agree to within \(2.3\times10^{-7}\) relative across the two starts. Several weights reach the limits, and the reported optimality diagnostics do not certify stationarity, particularly at the higher cap (\hyperref[supp-h]{Supplement H}). The selected candidates at both caps are the uniform starts, selected solely by training objective.

\Needspace{18\baselineskip}

Table S12. Joint-source nonlinear fitting at 1 ms. Training distance sums sources 0 and 2; held-out distance uses source 4. RMS discrepancies are for the held-out whole-field trajectories. Fixed controls have no fitted weights. The native regime is a distinct stress test with a different reference mobility, not a nested relaxation of the empirically informed target.

\begin{longtable}[]{@{}
  >{\raggedright\arraybackslash}p{(\columnwidth - 8\tabcolsep) * \real{0.2000}}
  >{\raggedleft\arraybackslash}p{(\columnwidth - 8\tabcolsep) * \real{0.2000}}
  >{\raggedleft\arraybackslash}p{(\columnwidth - 8\tabcolsep) * \real{0.2000}}
  >{\raggedleft\arraybackslash}p{(\columnwidth - 8\tabcolsep) * \real{0.2000}}
  >{\raggedleft\arraybackslash}p{(\columnwidth - 8\tabcolsep) * \real{0.2000}}@{}}
\toprule\noalign{}
\begin{minipage}[b]{\linewidth}\raggedright
Reference mobility and rival
\end{minipage} & \begin{minipage}[b]{\linewidth}\raggedleft
Training \(D^2\) (\(10^6\))
\end{minipage} & \begin{minipage}[b]{\linewidth}\raggedleft
Held-out \(D^2\) (\(10^6\))
\end{minipage} & \begin{minipage}[b]{\linewidth}\raggedleft
K RMS (mM)
\end{minipage} & \begin{minipage}[b]{\linewidth}\raggedleft
\(V_m\) RMS (mV)
\end{minipage} \\
\midrule\noalign{}
\endhead
\bottomrule\noalign{}
\endlastfoot
Reduced: total-weight matched & 8.68445 & 2.16666 & 0.13223 & 0.17023 \\
Reduced: 12 weights, cap 0.932 & 1.05136 & 10.70697 & 0.30262 & 0.14596 \\
Reduced: 12 weights, cap 3.108 & 0.97309 & 11.65268 & 0.31550 & 0.16278 \\
Native: total-weight matched & 4.35546 & 0.67027 & 0.07395 & 0.08610 \\
Native: 6 glial weights & 0.49367 & 0.51865 & 0.06621 & 0.05393 \\
Native: 12 weights & 0.12657 & 19.66986 & 0.41115 & 0.10324 \\
\end{longtable}

Under both reduced-mobility caps, freeing the twelve weights greatly improves training fit but worsens held-out K and the joint weighted distance relative to the simple strength-matched control. The held-out voltage RMS is slightly better than the control. Thus improving one observable or the training objective does not ensure improved joint prediction. Relative held-out discrepancies are 5.65\% and 5.89\% for K and 13.55\% and 15.12\% for voltage at the lower and higher caps, respectively. Forward reevaluation at 0.25 ms without any reoptimization yields RMS discrepancies of 0.30276 and 0.31564 mM (K) and 0.14625 and 0.16311 mV (voltage). At the higher cap the maximum absolute held-out differences are 0.77787 mM and 0.45344 mV at this finer resolution. The squared held-out distances change by less than 0.08\% under refinement. The discrepancies for these fitted candidates are therefore not a substantial timestep artifact.

\begin{figure}
\centering
\includegraphics[width=1\textwidth,height=\textheight]{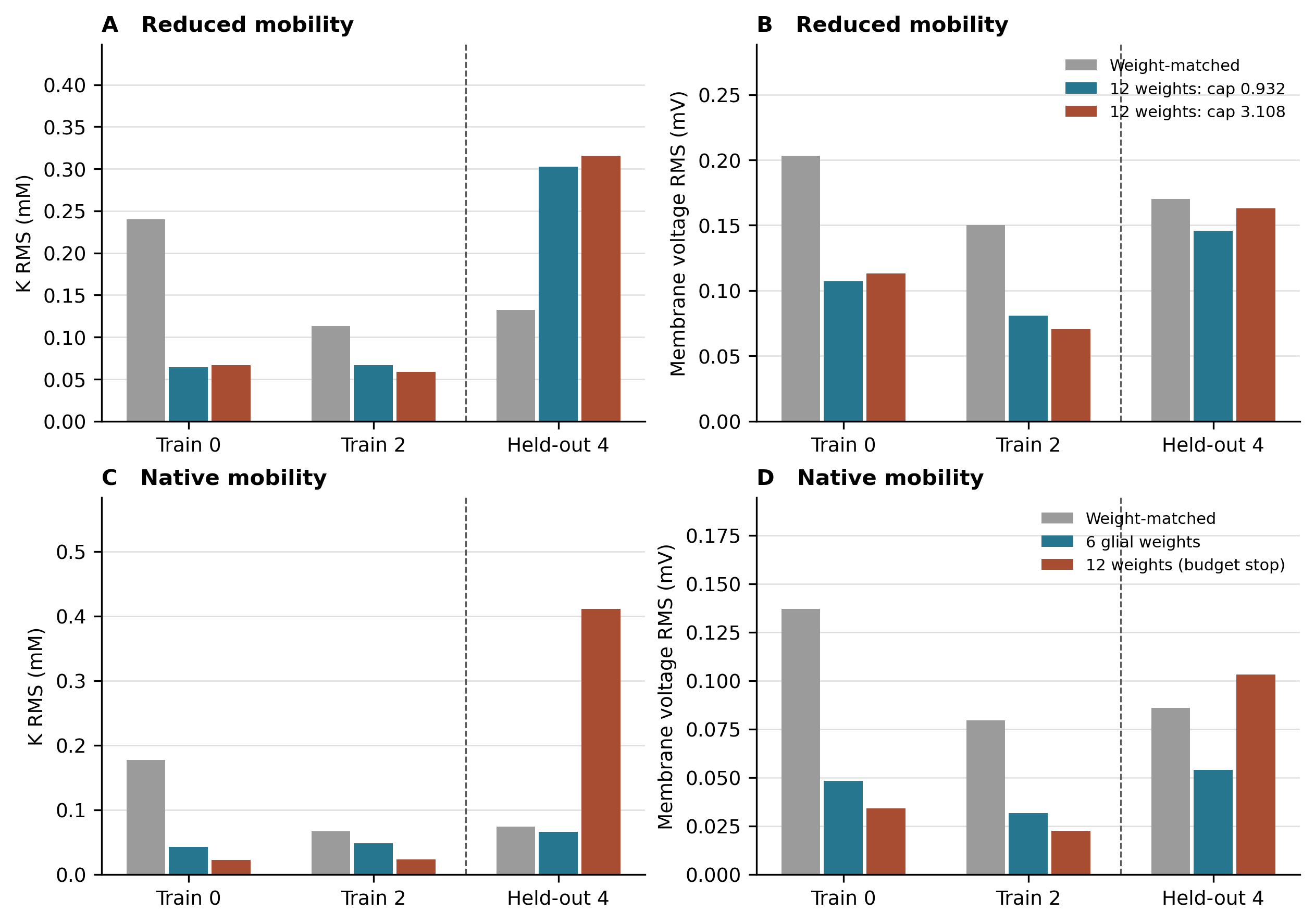}
\caption{Predictive discrimination in the neutral KCl experiment. Bars show whole-trajectory RMS discrepancies for separate training sources 0 and 2 and withheld source 4 at 1 ms. The upper panels use the reduced-mobility reference; the lower panels use the native reference. Candidate selection is by the sum of training distances only. The native twelve-weight optimization stopped at a computational budget; its bars describe the best obtained candidate, not a certified minimum. Distinct reference mobilities must not be interpreted as a nested comparison.}
\end{figure}

The secondary native-mobility stress test allows each fitted weight in \([0,500]\) with no preparation-specific conductance constraints. Six glial weights alone give smaller held-out residuals than the strength-matched control (Table S12 and \hyperref[supp-h]{Supplement H}), an important counterexample to any claim that greater flexibility must always impair generalization. With twelve adjustable weights, two warm-start continuations reproduce the earlier training checkpoints and complete a further 445 and 437 objective evaluations, respectively. Both stop at their preset runtime budgets rather than a convergence criterion; the final training distances are \(1.26574\times10^5\) and \(1.26735\times10^5\). Their largest glial weights are 158.22 and 104.28 in native units. Yet their frozen predictions on source 4 give nearly the same K residual (7.97\% and 8.00\%) and voltage residual (8.98\% and 8.99\%). The best training candidate, from the total-matched start, retains a 0.41139 mM K RMS and 0.10335 mV voltage RMS discrepancy at 0.25 ms; its maximum K discrepancy is 1.11745 mM. Despite the improved training objective, its held-out weighted distance is about 29 times that of the native total-weight-matched control. Both the modest discrepancies of the glial-only candidate and the poor transfer of the twelve-weight candidates are reported: the latter is not evidence that all less flexible rivals predict more accurately, nor is it proof that no other parameter vector can generalize.

These two-source experiments establish that accurate local fitting need not generalize to another site. Their frozen-candidate refinement establishes numerical stability of the reported discrepancies. The limited number of training sources, however, leaves open whether the observed mismatch is largely due to incomplete spatial coverage. We address that issue next using a broader source design, matched in-family controls and an all-source comparison. Such synthetic data motivate a prospective discrimination protocol. They do not identify a biological graph, show that the least possible held-out discrepancy is positive over all shared-support parameters, or provide an experimental significance level. A preparation-specific test should estimate local physiology, admissible edge weights and noise from independent measurements; use multiple independently characterized interventions; and assess predictive uncertainty without adjusting the rival on the withheld data.

\subsection*{Supplement M. Aggregate CA1 screening details}\label{supp-m}
\addcontentsline{toc}{subsection}{Supplement M. Aggregate CA1 screening details}

In a separate screening calculation, published pre-CBX means are used only to determine two effective neutral KCl source rates in a sealed 15-site, 5-µm one-dimensional path, with a recording proxy 10 µm from the source. Local EDNM physiology is unchanged, and the uncoupled responses use the fitted source rates without fitting to the blocker effects. The comparison is with the aggregate difference of the experimental CBX and no-drug group means in Fig. 5 of \cite{breithausen2020}, not a reanalysis of individual paired data.

\Needspace{13\baselineskip}

Table S13. Uncoupling-induced peak K change. Experimental differences are in percentage points; simulated changes compare the zero-glial-axial-coupling condition with its coupled control.

\begin{longtable}[]{@{}lrr@{}}
\toprule\noalign{}
Condition & Weak pulse & Strong pulse \\
\midrule\noalign{}
\endhead
\bottomrule\noalign{}
\endlastfoot
Experiment, CBX minus control & +8.18 & +16.88 \\
EDNM native, 1 ms & +3.140 & +4.199 \\
EDNM reduced mobility, 1 ms & +2.860 & +3.756 \\
\end{longtable}

The native weak-pulse fitted single-exponential decay time is 2.141 s, close to the reported pre-CBX mean 2.20 s. The native strong-pulse fitted fast/slow time constants are 0.321 and 4.629 s, compared with 1.20 and 6.79 s experimentally. The simulated spatial peak-decay lengths are 15.94 µm (weak) and 23.67 µm (strong), whereas the experimental overall decay estimate is 13.08 ± 0.79 µm. These spatial estimates are from different observation designs and are not a direct paired comparison. Both mobility conditions give closely agreeing results at 2 and 1 ms. The simulations reproduce the direction but underestimate the experimental difference in group means; neither a sharp experimental threshold nor quantitative physiological validation is established. In the separate current- and duration-ramp calculations, the largest coupled recorder peak is only about 9.31 mM, below the reported approximately 10--15 mM recorder-amplitude band at which significant uncoupling effects emerge. Thus these ramps do not test that threshold directly. The independent Fig. 6b near-source subgroup is especially instructive: at mean recorder distances of 9.20 and 9.86 µm, the reported CBX peak changes are 1.15 ± 9.49\% (weak, five observations) and 34.7 ± 10.9\% (strong, seven observations). Although pooled Fig. 5 amplitudes were used for the source fits rather than the near-source subgroup amplitudes, this subgroup emphasizes the magnitude of the unresolved high-load discrepancy. We do not interpret it as a matched quantitative prediction or test of statistical equivalence. The pooled distances, electrode geometry and delivered ion mass are not known; three-dimensional dilution and bath exchange are absent, native active neuronal physiology is retained rather than implementing the experimental activity blockade, neutral salt injection differs from iontophoresis, and complete axial glial uncoupling is not identical to CBX, which also has non-junctional neuronal effects \cite{rouach2003}. The discrepancy therefore cannot be assigned uniquely to underestimated glial transport. The individual traces are needed for a faithful calibration. These limitations concern this screening model and motivate more faithful preparation-specific modeling; they do not by themselves determine whether a distinct glial support is needed.

\subsection*{Supplement N. Gradient-only continuation of the four independent-target fits}\label{supp-n}
\addcontentsline{toc}{subsection}{Supplement N. Gradient-only continuation of the four independent-target fits}

The four runs in \href{https://github.com/ipsissima/SMM/actions/runs/46037902298}{workflow 46037902298}, commit \href{https://github.com/ipsissima/SMM/tree/df68b44c3f4c73f92058dd33203e587f37a7c251}{\texttt{df68b44c}}, resume the best evaluated training points of \hyperref[supp-k]{Supplement K}. All graph parameters, bounds, observations, solver, 1 ms timestep and objectives remain fixed. In dimensionless box coordinates \(x\in[0,1]^{12}\) with physical weights \(w=\ell+(h-\ell)\odot x\), minimize \(F(x)=D_{\mathcal T}^2(w)/(2D_{\mathcal T,0}^2)\). The scale \(D_{\mathcal T,0}^2\) is the starting training distance for each case. A centered second-order response Jacobian with coordinate step \(2\times10^{-4}\), one-sided second-order differences near bounds, and bounded trust-region least squares were used; objective and parameter-change termination were disabled, \texttt{gtol=1e-8}, \texttt{max\_nfev=120}, and a 265-minute soft deadline applied. The best training point was saved independently of termination. At the selected point, independent response Jacobians at steps \(10^{-4}\) and \(5\times10^{-5}\) were used to evaluate \(\nabla F=J^\top r\), the projected-gradient mapping \(x-\operatorname{clip}(x-\nabla F,0,1)\) and active-bound KKT sign violations. The acceptance threshold \(10^{-7}\) applied to both gradient audits and their agreement. The four selected points are interior.

Table S14. Continuation results (original unnormalized squared training and withheld distances). The gradient column is the maximum infinity norm across the two independently computed projected gradients. The disagreement column gives the infinity norm of the difference of the two gradient vectors. All four completed their output archives; none passed the numerical stationarity gate.

\begin{longtable}[]{@{}
  >{\raggedright\arraybackslash}p{(\columnwidth - 10\tabcolsep) * \real{0.1667}}
  >{\raggedleft\arraybackslash}p{(\columnwidth - 10\tabcolsep) * \real{0.1667}}
  >{\raggedleft\arraybackslash}p{(\columnwidth - 10\tabcolsep) * \real{0.1667}}
  >{\raggedleft\arraybackslash}p{(\columnwidth - 10\tabcolsep) * \real{0.1667}}
  >{\raggedleft\arraybackslash}p{(\columnwidth - 10\tabcolsep) * \real{0.1667}}
  >{\raggedleft\arraybackslash}p{(\columnwidth - 10\tabcolsep) * \real{0.1667}}@{}}
\toprule\noalign{}
\begin{minipage}[b]{\linewidth}\raggedright
Reference / sources
\end{minipage} & \begin{minipage}[b]{\linewidth}\raggedleft
Prior \(D^2\)
\end{minipage} & \begin{minipage}[b]{\linewidth}\raggedleft
Continued \(D^2\)
\end{minipage} & \begin{minipage}[b]{\linewidth}\raggedleft
Withheld \(D^2\)
\end{minipage} & \begin{minipage}[b]{\linewidth}\raggedleft
Gradient
\end{minipage} & \begin{minipage}[b]{\linewidth}\raggedleft
Disagreement
\end{minipage} \\
\midrule\noalign{}
\endhead
\bottomrule\noalign{}
\endlastfoot
native / four & 317671.666 & 317671.650 & 886082.099 & 2.825e-07 & 1.551e-07 \\
native / all & 826863.487 & 826861.496 & --- & 4.491e-07 & 5.077e-08 \\
reduced / four & 2214752.533 & 2214752.049 & 4777096.171 & 5.493e-07 & 5.526e-08 \\
reduced / all & 5069444.529 & 5069442.725 & --- & 2.118e-06 & 4.332e-08 \\
\end{longtable}

Table S15. Termination and artifact identifiers. The number of distinct evaluated candidates includes derivative evaluations.

\begin{longtable}[]{@{}lrlr@{}}
\toprule\noalign{}
Case & Evaluations & Termination & Artifact ID \\
\midrule\noalign{}
\endhead
\bottomrule\noalign{}
\endlastfoot
native / four & 393 & Evaluation limit & 15233686467 \\
native / all & 352 & Evaluation limit & 15232023666 \\
reduced / four & 468 & Evaluation limit & 15233571240 \\
reduced / all & 278 & Soft time limit & 15235623499 \\
\end{longtable}

The archived NPZ arrays were independently checked: every one of the 28 source-specific objective values, K RMS and voltage RMS values agrees with the JSON scores to relative tolerance \(10^{-12}\). The achieved training loss fell by less than 0.00025\% for each case. The native four-source held-out distance fell by about 0.025\%; the reduced four-source distance rose by about 0.038\%. The latter is not a selection criterion. All held-out physical residuals remain at their previous scales. The archived candidates, traces, solver patch, dependency versions and checksum manifest support exact inspection; the optimization stops are not convergence certificates. \hyperref[supp-o]{Supplement O} gives the completed implicit-sensitivity and curvature continuation; all four final fits satisfy the numerical acceptance criteria.

\subsection*{Supplement O. Implicit sensitivities, local curvature and final independent-target fits}\label{supp-o}
\addcontentsline{toc}{subsection}{Supplement O. Implicit sensitivities, local curvature and final independent-target fits}

\subsubsection*{Discrete tangent and curvature equations}\label{discrete-tangent-and-curvature-equations}
\addcontentsline{toc}{subsubsection}{Discrete tangent and curvature equations}

The final continuation (\href{https://github.com/ipsissima/SMM/actions/runs/46342826627}{workflow 46342826627}, \href{https://github.com/ipsissima/SMM/tree/0790a96bf5a876d500de0d47af669c1d06159a0a}{commit \texttt{0790a96b}}) starts from the four candidates of \hyperref[supp-n]{Supplement N}. It retains the exact pinned EDNM forward solver, graph and source patch, time step, source sets, observation map, residual weights and physical bounds. The objective normalization is unchanged from \hyperref[supp-n]{Supplement N}. Write the implicit step and gating update as

\[R_n(y_n,y_{n-1},q_{n-1},x)=0,\qquad q_n=Q_n(y_n,q_{n-1}),\]

where \(y\) collects the independent volume fractions, concentrations and potentials (238 unknowns at seven sites), \(q\) the six gating variables per site, and \(x\) the twelve dimensionless fitted graph coordinates. The state and gate sensitivities \(S_n=D_xy_n\) and \(T_n=D_xq_n\) are initialized at zero. When the current-state Jacobian \(A_n=D_{y_n}R_n\) is invertible on the smooth branch, implicit differentiation gives

\[A_n S_n=-\big(D_{y_{n-1}}R_n S_{n-1}+D_{q_{n-1}}R_n T_{n-1}+D_xR_n\big),\]

\[T_n=D_yQ_n S_n+D_qQ_n T_{n-1}.\]

The current-state Jacobian is constructed by complex-step differentiation of a vectorized residual reconstruction with the extracellular volume constraint included, using three spatial colors to recover the block-tridiagonal matrix for the path rival. It is not identified with the upstream Newton iteration matrix. For direction \(v\), let \(u_n=D_xy_n[v]\), \(p_n=D_xq_n[v]\), \(a_n=D_x^2y_n[v,v]\) and \(b_n=D_x^2q_n[v,v]\). Differentiating a second time gives

\[A_n a_n=-D_{y_{n-1}}R_n a_{n-1}-D_{q_{n-1}}R_n b_{n-1}-D^2R_n[\xi_n,\xi_n],\]

where \(\xi_n=(u_n,u_{n-1},p_{n-1},v)\). Similarly,

\[b_n=D_yQ_n a_n+D_qQ_n b_{n-1}+D^2Q_n[(u_n,p_{n-1}),(u_n,p_{n-1})].\]

The first directional derivatives use complex step \(10^{-20}\). The second directional terms are estimated by \(2\operatorname{Re}[f(z)-f(z+ih\xi)]/h^2\) at \(h=0.002,0.001\) with real base state \(z\), and the full Hessian is reconstructed by polarization from the twelve coordinate and 66 pairwise-sum directions. For the normalized residual vector \(r(x)\), \(F=\|r\|^2/2\), the full curvature is

\[\nabla F=J^\top r,\qquad H=J^\top J+\sum_k r_kD^2r_k,\]

not the Gauss--Newton matrix alone. The reported matrix is the Richardson combination \((4H_{0.001}-H_{0.002})/3\); inter-resolution spectral disagreement is \(\|H_{0.001}-H_{0.002}\|_2\). Sparse equilibrated linear solves are refined iteratively using extended-precision residual evaluation. The original componentwise backward-error criterion \(10^{-7}\) is retained; first- and second-tangent solve errors in the four final fits are all below \(10^{-8}\).

\subsubsection*{Checks against the forward implementation}\label{checks-against-the-forward-implementation}
\addcontentsline{toc}{subsubsection}{Checks against the forward implementation}

The numerical forward map is the pinned compiled EDNM residual, not an independently altered physiological model. Early implementation attempts showed that the checked-in source's dendritic lagged-state aliases did not reproduce the compiled residual. A separate diagnostic compared all current/previous-state combinations for membrane concentrations, potentials and volume fractions at deliberately perturbed states, without using the fitting objective. The retained reconstruction uses previous concentrations and potentials in both layers, previous soma volume fractions and current dendritic volume fractions in these membrane terms. All final training-source runs agree with the original compiled residual at the test states to absolute error below \(2.5\times10^{-14}\) (tolerance \(5\times10^{-12}+10^{-8}|R_{\rm compiled}|\)). Independent directional checks of the colored Jacobian and the gate updates are included. No forward equations were changed to obtain the final results. The preliminary failed workflows \href{https://github.com/ipsissima/SMM/actions/runs/46337901995}{46337901995} and \href{https://github.com/ipsissima/SMM/actions/runs/46338369764}{46338369764} are retained as implementation provenance; their residual-equivalence or linear-solve failures are not counted as accepted results.

The continuation uses the full-Hessian estimate for a local Newton-like step, with eigenvalue flooring, box-interior step restriction and a maximum dimensionless coordinate displacement of 0.03. A candidate can be accepted for smaller gradient if its training \(D^2\) does not increase by more than \(\max(0.002,10^{-9}D^2_{\rm continuation})\), or for training loss reduction beyond that tolerance. Each of the four final runs accepts one local polishing step; the final gradient is below the internal \(10^{-8}\) target. Thus the last selected point need not be the smallest raw floating-point objective ever evaluated. All physical residuals reported in the main article come from the final selected weights. No held-out site enters these operations.

For independent validation, complete forward trajectories are recomputed at \(x\pm hv\) for \(h=0.002,0.001,0.0005\), on each coordinate direction and three fixed-seed unit random directions. Centered differences of the residual vector and second centered differences yield first- and second-directional estimates, each Richardson-extrapolated on two nested pairs. This compares the reconstructed sensitivities with the original forward solver, rather than with a second evaluation of the same reconstruction. The stationarity gate requires all 15 audits, an interior candidate, \(\|\nabla F\|_\infty\le10^{-7}\), both independent coordinate-gradient estimates within \(10^{-7}\), their infinity-norm disagreement and disagreement with the tangent gradient at most \(10^{-7}\), and relative response-Jacobian directional errors at most \(10^{-5}\). The positive-curvature gate requires \(\lambda_{\min}(H)>\max(10\|H_{0.001}-H_{0.002}\|_2,10^{-7})\), and maximum disagreement with the independently estimated directional curvatures below \(\max(0.1\lambda_{\min}(H),10^{-6})\). All four candidates satisfy both gates. These are numerical first- and second-order diagnostics, not interval enclosures of the true derivatives. They support local minimization in the tested smooth branch and parameterization; they do not certify exact existence of a stationary point in a specified ball or exclude a remote lower minimum.

\Needspace{18\baselineskip}

Table S16. Final independent-target fits. The unnormalized distances are computed from the archived paired responses. Gradient norms and the smallest estimated full-objective Hessian eigenvalue use the dimensionless coordinates and objective normalization of \hyperref[supp-n]{Supplement N}. All four pass the stated numerical gates; a dash means no withheld set.

\begin{longtable}[]{@{}
  >{\raggedright\arraybackslash}p{(\columnwidth - 8\tabcolsep) * \real{0.2000}}
  >{\raggedleft\arraybackslash}p{(\columnwidth - 8\tabcolsep) * \real{0.2000}}
  >{\raggedleft\arraybackslash}p{(\columnwidth - 8\tabcolsep) * \real{0.2000}}
  >{\raggedleft\arraybackslash}p{(\columnwidth - 8\tabcolsep) * \real{0.2000}}
  >{\raggedleft\arraybackslash}p{(\columnwidth - 8\tabcolsep) * \real{0.2000}}@{}}
\toprule\noalign{}
\begin{minipage}[b]{\linewidth}\raggedright
Reference / sources
\end{minipage} & \begin{minipage}[b]{\linewidth}\raggedleft
Training \(D^2\)
\end{minipage} & \begin{minipage}[b]{\linewidth}\raggedleft
Withheld \(D^2\)
\end{minipage} & \begin{minipage}[b]{\linewidth}\raggedleft
Tangent gradient
\end{minipage} & \begin{minipage}[b]{\linewidth}\raggedleft
\(\lambda_{\min}(H)\)
\end{minipage} \\
\midrule\noalign{}
\endhead
\bottomrule\noalign{}
\endlastfoot
Native / four & 317671.650352 & 886082.411055 & 5.288e-11 & 0.045956 \\
Native / all & 826861.495755 & --- & 6.062e-11 & 0.236894 \\
Reduced / four & 2214752.049164 & 4777096.361534 & 5.081e-11 & 0.016920 \\
Reduced / all & 5069442.725382 & --- & 3.417e-10 & 0.407451 \\
\end{longtable}

Table S17. Independent forward-difference audit and archive identifiers. FD denotes the maximum infinity norm of the two independently estimated coordinate gradients. The inter-resolution curvature difference is the spectral norm defined above. The directional error is the largest absolute discrepancy between estimated Hessian quadratic forms and the independent directional estimates.

\begin{longtable}[]{@{}
  >{\raggedright\arraybackslash}p{(\columnwidth - 8\tabcolsep) * \real{0.2000}}
  >{\raggedleft\arraybackslash}p{(\columnwidth - 8\tabcolsep) * \real{0.2000}}
  >{\raggedleft\arraybackslash}p{(\columnwidth - 8\tabcolsep) * \real{0.2000}}
  >{\raggedleft\arraybackslash}p{(\columnwidth - 8\tabcolsep) * \real{0.2000}}
  >{\raggedleft\arraybackslash}p{(\columnwidth - 8\tabcolsep) * \real{0.2000}}@{}}
\toprule\noalign{}
\begin{minipage}[b]{\linewidth}\raggedright
Case
\end{minipage} & \begin{minipage}[b]{\linewidth}\raggedleft
FD gradient
\end{minipage} & \begin{minipage}[b]{\linewidth}\raggedleft
Curvature difference
\end{minipage} & \begin{minipage}[b]{\linewidth}\raggedleft
Directional error
\end{minipage} & \begin{minipage}[b]{\linewidth}\raggedleft
Artifact ID
\end{minipage} \\
\midrule\noalign{}
\endhead
\bottomrule\noalign{}
\endlastfoot
Native / four & 1.961e-08 & 2.853e-06 & 1.756e-04 & 15343809574 \\
Native / all & 8.035e-09 & 1.603e-06 & 5.030e-05 & 15347989304 \\
Reduced / four & 1.027e-08 & 3.527e-06 & 3.655e-05 & 15346148648 \\
Reduced / all & 3.119e-09 & 1.097e-06 & 2.006e-05 & 15346355115 \\
\end{longtable}

The maximum response-Jacobian relative directional error is \(1.19\times10^{-6}\) across the four final audits. Independently recomputing the saved seven source responses for each case reproduces all 28 sampled squared distances, whole-trajectory RMS discrepancies and maximum absolute discrepancies to relative tolerance \(10^{-12}\) (absolute tolerance \(10^{-12}\)). The archived Hessian matrices independently reproduce the reported spectra, and the independent trajectory-audit records reproduce the reported gradient and directional-curvature comparisons. The verification uses archived response arrays and numerical diagnostics; it does not rerun the underlying sensitivity or raw inventory calculations. The raw final fit archives, script \texttt{verify\_final\_fits.py}, derived \texttt{final\_fit\_scores.csv}, full unrounded \texttt{verified\_summary.json}, and figure regeneration script \texttt{make\_final\_figure.py} accompany the manuscript. The initial figures and summaries remain indexed in \hyperref[supp-k]{Supplement K}. Neither the positive achieved distance nor the local curvature estimates constitutes a global nonrepresentability theorem for the nonlinear family.

\clearpage
\pdfbookmark[0]{References}{bibliography}

\end{document}